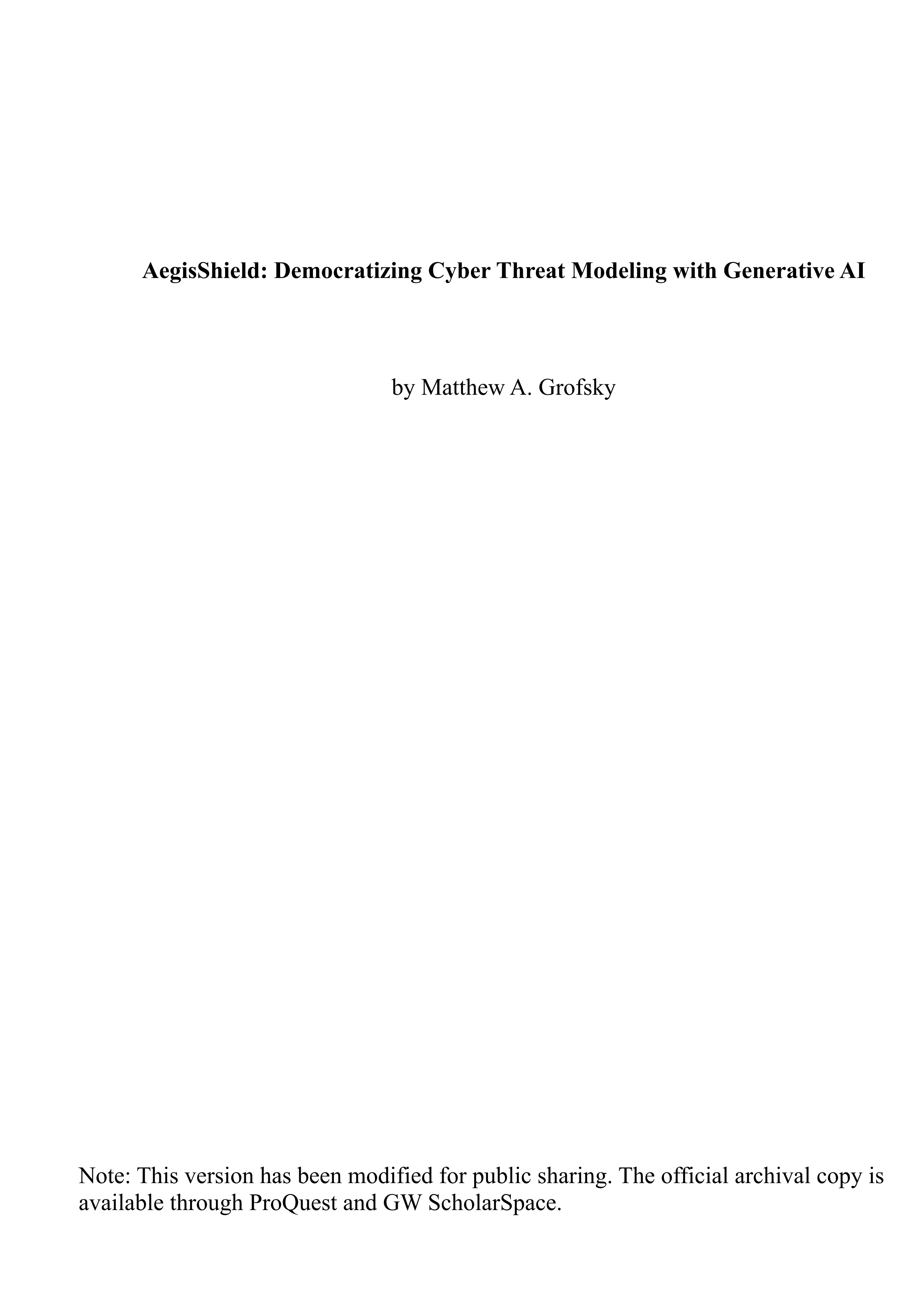

**AegisShield: Democratizing Cyber Threat Modeling with Generative AI**

by Matthew A. Grofsky

Note: This version has been modified for public sharing. The official archival copy is available through ProQuest and GW ScholarSpace.

**Abstract of Praxis**

The increasing sophistication of technology systems makes traditional threat modeling difficult to implement and scale. This is especially true for small organizations that lack resources and expertise. This research develops and evaluates AegisShield, a generative AI-enhanced threat modeling tool that implements frameworks such as STRIDE and MITRE ATT&CK to automate threat model generation and provide systematic assessments. By integrating real-time threat intelligence from sources including the National Vulnerability Database and AlienVault's Open Threat Exchange, AegisShield produces streamlined, accurate, and accessible threat descriptions. Our assessment of 243 threats from 15 case studies and over 8,000 AI-generated threats shows that AegisShield significantly reduces complexity ($p < 0.001$), produces outputs that are semantically aligned with expert-developed threats ($p < 0.05$), and achieves a statistically validated 85.4% success rate in mapping threats to MITRE ATT&CK techniques ($p < 0.001$). Simplifying threat modeling through automation and standardization helps under-resourced organizations get ahead of risks. As a result, this promotes a wider adoption of secure-by-design principles and encourages a more secure ecosystem.

## List of Acronyms

AI Artificial Intelligence

AmI Ambient Intelligence

API Application Programming Interface

ATT&CK Adversarial Tactics, Techniques, and Common Knowledge

BDD Behavior Driven Development

CAPEC Common Attack Pattern Enumeration and Classification

CI Confidence Interval

CI/CD Continuous Integration/Continuous Deployment

CLT Central Limit Theorem

CPS Cyber-Physical Systems

CTI Cyber Threat Intelligence

CVE Common Vulnerabilities and Exposures

CVSS Common Vulnerability Scoring System

CWE Common Weakness Enumeration

DaaS Drone as a Service

DREAD Damage, Reproducibility, Exploitability, Affected Users, Discoverability

DFD Data Flow Diagram

DoDCAR Department of Defense Cybersecurity Architecture Review

DSRM Design Science Research Methodology

EDA Exploratory Data Analysis

GANs Generative Adversarial Networks

HIPAAHealth Insurance Portability and Accountability Act

ICS Industrial Control Systems

IEC International Electrotechnical Commission

IoT Internet of Things

ISO International Organization for Standardization

MITRE Massachusetts Institute of Technology Research and Engineering

MQTT Message Queuing Telemetry Transport

NIPRNet Non-classified Internet Protocol Router Network

NIST National Institute of Standards and Technology

NVD National Vulnerability Database

OCTAVE Operationally Critical Threat, Asset, and Vulnerability Evaluation

OSINT Open-Source Intelligence

OTX Open Threat Exchange

PASTA Process for Attack Simulation and Threat Analysis

PDF Portable Document Format

PCI DSS Payment Card Industry Data Security Standard

RAG Retrieval-Augmented Generation

RBTM Relationship-Based Threat Modeling

SIPRNet Secret Internet Protocol Router Network

STRIDE Spoofing, Tampering, Repudiation, Information Disclosure, Denial of Service, and Elevation of Privilege

SSE System Security Engineering

| | |
|---|---|
| STS | Semantic Textual Similarity |
| SVA | Security Vulnerability Analysis |
| .govCAR | Government Cybersecurity Architecture Review |

## Glossary of Terms

**Attack Tree:** A hierarchical model used to represent potential attack paths or scenarios in cybersecurity, typically structured with possible goals at the root and attack steps as branches

**Common Vulnerability Scoring System (CVSS)**: A standardized framework for assessing the severity of software vulnerabilities based on characteristics such as exploitability and impact

**Cosine Similarity**: A metric used to measure the semantic similarity between two texts, where values range from -1 (completely dissimilar) to 1 (identical in meaning)

**Data Flow Diagram (DFD)**: A visual representation of the flow of data within a system, showing inputs, processes, storage points, and outputs

**Damage, Reproducibility, Exploitability, Affected Users, and Discoverability (DREAD):** A qualitative risk assessment model that provides a structured method to evaluate potential threats.

**Flesch-Kincaid Readability Score**: A readability test that calculates the grade level of text based on sentence length and syllable count

**Generative AI**: Artificial Intelligence systems that can create novel text, images, etc., after being trained on large sets of data

**Gherkin Syntax**: A development language for writing structured test cases used in behavior-driven development

**Internet-facing**: A system or application that is exposed to the internet

**JavaScript Object Notation (JSON)**: A simple data format that is easily readable and writable by humans and machines

**MITRE ATT&CK Framework**: A knowledge base documenting adversary tactics, techniques, and procedures used in cyberattacks (MITRE, n.d.)

**Message Queuing Telemetry Transport (MQTT)** A lightweight messaging protocol widely used in IoT applications to facilitate low bandwidth communication

**Open-source Intelligence (OSINT)**: Information collected from publicly available sources and used to enhance cybersecurity threat models

**Retrieval-Augmented Generation (RAG):** A method that improves a generative AI model's output by fetching relevant information from an external source before the model produces a response

**Semantic Textual Similarity (STS)**: The degree to which two pieces of text have the same meaning

**STRIDE**: A threat modeling methodology that identifies six types of threats, that is, spoofing, tampering, repudiation, information disclosure, denial of service, and elevation of privilege, and is commonly used to assess security vulnerabilities

# Chapter 1—Introduction

## 1.1 Background

Cybersecurity remains a problem in today's digital landscape. More than 71% of IT leaders indicate that workforce shortages have prompted their teams to adopt automated solutions for security processes within the software development lifecycle (Murtle, 2022). However, despite these automation efforts, only 10% of IT leaders feel confident that their threat modeling covers most applications. That points to a major gap in coverage (*The 2023 State of Threat Modeling*, n.d.). The increasing frequency and sophistication of cyber threats necessitate proactive measures to safeguard information systems and critical assets. Among these, threat modeling has emerged as an approach to enumerate potential security vulnerabilities. Using this approach, organizations can prioritize their security efforts by analyzing software design to address the threats before they materialize. This would naturally lead to enhanced software dependability and safety (Balamurugan et al., 2023; Suhas, 2023). This results in a win across the board.

Frameworks and methodologies such as the Operationally Critical Threat, Asset, and Vulnerability Evaluation (OCTAVE); Process for Attack Simulation and Threat Analysis (PASTA); Trike; and Spoofing, Tampering, Repudiation, Information Disclosure, Denial of Service, and Elevation of Privilege (STRIDE) offer structured approaches to threat modeling and allow organizations to evaluate and enhance their system security appropriately (Suhas, 2023). The Damage, Reproducibility, Exploitability, Affected Users, Discoverability (DREAD) model and the Common

Vulnerability Scoring System (CVSS) complement these methodologies by helping order threats based on impact.

As system complexity grows, traditional threat modeling methods including attack trees and attack graphs become more computationally intensive. This drives up both time and cost to implement. In this praxis, we use the term "attack tree" to describe hierarchical models representing potential attack paths. Attack trees align with the broader research on common cybersecurity threat modeling practices, though some researchers, such as Straub (2020), refer to similar artifacts as "threat trees." Many of these approaches and risk assessment tools can struggle in quick evolving environments and frequently rely on users' subjective judgment. Plus, we know adding security can slow things down, and even in an ideal world, people will make mistakes. (Granata & Rak, 2024; Zhang et al., 2022). A substantial number of methods are highly effective when introduced early in the development lifecycle, but they can often fail to scale effectively with more modern and complex software systems with faster development lifecycles. The challenges in applying these methodologies and tools show there is a need for more adaptive approaches to threat modeling.  A threat model approach that can handle the complexity and dynamism of contemporary cybersecurity landscapes. With all this in mind, organizations need to push for more automated methods to improve consistency, and reduce time and costs (Granata & Rak, 2024).

Modern day companies face growing vulnerabilities due to increased complexity and flexibility. The integration of cloud computing, Internet of Things (IoT), and mobile phones complicates the process of accurate vulnerability mapping to potential attack vectors and threats. Frequent changes in network configurations, software updates, and new applications also complicate threat identification and priority. These changes create a challenge to effective threat management (Adam et al., 2022). On top of this, traditional learning techniques are inadequate for vulnerability mapping at scale due to a lack of standardized vocabularies and labeled data. Because of this, advanced approaches such as machine learning (ML) and natural language processing are necessary. Complex organizations with multiple departments and security protocols can lead to gaps in consistency, and increased interconnectivity between systems can elevate risk. By utilizing systems including IBM Security X-Force Red Vulnerability Management Services and pulling in publicly available datasets and advanced analysis techniques, organizations can manage and sort vulnerabilities, achieving high accuracy and improving their overall security posture (Adam et al., 2022).

Recent, and high profile cyberattacks highlight the immediate need for effective threat modeling. In 2017, attackers breached Equifax. This breach exposed the personal information of 147 million individuals and occurred due to an unpatched vulnerability in the Apache Struts web application framework (Miyashiro, 2021). Then, in 2024, National Public Data suffered a breach. This exposed nearly 3 billion records, including Social Security numbers, addresses, and phone numbers. This National Public Data incident was worsened by the inadvertent exposure of back-end database passwords, which were freely available online (Dhaliwal, 2024; Krebs, 2024). Both attacks show how proactive threat modeling is necessary to reduce corporate vulnerabilities and protect sensitive data by quickly identifying and prioritizing critical security issues.

Meanwhile, we see the growing frequency and sophistication of cyber threats that point to an urgent need for skilled cybersecurity professionals. In the United States alone, about 600K cybersecurity positions remain open. The global total is a substantial 3.5 million. This highlights the severe shortage of skilled professionals in the marketplace exacerbating the vulnerability landscape (Smith, 2023). According to Vardhman (2024), by late 2025 cybercriminal activity will result in roughly 30,000 website breaches, and about 24,000 malicious mobile applications blocked daily. Tackling these challenges requires an increase in skilled professionals and the adoption of advanced threat detection and mitigation tools to merely keep pace with evolving threats. In response to this growing problem, regulatory requirements and industry standards, such as the National Institute of Standards and Technology (NIST) Special Publication 800-154 and the Executive Order on Improving the Nation's Cybersecurity, emphasized the value of threat modeling for resilient cybersecurity. They underscore the important role of threat modeling in solidifying cybersecurity resilience and meeting compliance standards (*Executive order on improving the nation's cybersecurity*, 2021; Shevchenko et al., 2018). Given the increasing complexity of cyber threats, companies are forced to adopt smarter methods for implementing these models.

Artificial intelligence (AI) has emerged as an enabler in current cybersecurity practices. AI plays a central role when it comes to scaling and accelerating the use of threat modeling and cybersecurity tasks. AI-enhanced tools draw upon technologies such as ML and neural networks to analyze large datasets, detect trends, predict potential security breaches in real time, and help create more accurate, efficient, and flexible models. These tools improve accuracy, speed, and accessibility, automating complex processes and reducing human error. These advancements highlight AI's advantage in cybersecurity. Neural networks demonstrate high performance with 98% accuracy in threat detection (Shahana et al., 2024).

Recent trends in AI-enhanced cybersecurity tools include the use of predictive analytics, automated response mechanisms, and resilient security frameworks (Kashyap, 2024; Shahana et al., 2024). In parallel, case studies across various industries demonstrate the effectiveness of AI in improving threat detection and response. For example, organizations have used AI systems to detect fraudulent financial transactions in real time and to protect HIPAA-protected data from cyberattacks in healthcare (Kashyap, 2024). Businesses adopting AI-driven security solutions have reported increased threat detection capabilities, reduced response times, and an improved security posture (Alevizos & Dekker, 2024; Shahana et al., 2024). Future efforts will likely focus on continued research and development to increase the scalability and reliability of AI algorithms and address the many ethical and privacy concerns in their use (Camacho, 2024; Shahana et al., 2024).

While the importance of threat modeling is evident, implementing it in a practical environment is challenging. Current threat modeling frameworks create significant hurdles due to their complexity, high costs, and the specialized talent required for effective application (Bokan & Santos, 2022; Master et al., 2022). These barriers often result in incomplete threat assessments that affect large and small organizations and pose risks to the overall cybersecurity ecosystem. As many cyber threats continue to evolve, the inefficacy of various organizations to model and mitigate them contributes to the growing global cybersecurity crisis.

## 1.2 Research Motivation

The primary purpose of this study is to democratize threat modeling and make it accessible for organizations of all sizes, particularly those that have struggled with implementing comprehensive models. As research has shown, organizations often lack skilled personnel and funding needed to use complex modeling tools (Master et al., 2022). Grosse et al. (2024) state that evolving threats add complexity, especially when organizations try to maintain continuous system updates. Addressing many of these challenges is necessary for improving organizational readiness and resilience. This research proposes a generative AI-enhanced tool, AegisShield, to simplify and speed up threat modeling, making it more accessible and effective for various organizations. By leveraging technologies such as ML and natural language processing, this tool aims to produce threat assessments while reducing the expertise and resource requirements traditionally associated with threat modeling. AegisShield's potential benefits extend beyond any one individual organization. It supports cybersecurity efforts by helping small businesses establish baseline protections (Schaad & Reski, 2019).

## 1.3 Problem Statement

*Existing threat modeling tools inadequately address organizational complexity (Schaad & Reski, 2019), and AI-focused frameworks often assume unrealistic conditions (Grosse et al., 2024). Together, these issues fuel a cyberattack cost projected to reach $10.5 trillion annually in 2025 (Aiyer et al., n.d.), highlighting the urgent need for more effective threat modeling practices.*

As previously mentioned, smaller entities often lack the years of experience and financial resources necessary for developing robust threat modeling and are highly susceptible to evolving cyber threats. This gap for businesses worsens with the growing complexity of modern software, which now includes a diverse mix of deployment models and emerging technologies, for instance, cloud computing and IoT (Kharma & Taweel, 2023).

We see threat modeling frameworks frequently fail to scale effectively with these complex systems, often exposing many smaller organizations to a wide range of threats (Bokan & Santos, 2022). Academic assumptions that overstate attacker capabilities exacerbate this blind spot, especially in AI-focused settings where researchers assume attackers can poison large portions of data or submit an unlimited number of queries (Grosse et al., 2024). In real world application, however, smaller organizations typically embed their AI components in complex pipelines or restrict inputs. These oversimplifications can inflate adversarial potential while discounting many domain specific checks. Addressing these challenges requires an approach that accounts for realistic data-access constraints, limited query opportunities, and multi-stage deployment processes to ensure solutions remain available, effective, and adaptable in the current cybersecurity landscape.

**1.4 Thesis Statement**

*A tool enhanced by generative AI will accelerate and innovate cyber threat modeling, significantly lowering the barrier to entry and democratizing the task.*

As a result, this research introduced a Python-based threat modeling tool, called AegisShield, that uses easily accessible generative AI technologies and is deployable in Streamlit. The tool simplifies complex ideas; automates lengthy processes; correlates outputs; and includes STRIDE categorization, DREAD analysis, test cases, and mitigation strategies. Integrating real-time data and organizational contexts, the script employs the MITRE Adversarial Tactics, Techniques, and Common Knowledge (ATT&CK)—a comprehensive framework for understanding adversary tactics and techniques based on real-world observations—and STRIDE frameworks for systematic threat evaluation. The generative AI component automates the creation of detailed reports; maps threats to MITRE ATT&CK techniques and STRIDE categories; and produces a DREAD analysis, mitigations, and test cases. Thus, AegisShield makes threat modeling accessible and cost-effective for organizations of all sizes; it enhances threat detection capabilities, democratizes cybersecurity, and improves security measures. The use of widely available generative AI technologies to automate this task will increase confidence. It also enables organizations to take control of their security posture.

## 1.5 Research Objectives

### *1.5.1 Democratize Threat Modeling*

The first objective of this study is to democratize threat modeling by developing a generative AI-enhanced tool that makes the process simpler and automated, and making it accessible for organizations with few resources. Known traditional methods require extensive expertise. This can lead to significant financial investment, which can then create difficult barriers for small organizations to overcome. The proposed tool will automate these complex tasks. AegisShield will reduce human error, correlate open-source intelligence (OSINT), and provide accurate threat assessments quickly. Also, by integrating frameworks such as MITRE ATT&CK and STRIDE, AegisShield will offer structured reports and effective identification and prioritization of potential security threats.

### *1.5.2 Automating Standardized and Streamlined Threat Modeling*

The second objective of this research was to address the challenges of scalability, adaptability, cost, and complexity in current traditional threat modeling techniques. These problems are even harder for small organizations that lack experienced staff. Conventional methods are labor-intensive and expensive. They also require frequent updates to keep up with emerging threats. This praxis introduces a Python-based threat modeling script built inside Streamlit that integrates real-time data and organizational contexts. The tool aims to simplify initial threat modeling phases and produce detailed quantitative analysis reports, including a DREAD analysis, mitigations, and test cases. This research promotes many secure-by-design principles by providing a scalable, adaptable, and cost-effective approach that helps organizations proactively address potential vulnerabilities while also attempting to comply with regulatory requirements and industry standards. This Generative AI enhanced approach will contribute to a more secure ecosystem by reducing the risk of cyberattacks and improving protection for assets.

## 1.6 Research Questions and Hypotheses

This practical approach examines the effectiveness of generative AI-enhanced tools in democratizing and improving threat modeling processes with the goal of making it more accessible and effective for a broader range of organizations. The approach addressed AegisShield's potential to accelerate threat identification, lower the barriers to entry, and impact cybersecurity responses. Listed below, the research questions explore how integrating generative AI with traditional threat modeling frameworks can improve upon the efficiency and accessibility of these processes, particularly for resource constrained organizations.

- RQ1: Does the integration of AegisShield with current threat modeling frameworks accelerate threat identification compared to conventional methods?
- RQ2: How can AegisShield lower the barrier to entry for organizations new to threat modeling?
- RQ3: Can simplifying threat modeling through AegisShield affect an organization's ability to respond to cyber threats?
- RQ4: Can AegisShield identify and incorporate emerging threats across diverse domains compared to established threat models?

The following hypotheses predict positive outcomes regarding the effectiveness of threat identification, simplification of descriptions, and compatibility with established cybersecurity frameworks.

- H1: AegisShield can significantly reduce the complexity of threat descriptions compared to expert-developed models.
- H2: AegisShield can generate outputs that exhibit semantic similarity with expert-developed models, as indicated by the cosine similarity using Sentence-BERT, regardless of the specific domain.
- H3: AegisShield can systematically map STRIDE-categorized threats to relevant MITRE ATT&CK Tactics, Techniques, and Procedures (TTPs), ensuring comprehensive alignment with established cybersecurity frameworks.

## 1.7 Scope of Research

This goal of this research is the development and evaluation an adaptive generative AI tool to enhance and simplify cyber threat modeling. It targets organizations that perceive themselves as under-resourced and have consequently avoided threat modeling. By using low cost and available generative AI models, this study created a tool that integrates detailed system level specifications into the threat modeling process, including the technologies used, industry specifics, and compliance requirements. It systematically applies the STRIDE and MITRE ATT&CK frameworks, incorporating open-source threat intelligence feeds to generate comprehensive threat assessments, including DREAD analyses. The tool considers organizational size and standard technologies and tailors its output to each organization's needs and constraints. This is done without regard to the resources available and is built for organizations without established threat modeling programs.

AegisShield as a tool had two main objectives. It aimed to democratize the task of threat modeling and tackle the challenges of scalability and automation. Quantitative methodologies were used, with case studies employed to evaluate the tool's effectiveness in reaching these objectives.

As noted earlier, the scope was limited to preemptive threat modeling and vulnerability assessment. Organizations that may have avoided such practices due to perceived resource constraints were targeted. This focus explicitly excluded large organizations with established threat modeling practices. It also excluded other aspects such as incident response or post-breach analysis.

In this praxis, STRIDE threat descriptions and MITRE ATT&CK framework mappings were analyzed. DREAD risk assessments, mitigation strategies, and test cases were included as supplementary outputs. They were not the focus of the analysis.

To evaluate the tool's effectiveness, threat descriptions made by AegisShield were compared to those from several case studies using semantic similarity scores and cosine similarity with Sentence-BERT (SBERT). Readability was also checked with Flesch-Kincaid scores to see how accessible the generated content was.

The dataset was obtained from various simulated organizational case study contexts and technology stacks. This enabled a comprehensive evaluation of the tool's ability to democratize threat modeling. It also helped assess its output for comprehensiveness, accuracy, efficiency, and accessibility compared to traditional approaches.

While the tool was developed for generalized use, the initial testing and evaluation focused on organizations in North America to ensure a manageable scope for the research. This allowed for a thorough assessment of the tool's impact on simplifying threat modeling, its potential to bridge the gap between perceived resource limitations and effective cybersecurity practices, and its ability to address limited region-specific regulatory challenges.

## 1.8 Research Limitations

While this research was performed to provide a comprehensive approach to simplify threat modeling through generative AI, there were several limitations:

- Geographical constraint: The evaluation focused on North American organizations may have limited the generalizability of findings to other regions with different regulatory requirements.
- Organizational size and sector: The study primarily targeted organizations that perceived themselves as under-resourced and therefore excluded insights from larger enterprises or those with more established cybersecurity practices. This focus may also partially reflect the actual resource disparities among different organizations, affecting the applicability of the findings in other contexts.
- Technological limitations: AegisShield's effectiveness depends on the rapidly evolving state of generative AI technologies (such as those offered by OpenAI, Google, and Anthropic). Future advancements may affect its relative efficacy, both positively and negatively. This study only utilized OpenAI's GPT-4o in its analysis.
- Data privacy and security: Using organizational data in the AI model raises potential privacy and security concerns, which may have limited the depth of information some organizations were willing to provide for threat modeling. The tool instructs users to not provide employee or organizational names and any identifying information. Only publicly available information was utilized.
- Bias in AI models: The underlying AI models used by OpenAI may contain inherent biases that could have influenced the threat assessments, requiring ongoing evaluation and mitigation strategies.

- Evolving threat landscape: The dynamic nature of cyber threats means that the tool's effectiveness may vary over time, requiring continuous updates and validation.
- User expertise: While the tool aims to lower barriers to entry, a basic level of cybersecurity knowledge would still be required for optimal use and interpretation of the results.
- Validation time frame: The study's time frame may have only partially captured the long-term effectiveness of the AI-enhanced threat modeling approach.
- Regulatory compliance: Real-world compliance needs may differ and evolve beyond the scope of the tool's capabilities and is solely dependent on the training within the model used.
- Dependence on external resources: AegisShield relies on the availability and accuracy of external data sources like the National Vulnerability Database (NVD) and AlienVault Open Threat Exchange (OTX). Any changes to access or quality of these data sources could impact AegisShield's effectiveness.

These limitations provide important context for interpreting all findings and help highlight areas for future study.

### 1.9 Organization of the Praxis

This paper has five main chapters. Each one focuses on a different part of developing AegisShield and its role in democratizing and streamlining cyber threat modeling. Here is what each section covers.

The second chapter, “Literature Review,” offers an analysis of existing research on current and past threat modeling methodologies. It also covers the challenges organizations face and the potential of generative AI to enhance cybersecurity practices. It identifies the gaps in current approaches. This sets the stage for the introduction of AegisShield.

The third chapter, “Methodology,” provides a detailed explanation of the research design and approach. Here, we focus on the development and integration of a Python-based threat modeling tool. It describes how low-cost generative AI models were employed. It also covers how integration with the STRIDE and MITRE ATT&CK frameworks enhanced the threat modeling process. The data collection and analytical techniques used are outlined. Finally, it discusses the ethical considerations that guided the evaluation of the tool’s performance in terms of the research hypotheses.

The fourth chapter, “Results,” presents the outcomes and statistics of the tool’s evaluation. Its focus was on its effectiveness in achieving the research objectives. It analyzes the tool’s ability to simplify threat descriptions, identify threats with high accuracy, and map them to the MITRE ATT&CK framework.

The fifth chapter, “Discussion and Conclusions,” provides the interpretation of the research findings. It goes over comparisons with existing literature, and their implications for democratizing threat modeling. We discuss whether the research objectives were met, and the hypotheses supported and highlight the study’s contributions to cybersecurity, limitations, and recommendations for future research and practical applications.

Finally, after a comprehensive list of references, the “Appendix” section provides the supplementary materials, such as the Python script, rubric tables, detailed case study reports, visuals, and samples of the tool’s output (e.g., DREAD reports, mitigation strategies, and test cases).

# Chapter 2—Literature Review

## 2.1 Introduction

This chapter presents a detailed literature review focusing on threat modeling methodologies and the emerging role of generative AI in cybersecurity practices. The primary objective was to establish a strong foundation for understanding the current landscape of threat modeling. It also explores the potential for AI-driven innovations in this field. This review provides a theoretical foundation and informs the methodology and tool development and offers insights to guide the design of generative AI-enhanced tools.

The aim was to comprehensively understand the current state of threat modeling and the potential for AI-driven enhancements. We will present findings from diverse areas and examine current practices and challenges. This foundation informed the research on integrating generative AI to democratize, accelerate, and simplify threat modeling processes. Together, this addresses the critical need for more effective and accessible approaches in cybersecurity.

Recent studies have emphasized the need for more effective threat modeling methods, especially to address high-risk vulnerabilities. As discussed earlier, although over half of the surveyed organizations in the U.S., U.K., and Germany can mitigate 70–89% of their vulnerabilities through threat modeling, only a tiny fraction (10%) are confident that they are protecting 90% or more of their systems (The 2023 State of Threat Modeling, n.d.). This highlights a growing challenge. Organizations must look for new ways to obtain comprehensive security coverage and need innovative approaches.

## 2.2 Historical Context and Evolution of Threat Modeling

Threat modeling has undergone significant evolution since its inception. Figure 2.1 illustrates a timeline highlighting the key milestones and developments in the history of threat modeling and a visual overview of the field's evolution.

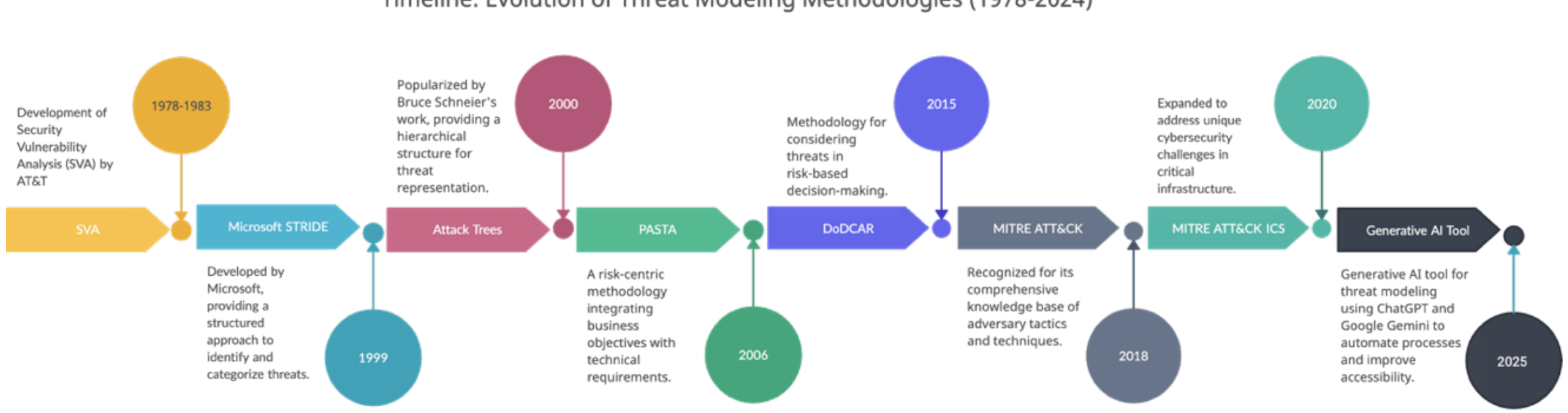

Figure 2.1. The Evolution of Threat Modeling

The Department of Defense made early attempts to formalize threat modeling for information systems in the late 1970s and early 1980s. One of the earliest dynamic threat analysis models was the Security Vulnerability Analysis (SVA) for the System Security Engineering (SSE) process. AT&T developed this for the Strategic Defense Initiative. This 10-step process was designed for the structured enumeration of system security requirements, and it uses "threat logic trees" for threat decomposition (Bokan & Santos, 2022).

Subsequently, Microsoft's development of the STRIDE framework in 1999 was a watershed moment in threat modeling. This framework provides a structured approach to identifying and categorizing threats where it significantly advanced the field toward a more systematic practice. STRIDE's influence remains strong. Today, it is one of the most mature and widely used methods (Bokan & Santos, 2022).

Next, in the early 2000s, attack trees gained traction. These provide a graphical notation for representing threats in a hierarchical structure. Developed by Bruce Schneier, this method allows for a detailed analysis of attack scenarios and their potential impacts (Bokan & Santos, 2022).

In 2015, the Department of Defense introduced a new threat modeling approach called the NIPRNet SIPRNet Cyber Security Architecture Review (NSCSAR), named for its integration of the Non-classified Internet Protocol Router Network (NIPRNet) and Secret Internet Protocol Router Network (SIPRNet). The method was later rebranded as the Department of Defense Cybersecurity Architecture Review (DoDCAR). This allowed the government to consider threats in its risk-based decision-making process. It also enabled a shift to an adversarial perspective. In 2018, the Department of Homeland Security copied and refined this model to create the Government Cybersecurity Architecture Review (.govCAR). This helped expand its application to federal, state, and public sector agencies (Bokan & Santos, 2022).

These progressions in threat modeling point to its ongoing evolution. They show the field's ever-changing nature and adaptability in response to a complex cybersecurity landscape. Why do challenges persist? Despite excellent progress, threat modeling practices are still considered off-the-cuff and spontaneous, often involving "whiteboard hacking" and being heavily dependent on the modelers' expertise (Yskout et al., 2020). According to Yskout et al. (2020), threat modeling as an engineering discipline is currently at a surface level of maturity in terms of research, tool support, and practice. Plus, many public and private organizations need help implementing system models and threats. This painful realization is attributed mainly to a lack of valuable and high-quality methodologies to assist developers in designing system models (Konev et al., 2022).

### 2.3 Theoretical Foundations and Methodologies

The current cybersecurity threat modeling field is built upon several fundamental theories and models and provide most of the conceptual basis for analyzing and reducing risk in complex systems. For example, data flow diagrams (DFDs) have long been a cornerstone of threat modeling. They visually represent system architectures and potential vulnerabilities. Some recent research suggests that while these diagrams are valuable, they can fall short in comprehensive threat modeling, especially for complex systems with multiple trust boundaries (Sion et al., 2020). Because of this, there is a need for more powerful approaches to threat modeling.

One of these approaches, and a significant advancement, is the concept of relationship-based threat modeling (RBTM). This method extends traditional threat modeling by incorporating threat relationships. RBTM provides a more nuanced understanding of threat propagation within a system and therefore allows for more accurate risk calculations (Verreydt et al., 2022). This helps give a holistic view of system vulnerabilities.

Next, we have the Cyber Kill Chain, developed by Lockheed Martin in 2011. It remains a popular threat modeling framework. The model consists of seven stages: reconnaissance, weaponization, delivery, exploitation, installation, command and control, and actions on objectives (Straub, 2020; Zografopoulos et al., 2021). Each stage represents a step in the attacker’s behavior. These steps are structured to help defenders understand and interrupt attacks.

Using this framework, defenders can design countermeasures for each stage. For example, reconnaissance can be addressed through traffic analysis and command and control can be mitigated by blocking outbound communications. The model enables more accurate detection and response by breaking the attack into stages. It is still especially effective in environments that face linear attacks.

However, the Cyber Kill Chain also has limitations. Its linear structure does not adequately capture the complexity of modern, multi-vector threats and is especially true for lateral movement or blended techniques. Even so, it remains a useful framework for mapping attacker behavior and informing security architecture.

The model's core strength is the idea that breaking any link in the chain can neutralize the threat. If delivery or installation is disrupted, for example, the attacker may fail to reach their goal. This reinforces the importance of layered defenses. It also highlights the value of timing in incident response (Lockheed Martin, n.d.).

Next, is MITRE ATT&CK, which builds off what the Cyber Kill Chain started. This framework offers a structured way to collect and categorize how attackers behave, and it breaks actions down into specific tactics and techniques. Strom et al. (2020) described ATT&CK as a knowledge base and model for cyber threat actor behavior. The framework reflects the various phases of an adversary's attack lifecycle and the many platforms they are known to attack. ATT&CK offers a separate matrix for each platform. These are broken down into enterprise, mobile, and industrial control systems (ICS) files. Finally, each matrix organizes tactics by attack stage.

The ATT&CK framework focuses on how adversaries compromise and operate within various systems and networks and originated from a project to document and categorize adversary TTPs against Microsoft Windows systems. This was done to improve the identification of threat actors. Since its inception, the framework's scope has grown to include behaviors that compromise various environmental and technological areas. These include mobile devices, cloud-based systems, and ICS (Strom et al., 2020).

The evolution of the ATT&CK framework shows its flexibility and need across diverse technology landscapes. A good example is its inclusion of the ATT&CK for ICS framework in 2020. That addition addresses unique cybersecurity challenges in critical infrastructure and cyber-physical systems (CPS). This ICS matrix incorporates tactics specific to industrial environments, such as Inhibit Response Function and Impair Process Control and it shows the growing need for specialized threat intelligence in these somewhat off-brand domains (Xiong et al., 2022).

Another great addition to the threat modeling toolkit is NIST Special Publication 800-154, "Guide to Data-Centric System Threat Modeling." This provides a structured methodology for identifying, analyzing, and mitigating threats in information systems. The guide zeroes in on protecting critical data assets. Most importantly, it advocates for integrating threat modeling throughout the system development lifecycle (Souppaya & Scarfone, 2016) and is a standout example of NIST doing what NIST does best.

When we examine all the insights from these frameworks—including the Cyber Kill Chain, the adversary behaviors in ATT&CK, and the lifecycle focus of NIST SP 800-154—it is clear these frameworks give organizations more options. Each one covers a different area or situation. Using these tools can provide a holistic view of potential threats, from initial access to impact and help companies focus on protecting critical data assets throughout the system lifecycle. With these tools, organizations can better anticipate, detect, and reduce cyber threats. This is valuable as technology and their threats evolve (Zografopoulos et al., 2021).

Within these approaches stands the STRIDE framework. STRIDE offers a structured way to identify threats during application design. In STRIDE, we can break down the acronym into its respective categories. These are spoofing, tampering, repudiation, information disclosure, denial of service (DoS), and elevation of privilege (Hernan et al., 2006/2019).

This framework helps to model software and systems clearly and consistently. Other modeling methods can take the perspective of the attacker or even the organization. Regardless, they are often paired with data flow diagrams (DFDs). DFDs are used to systematically assess each system or software component for a weakness in the six STRIDE threat categories. Using STRIDE means looking closely at each part of the system to identify which threats apply and helps designers demonstrate the system's security (Hernan et al., 2006/2019).

As well as STRIDE, there are other methodologies that offer alternative approaches to threat modeling. For example, PASTA is a risk-centric methodology and focuses on aligning business objectives with technical requirements. OCTAVE is another, and it heavily focuses on organizational risk management and offers a strategic assessment and planning technique (Suhas, 2023).

All these methods offer a unique perspective. They focus on areas that allow organizations to scope and tailor their approach to specific objectives and risk profiles. The choice of methodology often depends on various factors. These include the organization's size, industry, regulatory requirements, and even the system being analyzed.

This praxis is solely concerned with the STRIDE framework. However, the tool does prioritize the results using DREAD. Adding DREAD helps score threats and brings more structure to the STRIDE output and is typically paired with STRIDE because they are both straightforward and list threats clearly.

As we can see, threat modeling continues to evolve and leading this are approaches that use AI and ML. Recent research has explored the viability of AI in designing user interfaces for cybersecurity threat modeling. The goal is to increase the efficiency and effectiveness of the entire process (Ebenezer Taiwo Akinsola et al., 2022).

This intersection of AI and cybersecurity represents innovation in the field and holds hope for promising new threat detection capability. In the context of AI in cybersecurity and the need for new ideas, specialized frameworks have emerged to address specific domains. One such domain is Cyber Physical Systems (CPS).

More specifically, the increasing integration of physical and digital systems in industries such as energy, manufacturing, and transportation has necessitated frameworks tailored to the unique security challenges posed by CPS. For example, Zografopoulos et al. (2021) developed a comprehensive framework for threat modeling, risk assessment, and security evaluation in CPS to address the distinct vulnerabilities and operational challenges of hybrid environments.

Current advancements in generative AI have brought about both opportunities and challenges in cybersecurity. While it can enhance cyber defense automation, threat intelligence, and attack detection, it also provides a means of abuse through the creation of sophisticated phishing emails, disinformation, and malware (Neupane et al., 2023). Despite these risks, recent work shows that large language models are in fact being used more for defensive purposes. This includes tools for automating threat detection, streamlining incident response, and potentially aiding in secure code generation (Yigit et al., 2024).

The process of integrating AI and ML techniques into threat modeling represents a response to the growing technical challenges and threats to organizations. This shift highlights the importance of stronger research and development in the field of threat modeling. Research has increased in areas such as detection, deception, adversarial training, and explainable AI. These categories can help develop hardened defense mechanisms against generative AI-enabled cyberattacks (Yigit et al., 2024).

In addition to these widely used frameworks, the Threat Modeling Manifesto (Threat Modeling Manifesto, n.d.) outlines values and principles designed to make threat modeling more effective, inclusive, and iterative. The Manifesto was created by a working group of practitioners and academics. It emphasizes that anyone involved in a project can and should contribute to identifying potential threats and mitigations.

The manifesto lays out several core principles. These include aligning threat modeling with iterative design, prioritizing stakeholder value, and encouraging diverse perspectives. The principles align closely with this study's objective of simplifying and automating threat modeling. It advocates for continuous refinement and emphasizes actionable outcomes. It also serves as a guiding philosophy alongside established approaches such as STRIDE, ATT&CK, and NIST SP 800-154.

## 2.4 Practical Applications and Case Studies

It is important to apply threat modeling across different domains. Doing so shows its versatility and value in identifying and reducing security risks more broadly. This section examines several case studies that employ threat modeling methodologies.

The focus in these case studies is on STRIDE and DREAD. These are examined in contexts such as AI-ML systems, IoT-enabled smart manufacturing, medical contact tracing systems, and social media networks. We will use the case studies as a reference to compare against AegisShield's outputs.

Table 2.1 lists the case studies. It provides an overview that highlights their domain of application, threat count, gaps identified, and whether the study was peer reviewed.

Table 2.1. Summary of Threat Modeling Case Studies Across Various Domains.

| # | Authors | Peer-reviewed | Domain/Sector | Threats | Identified Gaps |
|---|---|---|---|---|---|
| 1 | Yuldasheva, 2024 | √ | Voice-based Applications | 16 | Does not consider physical attacks and real-world testing on the data flow between IoT devices |
| 2 | Simonjan et al., 2020 | √ | Visual Sensor Networks | 12 | Does not cover heterogeneous network setups and next-generation technologies |
| 3 | Sattar et al., 2021 | X | 5G Networks | 20 | Does not address physical security threats and focuses solely on the logical and virtualized aspects of 5G core slicing |
| 4 | AbuEmera et al., 2022 | √ | Smart Manufacturing Systems | 10 | Lacks the integration of threat mitigation techniques and |

| # | Authors | Peer-reviewed | Domain/Sector | Threats | Identified Gaps |
|---|---|---|---|---|---|
| | | | | | focuses on identification and assessment only |
| 5 | Kim et al., 2022 | √ | Industrial Control Systems/Oil and Gas | 13 | Has limited focus on real-world applicability and does not integrate AI/ML for enhanced threat detection and response |
| 6 | Sharma et al., 2023 | X | Social Media Networks | 15 | Has limited focus on emerging social media threats during pandemics and lacks proactive countermeasures |
| 7 | Hossain & Hasan, 2023 | √ | Ambient Intelligence | 29 | Does not involve real-world validation of mitigation strategies and comprehensive testing across diverse AmI environments |
| 8 | Das et al., 2024 | X | Infotainment Systems in Automobiles | 32 | Does not comprehensively evaluate hardware components and adherence to ISO/SAE 21434 standards |
| 9 | Mauri & Damiani, 2022 | √ | AI-ML Systems | 13 | Lacks a comprehensive security control framework for ML models |
| 10 | Hasan et al., 2022 | √ | Public Health | 13 | Does not include real-world testing and validation of the proposed threat model in diverse environments |
| 11 | Klein et al., 2022 | √ | Vehicular Fog Computing | 11 | Does not address physical attacks in vehicular fog computing, which is often neglected in existing threat models |
| 12 | Salzillo et al., 2021 | √ | The Open Energy Monitor | 15 | Involves limited automation in penetration testing and needs a more comprehensive threat catalog |

| # | Authors | Peer-reviewed | Domain/Sector | Threats | Identified Gaps |
|---|---|---|---|---|---|
| 13 | Chowdhury et al., 2023 | X | E2EE Messaging Applications | 11 | Fails to continuously adapt to evolving threats and new application features, leading to unaddressed vulnerabilities in contexts such as desktop clients |
| 14 | Salamh et al., 2021 | √ | Drone as a Service | 13 | Does not prioritize threats based on the specific operational context, potentially undervaluing critical threats such as spoofing compared to others such as DoS |
| 15 | Brown et al., 2022 | √ | CPS – Window Cleaning Business | 20 | Does not emphasize the specific vulnerabilities of real-time IoT data exchange, which are critical for SMEs adopting Industry 4.0 technologies |

Note. AI = Artificial Intelligence; IoT = Internet of Things; ML = Machine Learning; AmI = Ambient Intelligence; CPS = Cyber-Physical Systems.

Researchers have applied the STRIDE framework in various fields to tackle the cybersecurity issues highlighted in the case studies. In one study, STRIDE-AI illustrates an approach designed to assess threats in AI and machine learning environments where it adapts the original STRIDE model for ML systems (Mauri & Damiani, 2022).

In healthcare, STRIDE and DREAD are used in the risk evaluation of contact tracing apps. This is in the mobile health sector (Hasan et al., 2022). In the Drone-as-a-Service (DaaS) study, researchers adapted STRIDE into DIREST, a model that prioritizes denial-of-service (DoS) threats. Availability is the most important aspect of drone operation. The DIREST approach identified some unique vulnerabilities. One even included spoofing drone coordinates to attack the pilot that was operating the vehicle.

The adaptation of STRIDE into DIREST shows how threat models can be modified, scoped, and tailored as needs change (Salamh et al., 2021).

All these case studies reveal the applicability, challenges, and limitations in current practices. Many of these studies highlight a lack of standardized methods for evaluating security in specific environments. They also point to difficulties around the application of threat modeling tools in real-world scenarios and show an urgent need for more comprehensive reporting of results.

For example, Klein et al. (2022) reported fragmented security evaluations for Vehicular Fog Computing, showing the need for more standardization. Salzillo et al. (2021) looked at the challenges in achieving consistency in IoT threat modeling, particularly due to improvised methodologies. On a similar level, Hossain and Hasan (2023) stressed the importance of in-depth reporting in ambient intelligence systems to improve threat model reliability and adaptability. One key point here is that we can see how subjective these models are in their assessment. A very human trait that is guided by perspective. How will the tool match up?

**2.5 Challenges, Limitations, and Gaps**

Numerous technical challenges often impede the effectiveness and widespread adoption of threat modeling. Data quality and integration with systems are at the top of the list. This is because threat modeling typically relies on multiple sources to work through the process.

Generally, there is a security architect or group who gathers the data about a system to be modeled and typically involves creating DFDs. When revisiting the limitations of DFDs discussed earlier, we know that relying on overly simple models poses a significant challenge in capturing and integrating comprehensive security information. This is especially true when it comes to multi-trust boundary systems like an e-commerce web server, where scalability is a concern (Yskout et al., 2020).

When assessing systems, as size increases, techniques such as attack trees and graphs become exponentially more complex. This makes them impractical for large-scale environments (Yeng et al., 2020). Also, threat modeling work is mostly manual, with limited assurance of validations. Is it not better to have assurance before code is pushed into production? This is an indication that there is a need for more automated approaches to ease the burden (Xiong & Lagerström, 2019).

Research shows there are some major gaps in the threat modeling process. First, the practice of threat modeling cannot keep up with the rapid pace of technology growth. Second, empirical validation of current approaches is also difficult to find. Finally, despite the advancements in technology overall, threat modeling faces a lack of standardized process and general community (Yskout et al., 2020). These are ongoing issues. They reflect the evolving nature of the field and the need for maturity and raises the question of whether standardization is even possible.

Because of these gaps, many threat models within an academic setting make assumptions that are unrealistic in the real world. They often assume attackers have capabilities far beyond their scope (Grosse et al., 2024). Many of the assumptions made during threat modeling are implicit and not well documented. Literature shows that this process often lacks proper guidelines and best practices (Van Landuyt & Joosen, 2022).

AegisShield attempts to address this issue by documenting roles and assumptions for each threat. By integrating these structured assumptions into the STRIDE model,

AegisShield aims to enhance the clarity and effectiveness of threat modeling. This helps ensure a well-documented understanding of potential threats and their mitigations.

A disconnect also exists between different threat modeling methods. One study showed that there is no single method that covers all the aspects of threat modeling. Cloud computing is a perfect example of this (Granata & Rak, 2024). Cloud environments are dynamic and include multiple trust boundaries and shared responsibilities. This level of complexity requires threat models to be updated continuously. It poses a major challenge to organizations fighting to maintain security (Yeng et al., 2020).

Another area is ethics in threat modeling. This is a complex area with its own challenges. As we previously stated, AI will begin to play a bigger role in everything we do, and this is especially true of cybersecurity. We know attackers already use generative AI as a sword. Now it is time to look at ways to use it as a shield (Grosse et al., 2024). This duality of purpose highlights an important need for comprehensive ethical frameworks to guide AI implementation not just in the threat modeling process, but throughout cybersecurity.

Within ethics and security is privacy. Privacy adds yet another layer of complexity to the ethical landscape, especially in sectors that process sensitive data (Yeng et al., 2020). In sectors that manage a lot of personal information, such as healthcare, we need privacy-specific approaches. These should complement traditional security methods to protect sensitive data comprehensively (Yeng et al., 2020).

Because of this, AI-integrated threat modeling must address security concerns. It must also incorporate strong privacy protections that align with regulatory requirements and ethical standards. Focusing on both security and privacy is important if organizations want to build trust in AI-based threat modeling. Without that, adoption could stall.

The ethical side of this goes beyond technical fixes. It is also about how these tools are used. One real concern is insiders misusing the tools to map out weaknesses or find ways to exploit them. That kind of misuse shows why stronger governance and clearer ethical standards are necessary. Addressing this requires both technical safeguards and organizational policies that guide responsible use of AI.

To address these challenges, tools must use generative AI systems to alleviate privacy concerns proactively and transparently. While generative AI providers often claim built-in privacy controls, the ultimate responsibility for safeguarding sensitive information rests with the users of the system. This will be dependent upon the organization. A proper solution here would be the development of strict controls to prevent the input or processing of personally identifiable information or confidential organizational data. Such a proactive approach can protect individuals and organizations. It can build trust in AI-enhanced threat modeling practices and uphold ethical standards. In turn, this promotes wider adoption.

Organizational challenges in threat modeling are everywhere. There is constant demand for skilled personnel. Businesses struggle with resource limitations and with the complexity of quantifying the return on investment in security.

These challenges are critical barriers to transforming threat modeling practices (Yskout et al., 2020). It is also difficult to integrate threat modeling into broader corporate processes. The need for standardized metrics to assess the quality of threat modeling result sets further complicates the landscape.

The maturity levels observed in the field indicate a need for continued research and development. This will enhance the effectiveness and integration of threat modeling practices within organizational frameworks. It also reveals the constantly changing nature of the field and the ongoing need for innovation at every turn (Yskout et al., 2020).

### 2.6 Emerging Trends and Future Directions in Threat Modeling

AI and ML tool integration has influenced recent innovations in threat modeling. It is reshaping cybersecurity because of it. In recent studies, AI-enhanced threat modeling tools that capitalize on ML algorithms have demonstrated increased efficiency in processing large volumes of data. They are also identifying complex attack patterns that traditional methods might overlook (Shi et al., 2022).

What the author proposes makes complete sense. Computers are far better at automating repetitive tasks at high volume. They do this consistently better than a human could. However, these tools show a clear distinction when comparing commercial and open-source offerings.

Many commercial tools provide a full suite of features but may be cost prohibitive and overly complex for small organizations. Granata & Rak (2024) and Xiong & Lagerström (2019) state that open-source tools provide greater accessibility and customized options. Conversely, they also state that these tools may lack advanced functionality. The authors note the lack of real-time threat analysis, automated compliance checks, and in-depth AI-driven pattern recognition.

This is a major gap. The contrast between these two worlds reflects a need for solutions to democratize threat modeling. Organizations want advanced functionality, and it needs to be cheap.

Integrating common cyber threat intelligence (CTI) sources such as the widely used NVD and AlienVault's OTX with low-cost generative AI technology is a strong approach to bridge this commercial versus open-source gap. We suggest combining generative AI's capability to correlate and organize complex information with the detailed threat data from CTI. By mixing AI with CTI feeds, smaller teams can access reliable threat intel without spending too much.

This combination makes threat modeling practical. This kind of setup gives smaller organizations a better shot at doing secure-by-design work early on, even during planning.

Some recent advancements in threat modeling tools emphasize the use of standardized threat knowledge bases. Popular examples include systems such as the Common Vulnerabilities and Exposures (CVE), Common Weakness Enumeration (CWE), and Common Attack Pattern Enumeration and Classification (CAPEC). Some commercial threat modeling tools use resources like the NVD to incorporate detailed CVE data. Doing this enhances threat libraries and provides the AI a basis for evaluating and mitigating vulnerabilities.

Regardless of these advancements, many tools still rely on basic evaluation and prioritization methods. This suggests there is a good amount of room for improvement through the adoption of advanced AI-driven techniques to automate and fine-tune these processes (Shi et al., 2022; Yskout et al., 2020).

At the end of the day, the application of generative AI in specific cybersecurity domains is gaining traction. Newer approaches using generative adversarial networks (GANs) and common transformer-based models for cyber threat-hunting in 6G-enabled

IoT networks may demonstrate high accuracy in detecting IoT attacks. These have the potential to enhance security for emerging network technologies (Ferrag et al., 2023).

Several approaches to AI-enhanced CTI processing have also shown strong potential. These methods strengthen the alignment between AI and human expertise in producing timely, high-quality CTI. Other research examined automated mitigation recommendations, using AI to offer real-time, contextual, and highly predictive insights (Alevizos & Dekker, 2024).

Meanwhile, the readability of threat descriptions represents another area for innovation that this praxis will investigate. The analysis of AI-enhanced threat modeling tools like AegisShield needs to incorporate readability metrics, such as the Flesch-Kincaid readability score, to generate more accessible threat descriptions while keeping them comprehensive (Flesch, 1948). This analysis addresses a main challenge in the democratization of threat modeling and in reducing its complexity.

While studies have highlighted AI's capacity to enhance threat detection and automated response mechanisms, the complexity and cost of building and maintaining AI-driven security systems present challenges for organizations with limited resources. This praxis addresses the need for more reliable threat modeling solutions that pair generative AI technologies and open-source threat intelligence.

Future research directions in AI-enhanced threat modeling and cybersecurity practices point to several key areas. These include developing standards and baselines for quantitative tool evaluations. They also include refining AI models for improved accuracy and interpretability in CTI processing. Another area is exploring interdisciplinary approaches to bridge AI advancements and ethical considerations.

Other directions involve investigating the integration of open-source threat intelligence with generative AI, developing adaptive AI-enhanced tools, and studying the compatibility of AI-generated threats with established frameworks. There is also a need to address the ethical implications and potential biases in AI-driven processes. Finally, researchers are exploring advanced AI models to augment cyber defenses and optimize human-AI collaboration (Alevizos & Dekker, 2024). As this complex field evolves, this research will play a major role in shaping the future of AI-enhanced threat modeling and cybersecurity practices overall.

An important aspect of the proposed generative AI-enhanced threat modeling approach in this praxis is the application of Semantic Textual Similarity (STS) techniques. Because of this, this praxis uses cosine similarity with SBERT to compare the semantic similarity of the threat descriptions generated by the model to those from traditional or non-generative AI methods. This method will provide a sound measure of similarity by capturing the contextual meaning of the text at scale. It also enhances the identification and categorization of threats across different domains and frameworks (Reimers & Gurevych, 2019).

### 2.7 Synthesis of Literature and Research Implications

This literature review tracked the evolution of threat modeling in cybersecurity. It began with early formalized approaches from the 1970s. The review then moved to more structured methodologies such as STRIDE, introduced in the 1990s and beyond. Despite these advancements, the field still often relies on extemporary practices and expert knowledge. As a result, this praxis highlights the need for more standardized and accessible methods to address the complexities in modern cybersecurity tools.

While some structured and well-known foundations such as MITRE ATT&CK, NIST, STRIDE, and DREAD provide frameworks for identifying, categorizing, and mitigating threats, practical applications across domains reveal both their versatility and challenges. Technical issues such as data quality and scalability, along with organizational and ethical challenges, impede the implementation of stronger threat modeling, especially for smaller organizations.

As we have seen, traditional approaches demand substantial expertise and resources. They pose a barrier to widespread implementation, therefore can cause any resulting threat model artifacts to exclude critical components. Therefore, drawing from Van Landuyt and Joosen (2022), this praxis explicitly documented the assumptions, conditions, and roles alongside the produced STRIDE results.

This structured piece of documentation ensures that the threat modeling process is not just comprehensive, but also transparent while still maintaining practical applicability and reliability. Also, we constrained AegisShield to a standard format to simplify and accelerate the threat modeling process. This makes it accessible to a broader range of organizations. To summarize, by using technologies such as easily accessible generative AI models and open-source threat intelligence, AegisShield aims to lower entry barriers and aligns with the need for more automated, validated, and accessible threat modeling practices.

The praxis will show that integrating AI and ML in threat modeling represents a worthwhile method to tackle this ever-changing cybersecurity landscape. While this praxis will analyze its effectiveness against expert case studies, real-world effectiveness of such approaches was not tested. This is of particular concern for resource-constrained organizations and must be thoroughly investigated through long-term testing. This praxis contributes to the field by evaluating the inclusion of generative AI to democratize the practice of threat modeling, promote secure-by-design principles, and enhance them across many diverse organizational contexts.

# Chapter 3—Methodology

## 3.1 Introduction

The methodology in this praxis addresses two primary research objectives: democratizing threat modeling and automating the threat modeling process. The former focuses on making threat modeling accessible to small organizations by reducing the complexity and cost. The latter seeks to streamline threat modeling by automating essential tasks and providing detailed, standardized, and actionable outputs that enhance an organization's ability to detect and respond to potential threats. We also explored the tool's impact on accelerating threat identification, lowering barriers to entry, and enhancing the compatibility of AI-generated threat models with widely used and established cybersecurity frameworks. By focusing on these areas, the praxis aimed to show how generative AI can transform threat modeling practices, making them more effective and accessible to a broader range of organizations.

We chose a Design Science Research Methodology (DSRM) approach to guide the development and refinement of AegisShield. Why take this route? This approach is well suited for creating innovative tools in fields such as cybersecurity, where practical and effective solutions are highly beneficial. The process involved continuous and iterative design. We ran multiple rounds of development and evaluation cycles that pushed the tool to evolve in response to user needs and practical challenges (Hevner et al., 2004). AegisShield automates the initial phases of the threat modeling process and provides outputs such as STRIDE threats, attack trees, MITRE ATT&CK mapping, DREAD analysis, mitigation strategies, and Gherkin test cases. All these outputs help organizations, particularly those with limited cybersecurity expertise, build practical threat models and apply effective threat mitigation strategies in line with secure-by-design principles. Overall, the tool uses generative AI to automate complex tasks, reduce human error, and provide accurate, standardized, real-time threat assessments using established cybersecurity frameworks.

### 3.2 Research Design

This research used two distinct approaches. It combined quantitative techniques with a structured evaluation. Pulling information from the case studies in this way was somewhat unique and allowed us to score them numerically. Threat modeling is not an exact science. It is often subjective. Because of this, we needed a way to gather information from the case studies while also understanding the quality of what was extracted. In the end, we had a table filled with quantitative values that could be cross referenced and compared across case studies.

We used a quantitative analysis to evaluate several metrics that align with the research hypotheses. These include the complexity of threat descriptions, the semantic similarity of AI-generated outputs compared to expert-developed models, and AegisShield's effectiveness in mapping identified threats to the MITRE ATT&CK framework. Data for this analysis came from various simulated organizational case studies (see Section 3.4), which detail the approach and selection process. This gave us a precise way to evaluate the tool's capabilities in these areas. Part of the analysis was to check if AegisShield could generate at least one semantically similar threat per case study in 50% of its runs. Each execution generated 18 threats—three per STRIDE category. Cosine similarity was used to measure how closely these threats matched the expert-developed models.

The evaluation involved extracting data from case studies where domain experts used the STRIDE framework to model threats in specific systems. We scored these data points using a rubric that assessed the clarity, completeness, and specificity of the information obtained from each case study. The rubric evaluated the quality and reliability of the input data, focusing on the consistency and depth of the information on a scale of 1 to 5. This type of structured evaluation provided valuable insights into the variability and general comprehensiveness of the case study data across the different domains.

Next, we took the evaluated data and input it into AegisShield to generate a corresponding threat model for each case study. This allowed for a standardized and systematic examination of the tool's performance with inputs of varying quality and completeness. This helps it reflect real-world scenarios where often, the information supplied may present itself subjectively and with variations in detail level.

In this praxis, we used two main datasets: 243 extracted threats from the case studies, and 8,100 tool-generated threats. These datasets supported the analysis for H1 (readability scores), H2 (cosine similarity scores), and H3 (MITRE ATT&CK mappings). We created 8,100 tool-generated threats by extracting nine pieces of information from each of the 15 case studies. This resulted in 135 inputs for the prompts in AegisShield and helped with diversity and comprehensiveness.

For hypothesis 1, the larger sample size was needed for the Mann-Whitney U test. This test requires adequate observations to achieve high statistical power. This large dataset helped reduce the risk of Type II errors and supported stronger results (Brama &

Ahmed, 2023; Bürkner et al., 2022). The large number of observations also helped support a clearer detection of differences.

For hypothesis 2, we used the analysis of 21,870 cosine similarity scores. These were spread across 15 case studies and provided a solid foundation for the assessment. We used embeddings from the "stsb-roberta-large" model, which outputs 1024-dimensional vectors. The sample size made sure that there was a reliable detection of semantic patterns. We also ran several tests per case study to improve consistency and reduce random noise. The distribution across the case studies supported the generalizability of the findings. Of note, embedding based word similarities can be highly sensitive to changes in the body of data and the training conditions. This made multiple runs crucial for establishing stable, reliable results, whereas using larger datasets helped reduce the variability and ensured consistent, generalizable patterns (Antoniak & Mimno, 2018).

In hypothesis 3, we evaluated those 8,100 tool-generated threats across 15 case studies again. This provided a large dataset for mapping analysis, as this sample size far exceeds the minimum thresholds known to achieve a stable classification accuracy (>120 samples; Rajput et al., 2023). Overall, the wide number of observations distributed across multiple case studies and data points worked to create variation and enhanced consistency. It enabled reliable pattern detection and the generalizability of the mapping results. The statistical power supported the validity of the conclusions and enhanced confidence in the study's strength.

### *3.2.1 Development of the Threat Modeling Tool*

The development of AegisShield started with a core list of requirements. The main goal was to create a tool that organizations with limited cybersecurity resources could use. The difficult part was doing this while still producing outputs that consistently aligned with both the STRIDE and MITRE ATT&CK frameworks. The focus was on developing a "useful toolkit" that enhances productivity and supports continuous refinement rather than providing a one-time delivery (*Threat Modeling Manifesto*, n.d.). The key functionality of the tool includes automating the identification and categorization of threats using the STRIDE framework; mapping identified threats to the MITRE ATT&CK TTPs; and generating supplementary outputs such as DREAD scores, mitigations, Gherkin test cases, and attack trees. These features were essential to provide a starting point for a comprehensive threat modeling experience that simplifies the secure-by-design approach and enhances the analysis's depth and utility.

AegisShield is adaptable, scalable, and designed to meet the diverse needs of potential users. It integrates OSINT data from multiple sources, therefore, enhancing threats. This allows the tool to reach out and pull in additional and up-to-date information that would be labor intensive for the user to do otherwise. This process was coupled alongside frameworks such as MITRE ATT&CK data to bring deeper context into the attack techniques used. This pairing could potentially help identify emerging threats versus using AI alone.

In the design and planning phases, we worked to build and refine an interface that allows for a clean and intuitive navigation experience. The purpose was to build a user interface that reduced cognitive load to use. This aligned with the broader goal of making the threat modeling process accessible. There were no formal studies done on the usability of the interface. Future research could include a structured usability evaluation to validate and refine the interface.

Meanwhile, AegisShield's architecture is modular and scalable, as it integrates generative AI components with established cybersecurity frameworks such as STRIDE and MITRE ATT&CK. Its core architecture consists of several modules that work together to automate the threat modeling process; incorporate real-time threat intelligence; and generate comprehensive outputs with the capability to swap or add new modules.

Python 3.12 was used to build the tool. We also used the Streamlit.io service for hosting and running the application. Streamlit is framework that is used for building clean and easy to use UI interfaces in Python. This allowed us to focus on functionality rather than wasting time playing with stylesheets. We then built and used a batch script that performs the same functionality as the user interface. The script was used to create the research data at scale as doing this in the interface 450 times would have been too time consuming. The batch script created 30 models per case study. This generated 540 threats (18 per run) per case. 30 was chosen to conform to the central limit theorem (CLT) to help create a statistically significant sample and stabilize the results (Ifeanyichukwu et al., 2023). After verifying the integrity of the results, we went through the process of evaluating these batch files for Flesch-Kincaid grade levels, as well as for semantic similarity. For semantic similarity we use cosine similarity and set a threshold cosine score of ≥ 0.7. We configured the tool to use the default "medium" technical level to strike a balance between overly simplistic and overly technical outputs. In sections 3.2.2 and 3.3.3, we explain how the tool handles varying technical depth (low, medium, high) and discuss the reason for choosing "medium" to minimize bias and maintain clarity.

AegisShield is forced to output structured JSON in its responses. This process serves a few purposes. By enforcing this standard, it minimizes hallucinations, reduces overhead in parsing data, and helps ensure consistency. Using JSON also provides a commonly used standard that creates confidence the tool's ability to integrate easily with existing workflows (Liu et al., 2024). This also makes it flexible to add and remove modules more easily. Enhancements such as integrating the MITRE ATT&CK framework and additional user input options are available in both the Streamlit UI and in batch mode. We based the tool's initial development on the open-source project STRIDE-GPT (Adams, 2024), which is available under an MIT license. The original project provided a solid foundation to expand upon. AegisShield adds in an integration with MITRE ATT&CK, a PDF artifact, and enhances the workflow input. That goal here was to provide something tangible.

Even though the batch process had strong format checking and except handling instructions built into the script, we still went through a manual verification process of the resulting files. Each of the 450 files were opened and scanned manually. The purpose was to perform random checks on the data and confirm that the tool generated JSON formatted threats as expected. This gave us confidence in the consistency and reliability of the data generated.

In the user interface, we added several questions that prompt the user for details on various security, technology, and compliance aspects of their system. This helped us collect context outside of the system's description alone. The architecture of the tool allows for more questions if needed. The tool's modular approach allows for the ability to add more AI models down the road. Models in Google, Azure, and Anthropic, all have similar structure and capabilities to those found in OpenAI. We exclusively used OpenAI for this research, but the ability to add more models provides needed flexibility.

The Streamlit user interface guides users step by step in answering these prompts. Each step of the process is in a tab at the top of the interface. Users can input various details about the application, select relevant technologies, generate the threat model, and produce comprehensive outputs, including the STRIDE results and supplemental data. Visual representations such as the attack tree code can be obtained and rendered using tools such as the Mermaid diagramming tool. All results are then compiled into a single, downloadable PDF report. As well, scalability and adaptability are central to the tool's design so that it can be deployed via a transparent Streamlit CI/CD process across various environments and used by organizations of different sizes and technical capabilities. AegisShield's architecture is extensible and easily allows future enhancements such as the previously integration of additional AI models or data sources as they become available.

We did run across several technical challenges during the development process. For example, when evaluating various AI providers, the process of managing API rate limits, and optimizing performance and scalability had to considered. To address this, careful design and rigorous testing were applied. In a specific example, the MITRE ATT&CK TTP mapping process was optimized by tapping into a locally stored STIX JSON file from GitHub, which reduced the dependency on external API calls and minimized load and latency issues that were present. This enhanced overall performance and allowed the tool to remain powerful, accessible, and efficient as it avoids excessive execution times. It did not make sense to create all those excessive calls to MITRE, if it could be done locally.

Figure 3.1 illustrates the data flow and structure of AegisShield, emphasizing its modular architecture, key components, and their interactions. This figure shows how the generative AI model and OSINT sources are integrated to produce the comprehensive threat assessments. These data flow paths reveal how real-time data are processed and used to enhance accuracy.

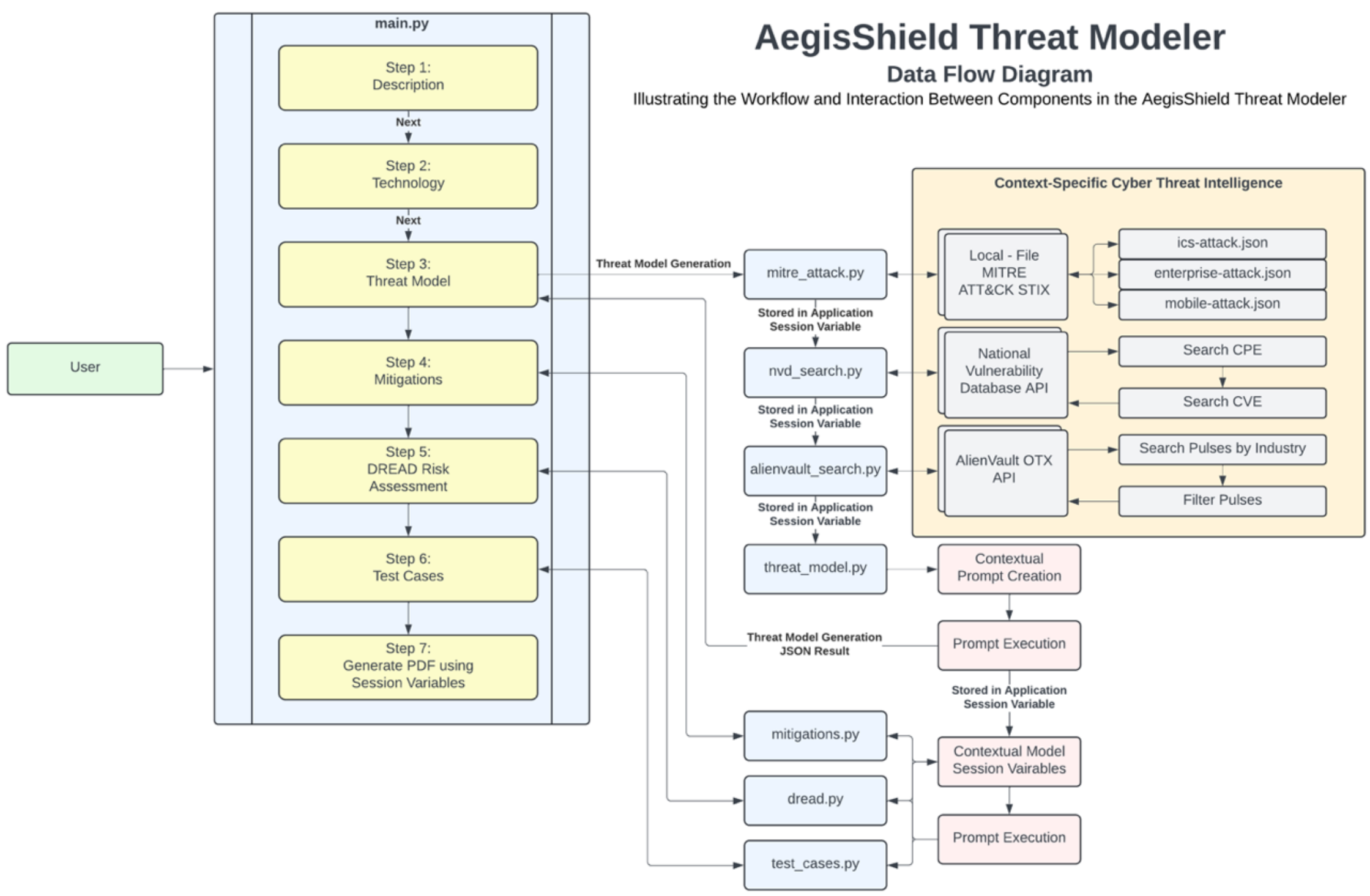

Figure 3.1. Data Flow Diagram of AegisShield

This figure also details the primary steps in the threat modeling process, from initial data collection, the acquisition of third-party data, to the generation of mitigation strategies and test cases. It highlights the integration of context-specific CTI, including sources such as MITRE ATT&CK, NVD, and OTX. It also shows how these inputs are processed to develop accurate and relevant threat models and ensures the most current data informs the strategies and test cases. For additional information, including a comprehensive list of the public APIs and resources AegisShield uses, the AI prompts, and subsequent requests embedded in the code, please refer to Table A2 in the appendix. For an illustration of the workflow, see Appendix A for screenshots of the initial user interface steps (Figures A7 and A8). Figure A7 shows the tool's first input screen, and Figure A8 shows the subsequent step where users provide the application details and constraints.

#### *3.2.2 Generative AI Prompt Creation*

The effectiveness of AegisShield is critically dependent on the quality and structure of the prompts used to engage the generative AI model. A generative AI prompt is a set of instructions provided to the model to create specific outputs. These prompts serve as the framework that shapes the AI's response. It also ensures that it generates relevant, contextually accurate, and task-specific information.

In this praxis, all the prompts used in the tool were built around the OpenAI's gpt-4o model. Several thoughts on democratizing threat modeling drove the selection of gpt-4o over other models such as Google Gemini and Anthropic Claude as the core generative AI model. OpenAI offers a powerful and scalable API with high TPS (transactions per second) rates and large token limits. This makes it particularly well-suited for supporting the research objectives (platform.openai.com). The decision prioritized development efficiency and tool functionality over broader AI capabilities.

This allowed the necessary features to be delivered promptly and effectively. AegisShield also abstracts the complexity of interacting directly with OpenAI's API, enabling organizations to adopt it more easily. After signing up with OpenAI, users may request an API key. The key can then be entered into the tool; this minimizes the need for extensive configuration. This process ensures that even non-technical stakeholders can use the threat modeling functionality.

The prompts were designed based on four principles: simulating expertise, contextual relevance, JSON structured output, and full coverage. These principles forced the AI to emulate the decision-making processes of an experienced cybersecurity professional. It then tailors its responses to specific contexts, generates easily interpretable outputs, and comprehensively addresses all aspects of the STRIDE methodology while integrating frameworks such as MITRE ATT&CK and DREAD. The design of the prompts is organized into the five components listed in Table 3.1. Each prompt was refined to direct the AI to perform a specific task. This helped ensure full coverage of the threat modeling process. Supplemental Figures A1–A6 in the appendix outline the prompts used in this praxis.

Table 3.1. AegisShield AI Prompt Architecture Components

| Component | Figure | Purpose | Methods and Techniques |
|---|---|---|---|
| MITRE ATT&CK | A1 | Select the single best attack pattern out of the 25 options presented. | Presents 25 attack patterns as a result of a keyword search, to select the single most relevant TTP |
| Threat Identification | A2 | Analyze and identify potential threats across STRIDE categories by inputting questions and external context in the tool. | Generates threat models by incorporating CIA triad considerations, MITRE ATT&CK keywords, relevant tool-specific questions, and external context such as known vulnerabilities |
| Risk Assessment | A3 | Evaluate the identified threats following a structured methodology | Applies the DREAD methodology to quantitatively assess the risk level of the identified threats, producing structured risk |

| Component | Figure | Purpose | Methods and Techniques |
| --- | --- | --- | --- |
| | | using the generated threats for context. | assessments in JSON format for integration into workflows |
| Mitigation Strategies | A4 | Develop actionable mitigations. | Generates specific and actionable mitigation strategies tailored to the identified threats, ensuring they are relevant to the context and align with the best practices in cybersecurity |
| Test Cases | A5 | Create test cases that integrate with behavior-driven development frameworks using the generated threats for context. | Uses Gherkin syntax to generate test cases that are directly applicable to behavior-driven development (BDD) frameworks, ensuring they are comprehensive and aligned with the threat models |
| Attack Trees | A6 | Visualize and map out potential attack vectors using the generated threats for context. | Constructs detailed and hierarchical attack trees that represent potential attack paths, helping to visualize and prioritize the mitigation of threats based on their likelihood and impact |

An iterative process helped guide the design of the AI prompts. The goal was to make them clear, accurate, and specific. This process is consistent with established research on prompt engineering (Velásquez-Henao et al., 2023) and integrated external domain knowledge drawn from sources such as the NVD and OTX. This aided the prompt design by adding contextual depth. The integration of third-party data into the prompts enhanced the clarity of systems being threat-modeled. This addition helped align the prompts with the unique vulnerabilities and attack vectors identified in each of the sources. The tool handled the more complicated tasks, and this allowed the user to answer a few simple questions. Meanwhile, the iterative development focused on refining the outputs across scenarios, with continual testing and revision of prompts ensuring precision and relevance. This process helped inject established cybersecurity practices into the initial draft prompts, reflecting the importance of domain expertise in applied prompt engineering (Velásquez-Henao et al., 2023).

An adaptive mechanism was also incorporated into the prompt design to adjust the complexity of the AI's explanations based on the user's technical proficiency (Ayad & Alsayoud, 2024). This feature helped guide the complexity of the resulting outputs. The complexity ranged from simplified overviews to detailed technical insights, but it was set to a default medium technical level for this research.

AegisShield employs an effective prompt engineering methodology that synthesizes iterative refinement approaches with a domain-specific focus. This method draws on established academic research while extending and adapting these concepts for cybersecurity threat modeling. The structure of the prompts—covering threat identification, risk assessment, mitigation strategies, test case generation, and attack tree generation—reflects a thoughtful application of these findings to a specialized context (Ayad & Alsayoud, 2024; Velásquez-Henao et al., 2023).

### *3.2.3 Data Collection*

In this study, the data we obtained was via a systematic analysis of 15 diverse case studies and followed established best practices for systematic reviews and thematic synthesis. These case studies spanned various industries and application types and ensured the tool's performance was assessed across various real-world scenarios. The 15 case studies used in this research were identified through extensive searches across multiple academic databases and search engines, including IEEE Xplore, Semantic Scholar, Google Scholar, and the George Washington University online library. The selection process involved reviewing a significant number of papers to locate studies that included detailed case studies resulting in a STRIDE threat model and were relevant to diverse cybersecurity domains.

A standardized set of questions (Table A1) about the system or application in focus and a comprehensive rubric (Table 3.2) were used to extract the details from each case study. The questions covered some fundamental aspects of the threat modeling process and provided a foundation for evaluating the outputs generated. As well, the previously described batch-processing script was used to ensure consistency and efficiency in generating the many required outputs for analysis.

Table 3.2. Rubric for Evaluating the Case Study Inputs

| Criteria | 5 Points | 4 Points | 3 Points | 2 Points | 1 Point |
|---|---|---|---|---|---|
| 1. Application/System Description | A thorough, clear, and detailed description of the application/system, including its purpose, functionality, architecture, and key components Enables a complete understanding of system operation | A clear and detailed description, possibly lacking depth in certain areas or omitting minor components Provides solid understanding and requires minimal clarification | A general description covering the main aspects but lacking detail and specificity Important components or functionalities may not be fully explained | A vague or incomplete description, missing significant information about the purpose, functionality, or architecture Requires substantial inference or additional research | Little to no description provided, or too vague to understand the system Critical details are absent, hindering effective analysis |
| 2. Application Type | Explicitly stated and precisely categorized (e.g., "Cyber-Physical System for Industrial Automation") Aids targeted analysis | Explicitly stated but lacks precise categorization Described in general terms | Can be inferred with reasonable confidence but is not explicitly stated Inference guides analysis but lacks confirmation | Ambiguous or vaguely referenced, making it difficult to determine the exact nature of the application Hampers targeted threat modeling | Not mentioned or cannot be reasonably inferred Prevents the application of specific threat modeling techniques |

| Criteria | 5 Points | 4 Points | 3 Points | 2 Points | 1 Point |
|---|---|---|---|---|---|
| 3. Industry Sector | Explicitly stated and relevant, with detailed context (e.g., "Healthcare sector focusing on medical device integration") Enhances the relevance of compliance and threat considerations | Explicitly stated but lacks additional context or specificity Indicates the general industry without sub-sector details | Can be inferred but is not explicitly mentioned Inference is sufficient for general analysis but may lack precision | Ambiguous or vaguely referenced, challenging the alignment of threats and compliance requirements | Not mentioned or cannot be reasonably inferred Limits the consideration of industry-specific threats and regulations |
| 4. Data Sensitivity | Explicitly described, including the types of data handled and any classification levels (e.g., PII, PHI) Informs threat modeling significantly | Explicitly mentioned but lacks detailed classification Types of data identified without specific sensitivity levels | Can be inferred based on the context but is not explicitly described | Vaguely mentioned without clear details Limits the accurate assessment of data-related threats | Not mentioned or cannot be reasonably inferred Challenges the prioritization of data confidentiality, integrity, and availability threats |
| 5. Internet Facing Status | Clearly stated, including the implications and potential exposure risks Critical for understanding external threat vectors | Mentioned but lacks detailed context or implications Clear exposure status with minimal discussion on risks | Can be inferred from the description but is not explicitly stated Inference allows general threat considerations | Ambiguous or vaguely referenced, making accurate assessment of external threats difficult. | Not mentioned or cannot be determined Hinders the identification of network exposure threats |

| Criteria | 5 Points | 4 Points | 3 Points | 2 Points | 1 Point |
|---|---|---|---|---|---|
| 6. Compliance Requirements | Explicitly stated, detailing the applicable laws, regulations, and standards (e.g., HIPAA, GDPR) and their influence on the system design<br>Ensures thorough consideration of compliance-related threats | Mentioned but lack detailed discussion or specific standards<br>Acknowledgment without in-depth analysis | Can be strongly inferred based on the industry and data but not explicitly mentioned<br>Allows the consideration of general compliance requirements | Vaguely referenced without specifics<br>Limits the assessment of compliance-related threats | Not mentioned or stated as not applicable<br>May overlook critical regulatory obligations and associated threats |
| 7. Authentication Methods | Detailed information provided on the authentication methods and access controls, including protocols, technologies, and multi-factor mechanisms<br>Aids in assessing authentication-related threats | Mentioned with some details but lacks full specificity | Mentioned but are generic (e.g., “uses passwords”) without implementation details | Vaguely referenced or can be inferred, such as “secure login,” without additional information | Not mentioned<br>Impedes the identification of unauthorized access threats |

| Criteria | 5 Points | 4 Points | 3 Points | 2 Points | 1 Point |
|---|---|---|---|---|---|
| 8. Technical Details (Database, OS, Languages, Frameworks) | Comprehensive technical details, including specific technologies and versions for databases, operating systems, programming languages, and frameworks Crucial for identifying technology-specific threats | Provided but may lack version information or completeness in one area | Mentioned but are generic or incomplete in multiple areas | Vaguely referenced with minimal detail | Not mentioned Challenges the assessment of technology-specific threats |
| 9. Threat Details | Well-defined, clearly organized by type, and presented in a structured format (e.g., tables or bullet points) Descriptions are detailed; acronyms are minimal or explained Facilitates effective analysis | Defined and organized but may lack some detail or have unexplained acronyms Overall organization aids understanding with minor clarification needed | Identifiable but not well-organized Descriptions may be brief or lack clarity Usable but may require additional effort to interpret DFD content from the paper | Hard to find, scattered in text, or lacking adequate definitions Disorganization hinders coherent threat analysis | Not clearly identified or absent Impossible to assess or model potential security issues effectively |

The rubric was developed following guidelines emphasizing both transparency and rigor in systematic data collection.  Our goal was to ensure a methodologically sound and reproducible evaluation process. These guidelines highlight the need for a trustworthy and auditable methodology to reduce bias and enhance reliability (Kitchenham & Charters, 2007). The rubric also incorporated thematic synthesis to systematically identify and classify any recurring patterns, themes, and insights essential for evaluating the tool. This approach allowed us to create a structured interpretation of the findings, focusing on dimensions such as clarity and completeness of system descriptions, security, and compliance requirements (Cruzes & Dyba, 2011).

The iterative refinement of the rubric in this research—driven by prompt-engineering testing—ensured both robustness and accessibility and struck a balance between comprehensiveness and usability during data extraction (Kitchenham & Charters, 2007). It established a standardized classification process and scoring protocol, thereby reducing free-form variability and minimizing subjective interpretation of category definitions. The rubric scores provided insights into the level of detail and range attributed to each case study's inputs, offering a consistent means to view data easily across the 15 case studies. This method supported a cohesive analysis and upheld the level of methodological rigor advocated in prior work (Cruzes & Dyba, 2011; Kitchenham & Charters, 2007).

## 3.3 Analytical Techniques

### *3.3.1 Initial Exploratory Data Analysis*

In this praxis's initial exploratory data analysis (EDA) phase, several steps were built using a Colab notebook to build a structured and comprehensive understanding of the dataset. The primary objective of the EDA was to assess the dataset for potential irregularities, such as bias or disproportionate weighting, and ensure that the data were suitable for continued analysis and reliability. As with most EDA processes, several standard Python libraries, such as Pandas, Numpy, and Matplotlib were installed. Next, the tool and case study threat datasets were imported into a Pandas DataFrame for streamlined data handling. Preliminary inspections were then conducted using basic functions such as .info() and .describe() to assess data completeness, identify missing values, and review the data types. This provided an overview of the dataset's structure and key attributes and formed the basis for a more a deeper analysis.

Subsequently, threat analysis was conducted to identify patterns, distributions, and potential anomalies. Data visualization techniques, including histograms, boxplots, word clouds, and radar charts, generated using Matplotlib and Seaborn, were used to explore the word count distribution and variability within the threat categories. The Anderson-Darling, Ryan-Joiner, and Kolmogorov-Smirnov normality tests were conducted to determine whether the data adhered to a normal distribution and to understand the underlying trends. A keyword frequency analysis was performed on the threat descriptions to identify any recurring themes. Further examination used the rubric in Table 3.2 for the case studies and compared the AI-generated threat models with the expert-curated ones. This comprehensive approach laid the groundwork for an in-depth analysis while supporting the study's objective of democratizing threat modeling.

### *3.3.2 Cosine Similarity Analysis*

Cosine similarity is the primary metric for evaluating the STS between tool-generated and expert-developed models. The batch-processing script results were loaded into a Colab notebook for analysis alongside the STRIDE threats extracted from expert case studies. A pre-trained sentence transformer (stsb-roberta-large) generated the embeddings for both the tool-generated and expert-extracted STRIDE threat scenarios. These embeddings were then compared using cosine similarity to assess the semantic closeness. While the stsb-roberta-large model was deprecated due to the production of lower-quality sentence embeddings in some cases, it outperformed other models tested.

Meanwhile, multi-qa-mpnet-base-dot-v1 (https://huggingface.co/sentence-transformers/multi-qa-mpnet-base-dot-v1) underperformed. It missed several obvious matches, but msmarco-bert-base-dot-v5 produced abnormally high cosine similarity scores for items with little semantic resemblance.

In contrast, stsb-roberta-large provided the most reliable scores for the threats measured. As noted in prior research, general-purpose models including multi-qa-mpnet-base-dot-v1 and task-specific models such as msmarco-bert-base-dot-v5 often struggle with semantic distinction in specific contexts, leading to inconsistent results when compared with the models fine-tuned for semantic similarity tasks (Vahtola et al., 2022). Although further investigation may be required to understand the underlying factors, stsb-roberta-large was selected due to its consistent and accurate performance for this specific use case and as it ensured that the comparison effectively captured contextual meaning and threat-specific nuances.

This method was selected for its proven effectiveness and computational efficiency in STS tasks, mainly using sentence embeddings generated by SBERT (Cer et al., 2017; Reimers & Gurevych, 2019). This reduced the computational overhead for similarity comparisons while maintaining high accuracy in capturing semantic meaning. To handle the large volume of data generated during batch processing, an A100 GPU was used in Colab to calculate the embeddings.  This sped up the process over using a standard CPU.  Then, AI-generated and expert-developed threats were converted into embeddings, so that each could be evaluated within a consistent vector space. Our research indicates that thresholds for determining semantic similarity using cosine similarity typically range between 0.5 and 0.7, depending on the task and dataset (Cer et al., 2017). To statistically validate this threshold, a one-proportion test was conducted for each case study to determine whether the proportion of successful batches exceeded 50%.

Each of the case study's generated threats were compared with the expert-developed threats. Cosine similarity scores determined the semantic alignment. As stated, a success threshold of a cosine similarity score ≥ 0.7 was used for at least one threat per batch in over 50% of the 30 executions per case study to obtain a majority vote. To reduce noise across STRIDE categories, only similar STRIDE threat types were scored between the tool-generated and expert-created threats. For example, spoofing threats generated by the tool were scored only against the spoofing threats in the case studies. Meanwhile, boxplots and violin plots were used to visualize the distribution of the cosine similarity scores for each case study. This helped visualize and compare the similarity scores between tool-generated and expert-generated threats. These visuals helped identify the spread and consistency of similarity scores across different domains as well. Next, heatmaps were used to visualize the number of semantically similar threats per batch and case study. These provided an easy-to-see overview of how batch success varied across different case studies, highlighting how often AegisShield generated threats that met the 0.7 similarity threshold. A correlation analysis was then performed to evaluate the relationship between the similarity scores and rubric scores, such as application type, industry sector, and threat count. This helped determine whether the tool's performance was influenced by the specific characteristic quality of the case studies or remained consistent.

Cosine similarity calculations were performed using the scikit-learn library, a well-established tool for ML and NLP tasks. This commonly used tool allowed us to objectively measure the tool's output against expert-developed threat models and provided consistent evaluations across multiple case studies and domains. Scikit-learn enables a fast, reliable computation of cosine similarity scores, which was necessary to evaluate the tool's performance in generating meaningful and comparable threats to those produced by experts.

### *3.3.3 Readability Assessments*

For readability metrics hypothesis 1, this research used the Flesch-Kincaid test to ensure that the outputs generated by the tool are accessible to users with varying levels of cybersecurity expertise. The Flesch-Kincaid test measures readability with a focus on sentence length and syllable count. The resulting statistic indicates the grade level required to comprehend the text. This test is frequently applied across various fields and subjects, including education, government publications, and technical documentation, to evaluate the ease of reading complex materials (Flesch, 1948). Using the "textstat" Python library and a Colab notebook, our test was done to measure the grade level of both tool-generated and expert-extracted threat descriptions. The results from the test helped validate whether the tool's outputs were readable and accessible to a broader audience. In other words, the main goal was to demonstrate that the AI-generated outputs had statistically significant lower Flesch-Kincaid grade-level scores when compared to expert-developed models. This would indicate a reduced complexity. This test supported the specific goal of democratizing threat modeling by making it less complicated for organizations new to the process to understand the generated threat descriptions. These assessments also served as a critical criteria point for Hypothesis 1, which ultimately examines whether the threats from AegisShield are considerably less complex than those from the case studies. Finally, to assess whether the reduced complexity was indeed statistically significant, a Mann-Whitney U test with a 0.05 significance level was conducted in Minitab due to the non-parametric nature of the data.

Next, a “medium” technical threshold was established in the tool to quantify the reduced complexity. In practice, the “medium” mode is designed to balance readability with technical depth. The methodology evaluates whether this setting results in descriptions that are less complex than expert-developed descriptions while at the same time retaining sufficient detail for the detailed threat similarity analysis conducted in hypothesis 2. This approach helps ensure an objective assessment of readability, helping drive clearer communication and decision-making capabilities among diverse teams. AegisShield also allows for adjustments in technical depth, either higher or lower, to align with user needs, contextual requirements, and potential future demand.

Despite the ability for Flesch-Kincaid to score readability, it does not capture the technical depth and nuance and risks oversimplification if used in isolation. As a result, this research also evaluated AegisShield’s threats for technical accuracy (via semantic similarity to expert models, H2) and framework alignment (via MITRE ATT&CK TTP alignment, H3). This ensured that the lower Flesch-Kincaid grade levels did not occur at the expense of either precision or breadth of threat coverage. This made the threats both accessible and technically sound. This praxis explores the readability tests through visualized through boxplots and histograms to allow a clear comparison of the grade levels between the two sets of threats while at the same time providing a comprehensive view of the distribution and variance of those scores.

### *3.3.4 MITRE ATT&CK Framework*

The integration of AegisShield with the MITRE ATT&CK framework was an important step in accurately mapping STRIDE-categorized threats to relevant TTPs. But how was this done? First, three descriptive keywords were generated for each threat scenario to query the MITRE ATT&CK database. This search produced up to 25 relevant TTPs. The keywords were then matched against both the names and the descriptions of the attack-pattern objects in the MITRE ATT&CK database. Next, the AI refined this list by evaluating the application's context, including factors such as application type, industry, authentication mechanisms, internet-facing components, and sensitive data and chose a single mapping to apply to the threat.

The batch-processing script generated JSON outputs of STRIDE threats mapped to MITRE ATT&CK TTPs. A new column labeled "Mapped to MITRE TTP" was created for each threat to indicate whether it had a valid mapping. A threat was considered as mapped if its associated IDs were valid; unmapped threats with the ID "attack-pattern--00000000-0000-0000-0000-000000000000" and technique TID "N/A" were excluded. Bar charts were used to compare the mapped versus unmapped threats, pie charts to show the distribution of the mapped threats, and radar charts to compare the mapping success across the STRIDE categories.

Ultimately, the context of the case studies drove the tool's dataset selection. For example, when analyzing a mobile application, it used the MITRE ATT&CK dataset specific to mobile threats. Similarly, for desktop, web, or cloud services, the tool referred to the Enterprise ATT&CK dataset. For ICS and SCADA environments, the ICS ATT&CK dataset was applied. This works well for very specific and defined systems, but what about systems that are multi-platform? When dealing with applications that span multiple platforms, such as IoT systems with both ICS and enterprise components, the tool combined data from relevant datasets to provide a more accurate threat model (*Mitre-Attack/Attack-Stix-Data*, 2024). This contextual blending enables the AI to align each threat with the system's unique characteristics and vulnerabilities and eventually select the most relevant TTP for the threat. The result ends in a tailored threat model that provides an appropriate framework for threat modeling.

Let us now discuss how we calculated the mapping percentage and why this was important. The AI batch tool was executed multiple times across different case studies in the evaluation process, generating 8100 threats across all runs. The mapping accuracy was determined by calculating the ratio of AegisShield threats successfully mapped to one or more corresponding MITRE ATT&CK TTPs. A threshold of 80% successful mapping was used as the criteria for validation. This threshold reflects an application of the Pareto Principle as discussed by Koch (2011), which summarizes that addressing 80% of the most critical issues often yields a disproportionately large impact. A one-proportion test was applied to statistically assess whether the observed mapping rate exceeded this 80% benchmark. This was done with the understanding that complete coverage (100%) may require extensive effort and cost with diminishing returns.

***3.3.5 Real-time Threat Intelligence Integration***

AegisShield also integrates data from the NVD, AlienVault OTX, and as stated previously, MITRE ATT&CK, to enhance the accuracy of its threat models by providing real-time intelligence. AegisShield uses the "NVDLib" library to search for related CVE entries based on the technologies and versions used provided by the user. This process involves invoking the NVD API to prioritize any related vulnerabilities. The returned results are then filtered by CVSS score and recent publication dates, so the prompt is feed only the most recent and critical threats. By having AegisShield run using this real-time data, we can make sure that the models generated reflect the latest security risks relevant to the technologies in use.

The tool also uses OTX to gather community supplied threat intelligence. This helps to refine the threat models about emerging threats. It executes queries tailored to specific industries, filtering and sorting OTX pulse responses by relevance and recency. Because of this, we can make sure only the most pertinent threat intelligence, including historical indicators of compromise (IOCs) and insights on malware families and adversaries are used. In summary, by combining structured vulnerability data from the NVD with real-time threat intelligence from OTX, AegisShield provides a thorough and up-to-date threat model that accounts for known vulnerabilities and risks across various sectors.

### 3.4 Case Study Methodology

This study employed a method called qualitative comparative analysis (QCA) to examine the 15 selected case studies. QCA is well suited for small-to-medium N samples, typically ranging from 5 to 50 cases. QCA is used to conduct structured assessments of complex configurations and outcome patterns within a limited number of cases. (Cairns et al., 2017). For this praxis, we chose the current sample size of 15 based on the following factors: data availability, completeness, correctness, and proper structuring. Each case represents an acceptable source where the required data can be extracted and scored using a rubric. This helped us with the integrity and consistency of the data across all cases and is an important factor for the validity of QCA results. While we feel it is possible that a larger sample size might provide broader coverage, our preliminary analysis suggested that increasing the number of cases beyond 15 could also lead to data overload. This could produce diminishing returns. In the end, the rubric scores showed no significant variation with the addition of cases, indicating that the main reason things happened the way they did was captured effectively within the used sample size.

The case studies were selected to help assure that AegisShield could be adequately evaluated across a mixture of real-world scenarios. We used various criteria to capture the complexities and variations that existed in these different case study application and systems.  This provided a solid foundation for the analysis. To start, each case study involved a researcher going through and writing out a threat modeling exercise on a specific application or system. As we see in Table 2.1, most of the case studies were published in well-respected journals or conferences, and each one resulted in a STRIDE model. This allowed us to assume that the threat model outcomes in the papers we both credible and reflective of expert practice. This approach provided a well-vetted and reliable source of threat modeling data from a researcher's perspective. When this was paired with the resulting analysis, it ensured that the case studies were directly relevant to the research objectives and allowed for a targeted evaluation of AegisShield's effectiveness. Another critical criterion for the selection was the availability of detailed descriptions of the applications or systems being modeled. The availability of this data was essential for extracting the information necessary for the subsequent testing of AegisShield.

As previously stated, another important data point was that each case study must include the outcome of a STRIDE threat model. This allowed for an equal comparison of the outputs within a defined framework that were created by AegisShield and extracted from the case studies. Another point, the case studies also spanned a wide array of domains, such as healthcare, finance, ICS, IoT, and mobile. This diversity in sectors tested the tool's applicability across industries with varying regulatory requirements or guidelines.

The selected case studies varied in the amount of STRIDE threats they contained. This variation helped provide a comprehensive basis for evaluating AegisShield against case studies that varied in their threat model complexity. The case studies also varied in terms of their extractable information. Some studies offer extensive data in some areas while others provide limited content. With this in in mind, we were able to test the tool's adaptability to different data availability levels by simulating imperfect real-world conditions.

Case study age and source diversity was also a major consideration. The case studies spanned different years and originated from various academic and professional sources. This allowed us to test against a broad range of threat modeling research inputs and outputs and increased the generalizability of the findings. To build on this, these case studies also varied in terms of gaps identified, obstacles, and objectives within threat modeling. In the end, all these case studies provided a rich context for assessing how AegisShield could address diverse scenarios. This helped contribute to a more nuanced understanding of its effectiveness.

It's important to note that in cases where the industry sector or compliance requirements were not directly specified, they were inferred based on the descriptions of the system or the technology in question. This approach is supported by well-established practices in systematic reviews, which often involve the process of synthesizing incomplete data to form a comprehensive analysis (Kitchenham & Charters, 2007). For instance, if a case study described a system that stored hospital patient data but did not mention healthcare, we inferred the sector as healthcare and applied the Health Insurance Portability and Accountability Act (HIPAA) to it in the rubric. In a similar fashion, International Organization for Standardization (ISO) 13485 compliance was inferred for systems related to manufacturing medical devices. This process continued for all case studies as needed.

Ultimately, these inferences were critical for applying domain-specific considerations during the threat modeling. The implicit or explicit industry sector was used to conduct targeted searches within OTX for relevant threat intelligence within that sector. AlienVault refers to these broad CTI dumps as "pulses." The compliance requirements and industry mappings that were obtained and integrated into the prompts were used by the AI to generate detailed threat models that considered both the specific regulatory frameworks and standards applicable to the domain. As a final example, in the e-commerce sector, where compliance with regulations such as the Payment Card Industry Data Security Standard (PCI DSS) is important, the tool generated threats and mitigation strategies aligned with these standards. Much like in the approach outlined by Kitchenham & Charters (2007), integration of compliance considerations in this praxis is consistent with thematic synthesis methodologies, which involve making informed judgments to interpret and analyze sometimes incomplete data (Cruzes & Dyba, 2011). Overall, the AegisShield's ability to adapt to different sectors and compliance needs, even when not explicitly provided in the case study, demonstrated a high degree of flexibility and potential for it to be applied broadly across various industries.

## 3.5 Ethical Considerations

### *3.5.1 Privacy and Data Security*

Given the sensitive nature of cybersecurity research, particularly in the context of AI-enhanced threat modeling, this research implemented strict measures to safeguard privacy and data security. Consistent with the Threat Modeling Manifesto, which prioritizes the enhancement of system security and privacy through structured threat modeling practices, stringent privacy standards were upheld in the tool while mitigating bias in the AI models used (*Threat Modeling Manifesto*, n.d.).

First, the tool was designed to prevent the input of personal or organizational data that could help identify individuals or entities. Users were instructed to not provide any identifying information in the descriptions used for threat modeling, ensuring that the data remained anonymized and privacy was preserved. Second, when the application began, the tool requires the entry of tokens for the generative AI model and external services. We provide Streamlit stored environment variables for the NVD and OTX services. Once used, these tokens were securely kept in session memory and destroyed immediately upon the application's termination. This ensured no sensitive data were retained after use. The user also could use their own NVD and OTX tokens. Third, all the data transmitted between the user and the tool and between the tool and external services used TLS encryption in transit. This protected the data from interception during communication. Fourth, the OpenAI model used for generating these threat models operate under the account associated with the token provided. Any account details or data used during these sessions therefore are inaccessible to the researcher. This helps ensure data remains confidential and under the user's control. As a final point, the tool is designed to conduct all operations in session memory, and data is not stored on the server. This architecture reduces the risk of data breaches as no residual data is left once the session ends. This contributes to a strong overall data security and privacy posture.

#### *3.5.2 Bias Mitigation*

While developing and testing AegisShield, care was taken to minimize biases in its logic and in any inferred system descriptions. Specifically, references to personal pronouns (e.g., "I think" or "you must") were avoided. We also took care to avoid sentences with emotive or value-laden words (e.g., "dangerous," "catastrophic," or "trivial") unless specifically pulled in via external sources such as the NVD or AlienVault OTX. This also applied to culturally or socially charged phrases and personal viewpoints on politics, religion, or other external factors. All of this was done so that the prompts were presented in neutral, technical terms. Finally, the AI was directed to produce responses in a JSON format. This templating request reduced the likelihood of unintended bias or subjective commentary. Overall, we recognized the potential limitations of OpenAI, which is trained on diverse data. We also took some smaller additional measures to refine the prompts so that they remained objective and focused on cybersecurity. For example, prompts were carefully structured to avoid ambiguity and guide the AI toward generating responses that were accurate and impartial. The restriction of words like "secure" or "insecure" were important in not tilting its response in one direction. This involved asking specific, well-defined questions, validating outputs and explicitly requesting multiple perspectives during the design process. These measures are consistent with recent research emphasizing the importance of addressing biases inherent in AI systems. This happens in cases where pre-existing datasets and black-box algorithms can inadvertently amplify biases if not carefully managed. In particular, as Kitchenham and Pfleeger (2025) highlight that in summary, understanding the risks of AI bias requires thorough techniques that ensure fairness, accountability, and trustworthiness, especially as the reliance on AI systems grows across various domains. This is why we must make sure to acknowledge neutrality and transparency in prompt design so we can rectify the risk of bias influencing outputs.

## 3.6 Limitations of the Methodology

### *3.6.1 Constraints*

This praxis faced several challenges in its research methodology that may have shaped the study's comprehensiveness. First, resource limitations restricted the depth of threat intelligence integration. AegisShield's AlienVault OTX searches were capped at the five most recent pulses, which could have excluded broader or emerging threats. In cybersecurity, new threats appear constantly. Second, the NVD search was limited as well; only the 10 most recent active results per technology were used. If a user did not select any technologies, this part of the contextual threat analysis was skipped and did not contribute to the prompts at all. Any of these limitations might have caused older, but still relevant, vulnerabilities to be omitted, which could have influenced the accuracy and completeness of threat models. The main reason for these constraints was resource availability: limited computational resources, OpenAI token limits, and time to process large volumes of intelligence data. These limitations may change in the future as technology improves.

### *3.6.2 Impact on Findings*

The constraints described above could have had some implications for the findings, though their actual impact was likely minimal. The OTX (limited to 5) and NVD (limited to 10) searches did not significantly affect the tool compared to manually testing a much larger set of data. In other settings, though, or when using different systems and datasets, these limitations could matter more for finding specific threats.

The strategies used were intended to strike a balance and reduce bias in the resulting threat models, yet some inherent biases might still have remained. The model's training data is opaque; OpenAI provides no details about what data it is trained on. The study tried to offset this by using current threat intelligence and incorporating the most critical threats into models. However, these limitations are real and should be kept in mind when interpreting results, since they reflect the boundaries of the research environment.

# Chapter 4—Results

## 4.1 Introduction

Chapter 4 presents the findings obtained by analyzing the data gathered across the 15 case studies and batch runs conducted in this research. To reiterate, the core research problem was that existing threat modeling tools struggle to effectively identify threats early due to increasing system complexity and evolving dynamics. This challenge especially affects organizations that have limited expertise and budget. The broad objective was to democratize threat modeling by developing a tool that simplifies and automates the threat modeling process, making it accessible to these types of organizations with limited resources.

The results presented here are organized to assess the validity of the three research hypotheses.

- H1: AegisShield can reduce the complexity of threat descriptions compared to expert-developed models while maintaining a complete coverage of threat categories.
- H2: AegisShield can generate outputs that exhibit semantic similarity to expert-developed models, as indicated by the cosine similarity using Sentence-BERT, regardless of the specific domain.
- H3: AegisShield can systematically map STRIDE-categorized threats to relevant MITRE ATT&CK Tactics, Techniques, and Procedures (TTPs), ensuring comprehensive alignment with established cybersecurity frameworks.

The rest of the chapter is structured as follows: Section 4.2 presents a high-level overview of the data, highlighting the diversity and completeness of the case studies. Sections 4.3 - 4.5 provide detailed analyses and results corresponding to each hypothesis. Section 4.6 examines the time taken by the tool to process threat results. Finally, Section 4.7 summarizes the findings and draws conclusions from the results.

## 4.2 Data Overview

### *4.2.1 Case Study Data*

The data collected from the 15 case studies, each scored using a standardized rubric, provided a comprehensive foundation for evaluating the performance of AegisShield. These case studies, taken from sectors such as healthcare, energy, manufacturing, and IoT, represent diverse architectures and security challenges. The variation provided a rich dataset for the analysis. The following case study rubric data provides background for the upcoming hypothesis tests, and the scores reveal the variability and complexity of the systems modeled. Table 4.1 contains the rubric scores, and Table 4.2 the descriptive statistics.

Table 4.1. Rubric Scores for Evaluating the Case Study Inputs

| Case# | Crit. 1 | Crit. 2 | Crit. 3 | Crit. 4 | Crit. 5 | Crit. 6 | Crit. 7 | Crit. 8 | Crit. 9 | Threat Count |
|---|---|---|---|---|---|---|---|---|---|---|
| 1 | 5 | 3 | 3 | 3 | 4 | 1 | 1 | 1 | 3 | 16 |
| 2 | 1 | 5 | 3 | 5 | 5 | 1 | 1 | 2 | 3 | 12 |
| 3 | 5 | 5 | 5 | 5 | 5 | 5 | 1 | 1 | 4 | 20 |
| 4 | 5 | 5 | 5 | 5 | 5 | 3 | 1 | 1 | 5 | 10 |
| 5 | 5 | 5 | 5 | 5 | 5 | 3 | 5 | 2 | 5 | 13 |
| 6 | 5 | 5 | 3 | 5 | 5 | 1 | 1 | 1 | 4 | 15 |
| 7 | 5 | 5 | 3 | 5 | 5 | 1 | 5 | 1 | 5 | 29 |
| 8 | 5 | 5 | 5 | 5 | 5 | 3 | 1 | 1 | 5 | 32 |
| 9 | 5 | 5 | 3 | 5 | 5 | 1 | 1 | 1 | 5 | 13 |
| 10 | 5 | 5 | 5 | 5 | 5 | 3 | 1 | 1 | 4 | 13 |

| Case# | Crit. 1 | Crit. 2 | Crit. 3 | Crit. 4 | Crit. 5 | Crit. 6 | Crit. 7 | Crit. 8 | Crit. 9 | Threat Count |
|---|---|---|---|---|---|---|---|---|---|---|
| 11 | 5 | 5 | 5 | 5 | 5 | 3 | 1 | 1 | 4 | 11 |
| 12 | 5 | 5 | 5 | 5 | 5 | 3 | 1 | 3 | 4 | 15 |
| 13 | 5 | 5 | 5 | 5 | 5 | 3 | 5 | 3 | 4 | 11 |
| 14 | 5 | 5 | 5 | 5 | 5 | 3 | 1 | 3 | 5 | 13 |
| 15 | 5 | 5 | 5 | 5 | 5 | 1 | 5 | 3 | 3 | 20 |

Note: "Crit." = Criteria as defined in Table 2.1.

Table 4.2. Descriptive Statistics for the Case Study Variables

| Variable | Mean | StDev | Minimum | Maximum | Range |
|---|---|---|---|---|---|
| Application/System Description | 4.73333 | 1.03280 | 1 | 5 | 4 |
| Application Type | 4.86667 | 0.516398 | 3 | 5 | 2 |
| Industry Sector | 4.33333 | 0.975900 | 3 | 5 | 2 |
| Data Sensitivity | 4.73333 | 0.703732 | 3 | 5 | 2 |
| Internet-facing Status | 4.93333 | 0.258199 | 4 | 5 | 1 |
| Compliance Requirements | 2.46667 | 1.40746 | 1 | 5 | 4 |
| Authentication Methods | 2.06667 | 1.83095 | 1 | 5 | 4 |
| Technical Details | 1.66667 | 0.899735 | 1 | 3 | 2 |
| Threat Details | 4.2 | 0.774597 | 3 | 5 | 2 |
| Threat Count | 16.2 | 6.52687 | 10 | 32 | 22 |

Most of the case studies had strong system descriptions (mean: 4.73). This suggests they provided enough detail for effective threat modeling. Similarly, the application types scored an average of 4.87, indicating that the diversity of the systems was largely identifiable. They ranged from straightforward IoT-based solutions to more complex setups, such as ICS and vehicular fog computing.

The data sensitivity scores were also notably high (mean: 4.73), indicating that the sensitivity levels were identified in many case studies. The internet-facing score (mean: 4.93) shows that nearly all the systems provided clear information, either stated or inferred, about their internet-facing status, including the exposure implications, with minimal variability in data quality (standard deviation: 0.26). The identification of whether a system's trust boundary interacts with the public internet can absolutely affect its security posture. An internet facing service would inherently have a larger attack surface than an air gapped system.

The compliance requirement scores showed much more variability across the case studies (mean: 2.47, standard deviation: 1.41). This was due to the diversity of the case studies and their respective domains. The Healthcare and the auto industries face unique compliance requirements (e.g., HIPAA and ISO/SAE 21434). Where an industry like social media may not. Meanwhile, the authentication methods also varied (mean: 2.07, range: 4); some systems lacked identifiable authentication mechanisms while others explicitly listed protocols.

The threat counts also varied (min: 10, max: 32, mean: 16.2, standard deviation: 6.53), reflecting the number of case study threats per system. This indicates different levels of system complexity and risk exposure. All these case studies varied in complexity. Because of this, the number of threats varied as well. The studies on vehicular fog computing and infotainment systems were more complex systems and therefore had more threats defined in the research. Simpler systems, such as those in voice-based applications, have fewer threats. The variation in the number of threats pulled from each case study highlights how in-depth each researcher modeled each system.

Although real-world threat modeling is typically integrated into an organization's development workflow and conducted iteratively to match design changes in manageable system components (*Threat Modeling Manifesto*, n.d.), the case studies in this research modeled entire systems. This broader scope also explains why larger, more complex systems produced a higher threat count with a more diverse context. This difference in threat counts across all the case studies showcases how varied threats in different systems are profiled by different researchers. The practice of threat modeling needs a tool that can adapt to this type of environment.

These findings are illustrated in Figure A9. This radar chart shows a comparison of rubric scores for each case study, broken down by the item scored. This figure helps tie the quantitative values to the analytical insights. It offers a clearer picture of how security requirements differ among the case studies.

The diversity in the case studies was substantial. It allowed us to test AegisShield across a broad and generalized range of environments and supported the need for a flexible approach. The rubric scores helped visualize this variation and showed that AegisShield was tested with a truly varied pool of information.

The rubric analysis helped identified security gaps and opportunities for the tool to effectively address. These findings inform the subsequent sections where the tool's performance in threat identification will examine how it compares with expert-generated threat models.

#### *4.2.2 Threat Scenario Data*

This section analyzes the threats identified from the case studies and AegisShield using word count as a primary metric for comparison. AegisShield was configured to generate three threats for each STRIDE category resulting in 18 threats per tool execution. The tool was run in 30 batch executions. Each execution generated 540 threats per case, resulting in 90 threats per STRIDE category.

243 total threats were extracted from the case studies. Figure 4.1 below provides a breakdown of the threat counts with Tampering the most prevalent, and Repudiation the least.

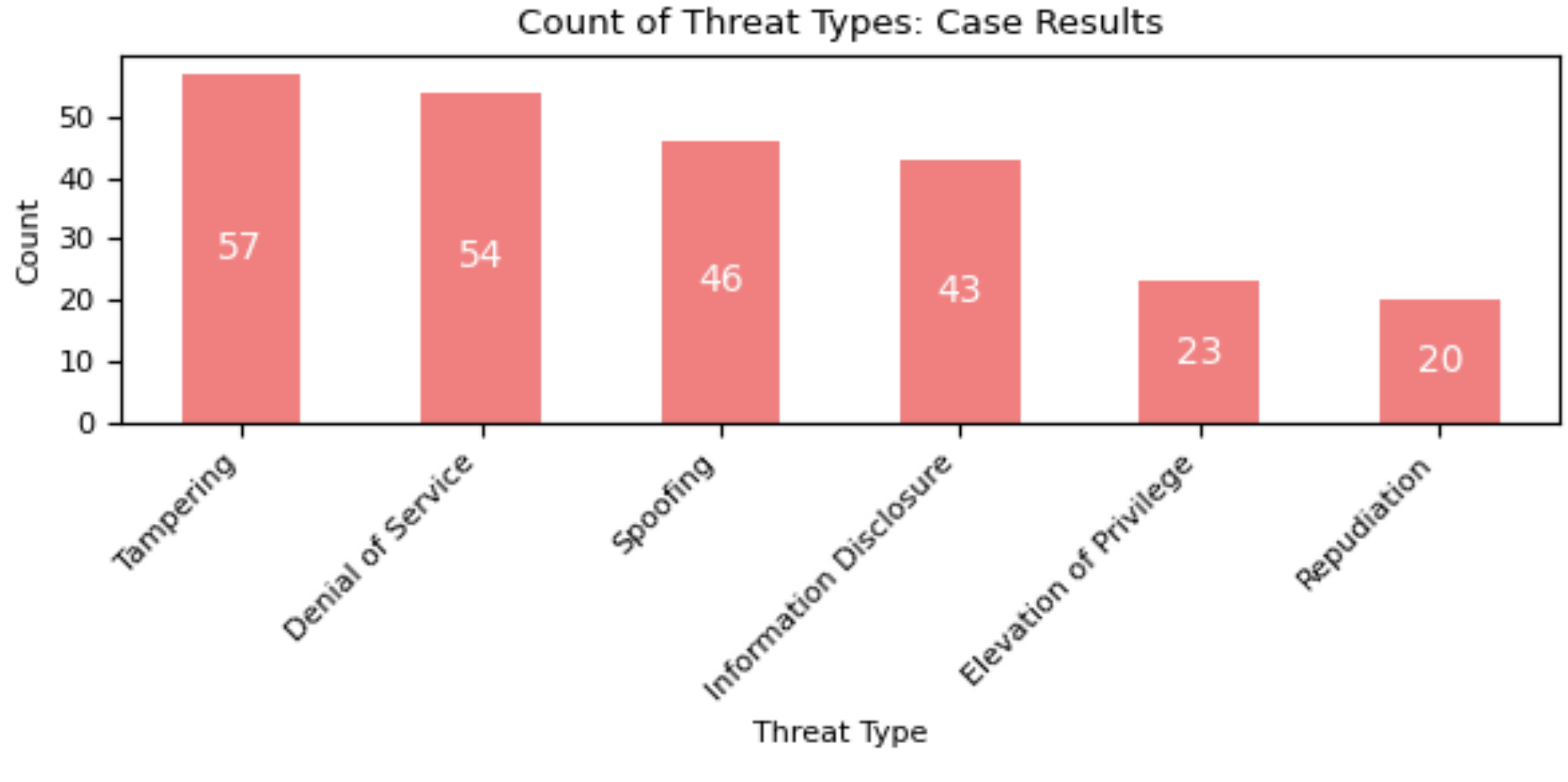

Figure 4.1. Case Study Threat Counts by STRIDE category

The substantial number of DoS and tampering threats versus repudiation threats shows how threats in certain categories manifest in the real world. DoS and tampering are more common in complex systems and industries that are generally exposed to more breaches and availability issues. Public administration, healthcare, transport, and communications are some of the most crucial industries affected by these threats (Robinson, 2024). In contrast, repudiation threats show the lowest count of represented threats. This is because attacks against repudiation are generally associated with legal, transactional, or intellectual property. Attacks in these areas are typically not as time sensitive as attacks in other domains and may also be underreported.

Next, the word count of the threats from the case studies and tool outputs were measured across the six STRIDE categories. A radar chart (Figure A10) illustrates the average word count by threat type for both the tool results and the case study findings. Notably, the word count differences between the two sources were significant. The case studies produced longer descriptions, particularly in the DoS and spoofing categories, while the tool-generated threats had more uniform word counts across categories (Figure 4.2). Generative AI consistently produces responses that fall within a constrained character range.

Figure 4.2 shows a comparison of the word counts broken out by STRIDE category. The threats in the case studies had longer descriptions in categories like DoS and spoofing. This suggests that the researchers provided slightly more detailed analysis of these threats than generative AI. The tool-generated descriptions are shorter and more uniform across all categories. This shows the capped and predictable nature of AI-generated content.

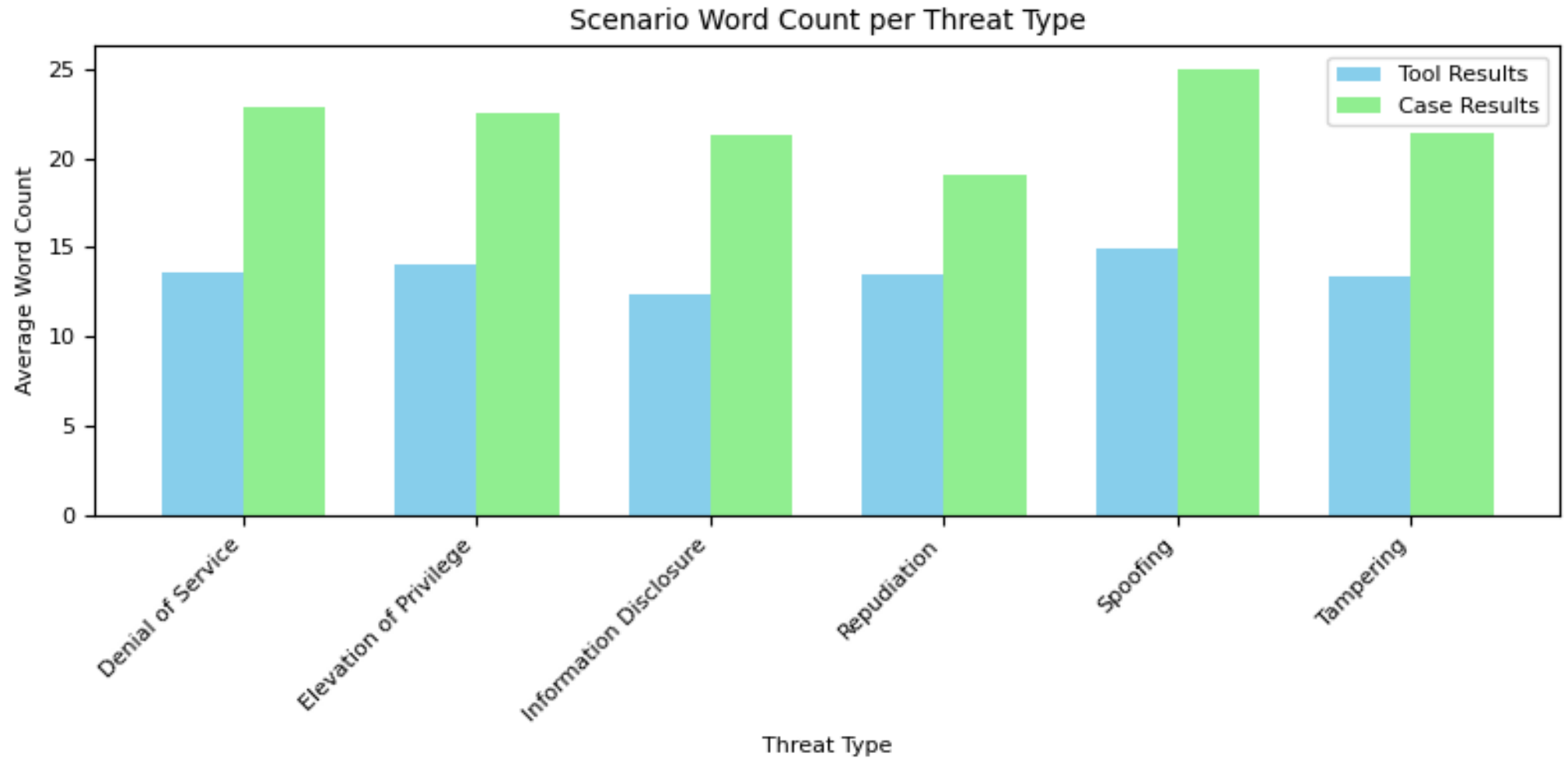

Figure 4.2. Bar Chart of the Word Count by Threat Type and Source

Figures A11 and A12 provide boxplots highlighting the differences in word count variability between the tool-generated and case study threats. Figure A11 shows that the tool-generated threats had relatively consistent word count distributions, with most threat types having similar interquartile ranges (IQR) and median word counts across the STRIDE categories. In contrast, as per Figure A12, there was more variability in the case study threats, particularly in categories such as DoS and spoofing, with larger and more dispersed word counts. This again shows that sentences created using generative AI are more uniform in length compared to the expert-created threats.

The word clouds in Figures A13 and A14 show the word prevalence for each threat dataset. In the case studies, the words "attacker", "device", and "system" are more prominent.  This shows the core focus of the researchers when creating these threats.  The tool-generated threats, however, emphasize the words "vulnerability", "cloud", and "unauthorized access" and align with generalized risks and conditions.

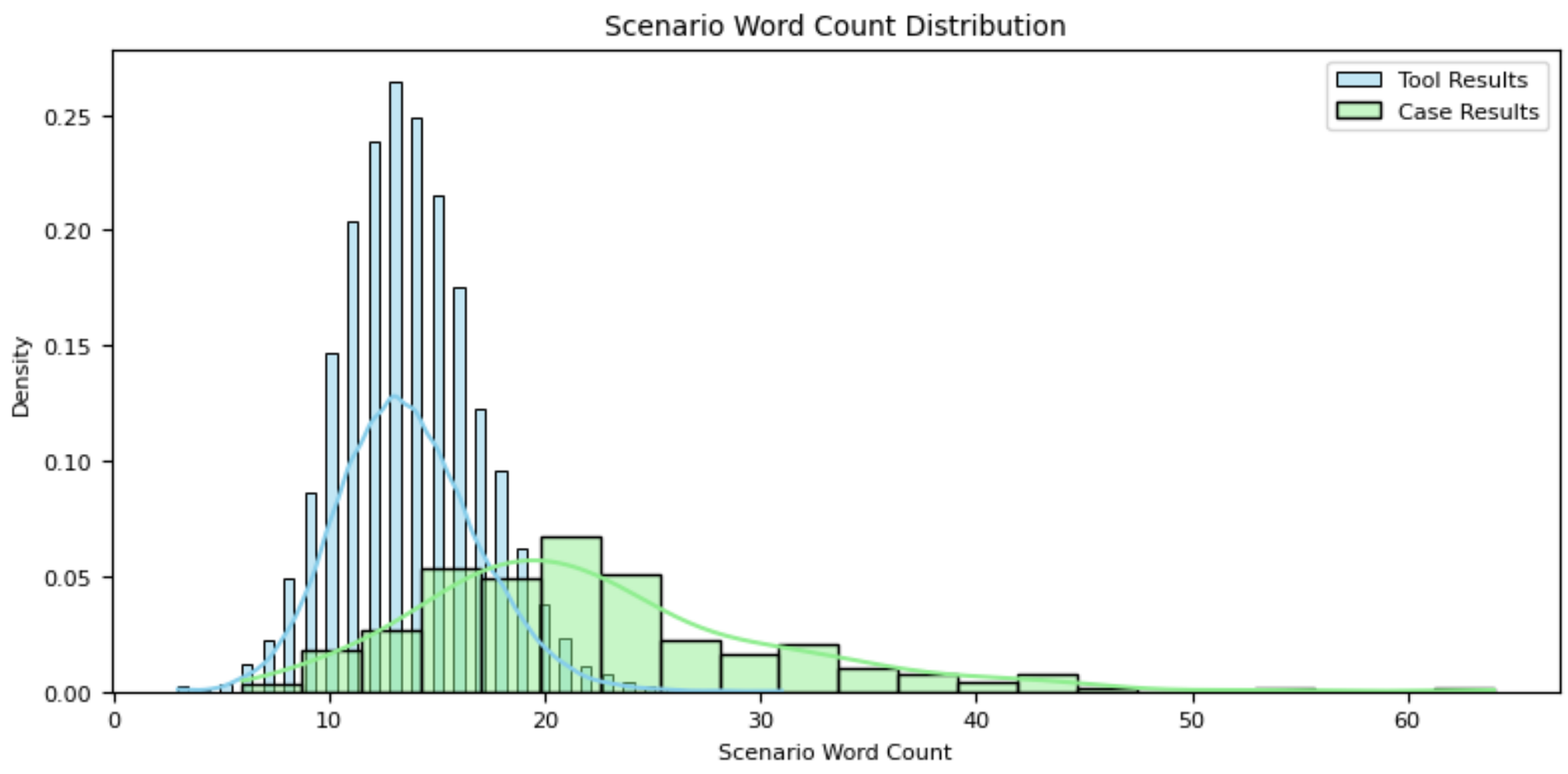

Figure 4.3. Word Count Distribution by Case Study and Tool-generated Threats

Figure 4.3 illustrates the word count distribution for case study and tool-generated threats. The case studies showed a broader distribution, with some threats containing much longer descriptions. Both datasets failed the normality test (Anderson–Darling, $p < 0.005$), so a Mann–Whitney U test was conducted to compare their median word counts. The case study threats had a significantly higher median word count (21) compared to the tool-generated threats (13; $W = 33{,}096{,}358$, $p < 0.001$). This suggests that the case study threats often included more detailed, nuanced information. This may reflect the manual process used to identify and describe the threats by the researchers.

In contrast, the tool-generated threats had a tighter distribution, clustering around a lower word count. This consistency was because, as noted earlier, generative AI prefers to produce capped outputs and focuses on clarity and brevity. This can be beneficial for generating streamlined and actionable threats, but it may also limit the detail captured compared to the expert-driven case study threats.

Nevertheless, the generative AI tool produced threats that showed a degree of standardization, which could be ideal for automating the threat modeling process in environments where scalability and repeatability are considerations. It may also enable quicker deployment of threat models across diverse systems. This makes it useful for organizations looking to scale their security efforts.

### 4.3 Reduction in Complexity (H1)

H1: AegisShield can reduce the complexity of threat descriptions compared to expert-developed models while maintaining a comprehensive coverage of threat categories.

To address this, complexity was assessed using the Flesch-Kincaid readability test, which quantitatively measured the grade level of the threat descriptions generated by both the tool and the experts. Lower scores reflect simpler, more accessible language. This analysis aimed to determine if the tool, operating in its default medium setting, consistently produced threat descriptions that were easier to read.

### *4.3.1 Normality Tests*

Before conducting comparative analyses of the tool- and expert-generated threats, it was necessary to assess the normality of the readability score distributions. Establishing whether the data followed a normal distribution helped determine the appropriate statistical tests for analysis. Figure 4.4 compares the readability score distributions and Kernel Density Estimation (KDE) plots for the tool- and expert-generated threats. The former had a more concentrated distribution around the mean, with lower variability, while the latter displayed a broader, more symmetrical distribution.

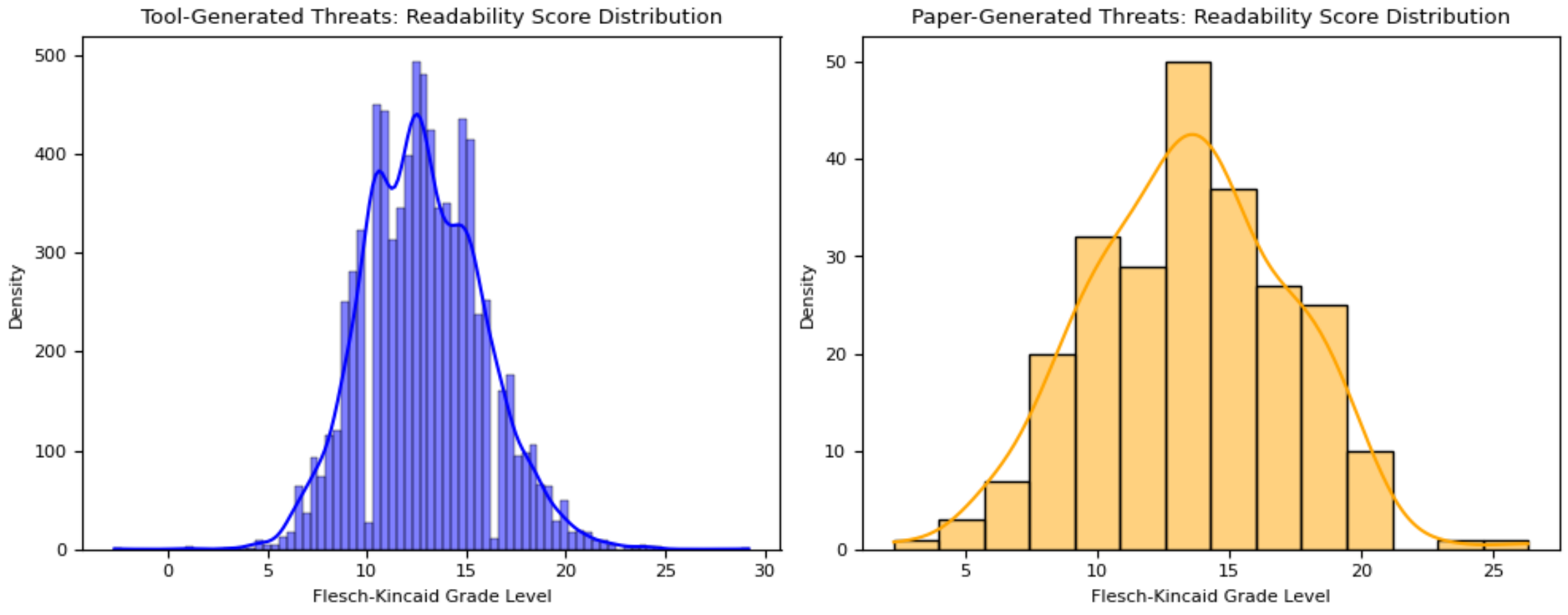

Figure 4.4. Readability Score Distributions and KDE for Tool- and Expert-generated Threats

**4.3.1.1. Tool-generated Threats.** The Anderson-Darling (AD) test yielded a statistic of 13.486 with a p-value of less than 0.005, indicating a significant deviation from normality. The Kolmogorov-Smirnov (K-S) test returned a statistic of 0.053 and a p-value of less than 0.010. This confirms that the AegisShield's readability scores did not follow a normal distribution. These results suggest that a non-parametric statistical test would be more appropriate for comparing the datasets.

**4.3.1.2. Expert-generated Threats.** On the other hand, the expert-generated readability scores were consistent with a normal distribution. The Ryan-Joiner (RJ) test returned a statistic of 0.997 and a p-value greater than 0.100, showing no significant deviation from normality. Also, the Kolmogorov-Smirnov (K-S) test produced a statistic of 0.041 with a p-value greater than 0.150, confirming that the readability scores of the case studies met the assumptions of normality.

These results indicate a marked difference in the distributional characteristics of the two datasets. Table 4.3 provides the results of the normality tests for both datasets.

Table 4.3. Normality Test Results for Tool- and Expert-generated Threats

| **Test** | **Tool-generated Threats (*p*-value)** | **Expert-generated Threats (*p*-value)** |
|---|---|---|
| Anderson-Darling | <0.005 | N/A |
| Kolmogorov-Smirnov (K-S) | <0.010 | >0.150 |
| Ryan-Joiner | N/A | >0.100 |

### *4.3.2 Readability Statistics*

The Flesch-Kincaid readability scores provide insights into the complexity of the threat descriptions generated by the AI tool and human experts. A comprehensive comparison of these scores revealed significant differences in their readability levels and the consistency of the language used in the threat descriptions. Table 4.4 provides the descriptive statistics for the tool-generated threats.

Table 4.4. Descriptive Statistics: Readability Score for the Tool-generated Threats

| **Variable** | **Total Count** | **Mean** | **SE Mean** | **StDev** | **Variance** | **CoefVar** | **Minimum** |
|---|---|---|---|---|---|---|---|
| Flesch_Kincaid_Score | 8100 | 12.8059 | 0.0335203 | 3.01683 | 9.10124 | 23.56 | -2.7 |

| **Variable** | **Q1** | **Median** | **Q3** | **Maximum** | **Range** | **IQR** | **Mode** | **N for Mode** | **Skewness** |
|---|---|---|---|---|---|---|---|---|---|
| Flesch_Kincaid_Score | 10.7 | 12.7 | 14.7 | 29.2 | 31.9 | 4 | 12.3 | 494 | 0.23 |

| **Variable** | **Kurtosis** |
|---|---|
| Flesch_Kincaid_Score | 0.25 |

As seen above, the mean Flesch-Kincaid score was 12.81 and the median was 12.7 for the tool-generated threats. That is, these descriptions were generally accessible to and relatively easier to read for individuals with a high school education level. Interestingly, the standard deviation of 3.02 indicates that most threats were consistent. Some scores deviated from the central tendency. Simultaneously, the IQR of 4 suggests that most scores were clustered closely around the median. The skewness value of 0.23 points, a slightly positive skew, shows that while most tool-generated descriptions fell within a consistent range, there were occasional outliers with higher complexity. Meanwhile, the kurtosis value of 0.25 suggests that the tool-generated data followed a distribution close to normal, with slightly fewer extreme values than expected. Next, Table 4.5 presents the descriptive statistics for the expert-generated threats.

Table 4.5. Descriptive Statistics: Readability Score for the Case Study Threats

| **Variable** | **Total Count** | **Mean** | **SE Mean** | **StDev** | **Variance** | **CoefVar** | **Minimum** |
|---|---|---|---|---|---|---|---|
| Flesch_Kincaid_Score | 243 | 13.4675 | 0.241467 | 3.76409 | 14.1684 | 27.95 | 2.3 |

| Variable | Q1 | Median | Q3 | Maximum | Range | IQR | Mode | N for Mode | Skewness |
|---|---|---|---|---|---|---|---|---|---|
| Flesch_Kincaid_Score | 10.7 | 13.5 | 16.2 | 26.3 | 24 | 5.5 | 12.7 | 15 | 0.02 |

| Variable | Kurtosis |
|---|---|
| Flesch_Kincaid_Score | -0.02 |

The threats extracted from the case studies had a higher mean Flesch-Kincaid score of 13.47, with a median of 13.5. This suggests that the expert-developed descriptions were generally more complex than the tool-generated ones and required some level of college education to comprehend fully. The standard deviation of 3.76 indicates more significant variability in the readability of the threats from the case studies. The IQR of 5.5 suggests a wider spread of readability scores and more diversity in how experts write their threat descriptions. The near-zero skewness of 0.02 indicates a nearly symmetrical distribution of the scores around the median. Finally, the slightly negative kurtosis of -0.02 suggests a flatter distribution, with a broader range of values on both the simpler and more complex ends of the scale. Next, Figure 4.5 provides a side-by-side boxplot comparison of the Flesch-Kincaid grade levels for both tool- and expert-generated threats.

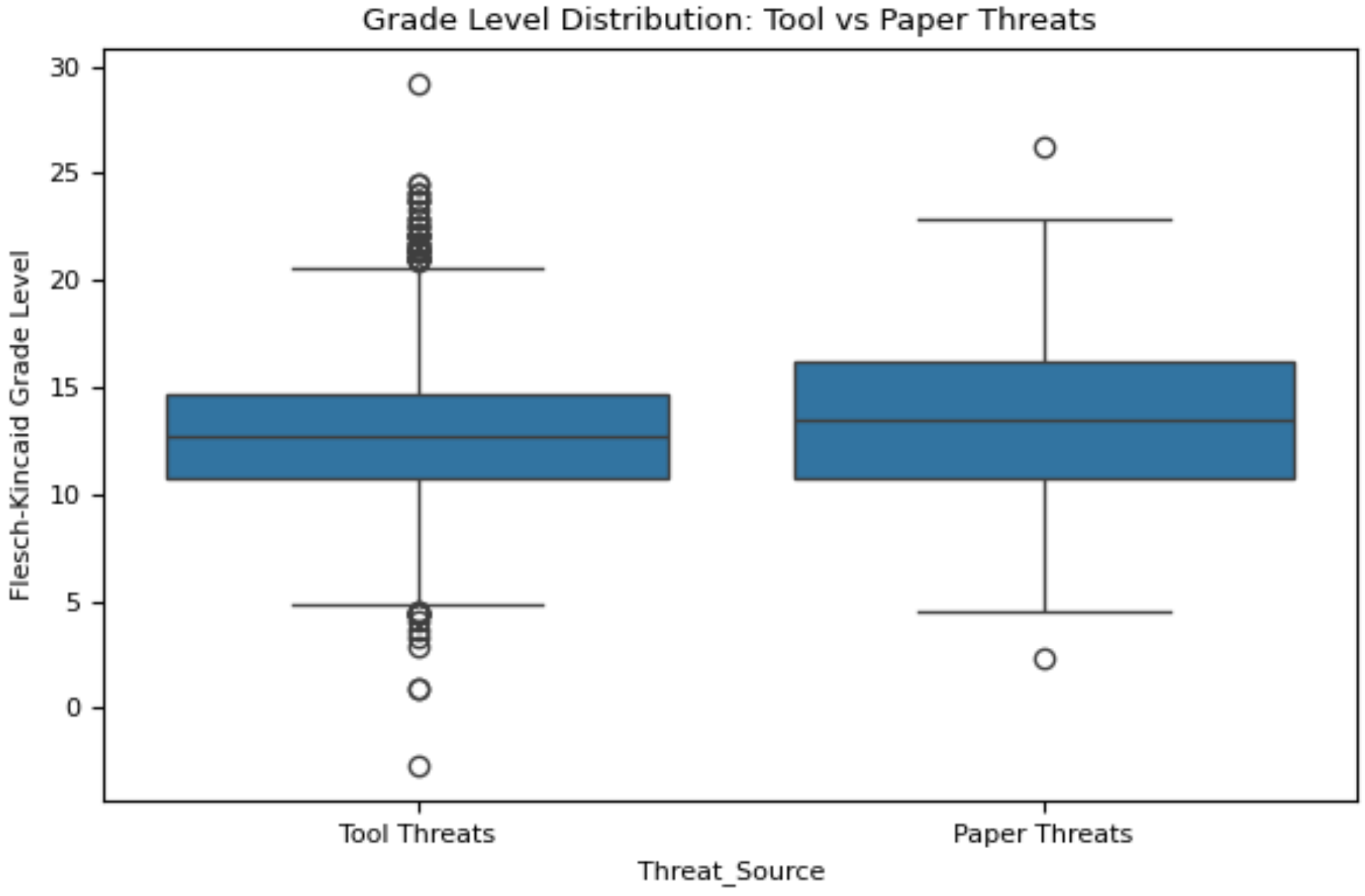

Figure 4.5. Boxplot of Readability Scores for Tool- and Expert-generated Threats

A more tightly clustered readability score distribution can be seen for the tool-generated threats. In contrast, the expert-generated threats had a broader range of grade levels. While outliers were present in both data sets, they appeared to be symmetrically distributed, suggesting that these deviations did not heavily influence the overall trends in readability. This comparison suggests that while expert-generated threats may have incorporated more complex language, the AI tool consistently produced accessible, more standardized descriptions. The relatively narrow distribution of the readability scores for the tool-generated threats reflects AegisShield's tendency to standardize language. This helps in creating more universally accessible threat descriptions.

### *4.3.3 Transforming Non-normal Data*

Given the non-normality of the tool-generated readability scores, two methods were attempted to normalize the data before proceeding with the statistical comparisons.

These included log transformations and a Box-Cox transformation, standard techniques for addressing skewed or non-normal data. Table 4.6 presents the results of these transformation attempts and shows how each method affected the normality. All the normality tests and transformations were performed using the SciPy library.

Table 4.6. Normality Test Results After Transformations to Tool-generated Threats

| **Transformation Type** | **Anderson-Darling Statistic** | **K-S *p*-value** |
|---|---|---|
| Normality test results after filtering and log transformation (tool-generated threats) | 21.7484 | < 0.001 |
| Normality test results after adjusting and log transformation (tool-generated threats) | 22.8399 | < 0.001 |
| Normality test results after box-cox transformation (tool-generated threats) | 7.6851 | < 0.001 |

**4.3.3.1 Log Transformation.** The first log transformation was applied after filtering out the single negative value. However, this transformation did not normalize the data; the Anderson-Darling statistic remained high at 21.7484 and the K-S test returned a *p*-value of 0.0000. Even after adjusting the log transformation by replacing the single negative value with a small constant (0.1), the data remained highly non-normal, with an Anderson-Darling statistic of 22.8399. The negative value was due to the tool outputting the overly simplistic threat of a "Spoofed Fog Node" for the Vehicular Fog Computer case study.

**4.3.3.2 Box-Cox Transformation.** The Box-Cox transformation was more successful in reducing the non-normality, though it did not fully normalize the data. After the transformation, the Anderson-Darling statistic dropped to 7.6851, but the K-S *p*-value remained at 0.0000, indicating that significant deviations from normality persisted. The failure to normalize the tool-generated data through these transformations confirmed the need to rely on non-parametric statistical methods for subsequent comparisons.

#### *4.3.4 Mann-Whitney U Test Results*

A Mann-Whitney U test (Table 4.7) was conducted in Minitab to evaluate H1, which posits that tool-generated threats are simpler (i.e., have lower Flesch-Kincaid readability scores) than expert-generated threats. This test was selected due to the non-normality of the tool-generated readability scores, as discussed in Section 4.3.3. The Mann-Whitney U test is a non-parametric alternative to the t-test and well-suited for comparing two independent samples when the normality assumption is violated. The null hypothesis ($H_0$) for this analysis is that there is no significant difference in the readability scores between the tool- and expert-generated threats and those extracted from the case study.

Table 4.7. Mann-Whitney U Test Results

Descriptive Statistics

| Sample | N | Median |
|---|---|---|
| Case_Readability_Scores | 243 | 13.5 |
| Tool_Readability_Scores | 8100 | 12.7 |

Estimation for Difference

| Difference | Lower Bound for Difference | Achieved Confidence |
|---|---|---|
| 0.7 | 0.4 | 95.00% |

Test

Null hypothesis $H_0$: $\eta_1 - \eta_2 = 0$

Alternative hypothesis $H_1$: $\eta_1 - \eta_2 > 0$

| Method | W-Value | P-Value |
|---|---|---|
| Not adjusted for ties | 1125215.50 | 0.001 |
| Adjusted for ties | 1125215.50 | 0.001 |

***4.3.4.1. Descriptive Statistics.*** The following descriptive statistics were obtained from the data:

- The median Flesch-Kincaid readability score for expert-generated threats was 13.5 and the median for tool-generated threats was 12.7.
- The estimated difference between the medians was 0.7, with a 95% confidence interval for the difference ranging from 0.4 to infinity. Under the default tool settings, the expert-generated threats were generally more complex than the tool-generated ones.

**4.3.4.2. Mann-Whitney U Test Results.** The Mann-Whitney U test yielded the following results:

- W-value (U statistic): 1,125,215.50
- *p*-value: 0.001 (both adjusted and unadjusted for ties)

Given that the *p*-value was well below the significance threshold of 0.05, $H_0$ was rejected. These findings prove that a significant difference exists between the readability scores of the tool- and expert-generated threats. Specifically, the former were statistically simpler, as reflected in their lower median Flesch-Kincaid grade level of 12.7 compared to the expert-generated threats' median score of 13.5. While the readability improvement was statistically significant (p = 0.001), the rank-biserial correlation ($r_e$ = -0.143) suggests a small effect size, indicating that while AI-generated threats were generally more readable, the practical impact is modest.

The simplicity of the tool-generated threats arises from AegisShield's ability to distill complex data into brief, standardized outputs. This design aligns with the broader demand for succinct communication, as AI models are increasingly used to streamline the creation and consumption of information (Kawsar, 2023). In contrast, human experts often provide nuanced and detailed descriptions from their extensive experience. The AI tool, however, emphasizes uniformity and clarity and is advantageous for scaling the threat modeling efforts across organizations with varying levels of expertise. By producing more accessible and easily understandable threat descriptions, it offers a practical solution for organizations aiming to democratize and simplify the threat modeling process.

Of the 8,100 tool-generated threats, 71 (less than 1%) were duplicates that appeared in multiple batches. Excluding these (N = 8,029) yielded nearly identical median readability scores (12.7 vs. 13.5) and a comparable *p*-value ($p = 0.002$), indicating no material impact on the test outcome. Consequently, the results presented here include duplicates to reflect the tool's real-world behavior.

#### *4.3.5 Conclusion*

The results of this analysis confirm H1, that AegisShield produces simpler (lower grade level) threat descriptions than the expert-developed models. The Mann-Whitney U test revealed a statistically significant difference in readability, with tool-generated threats being more accessible on average. These findings suggest that the tool effectively generates streamlined and standardized threat descriptions while maintaining a comprehensive coverage of all the STRIDE categories. The consistency and simplicity of the tool-generated threats could make them more practical for organizations looking to scale threat modeling across diverse teams with varying levels of expertise. While the Flesch-Kincaid readability score provides a quantitative measure of surface-level complexity, it does not account for the technical nuance or accuracy critical in cybersecurity contexts. This limitation highlights the importance of evaluating statistics from multiple angles. This is why we use both semantic similarity and MITRE ATT&CK mappings, to ensure that reduced complexity does not harm technical depth. Another point is, Flesch's (1948) original research highlights the limitations of readability metrics in assessing conversational or technical texts. While AegisShield outputs are standardized and concise, the Flesch-Kincaid scores may not fully reflect their accessibility to technical audiences. This is why complementary metrics were used to validate both technical accuracy and comprehensiveness, with readability.

## 4.4 Semantic Similarity (H2)

H2: AegisShield can generate outputs that exhibit semantic similarity to expert-developed models, as indicated by the cosine similarity using Sentence-BERT, regardless of the specific domain.

This analysis aimed to evaluate whether the AI-generated threats aligned with the expert-generated ones. The alignment in over 50% of the executions per case study indicates that the tool can reliably generate meaningful threat descriptions comparable to those an expert would produce, thereby validating the tool's efficacy in replicating expert-level insights.

### *4.4.1 Results Across All Case Studies*

This research used a Colab notebook and compared 21,870 threat scenarios across 15 case studies. The cosine similarity scores for these threat descriptions were calculated and analyzed. Tables 4.8 and 4.9 and Figure 4.6 summarize the overall statistics for the cosine similarity scores across all the case studies.

Table 4.8. Descriptive Statistics for the Similarity Scores

| Variable | Total Count | Mean | SE Mean | StDev | Variance | CoefVar | Minimum |
|---|---|---|---|---|---|---|---|
| Score | 21870 | 0.508789 | 0.0009076 | 0.134216 | 0.0180139 | 26.38 | -0.0025464 |

| Variable | Q1 | Median | Q3 | Maximum | Range | IQR | Mode |
|---|---|---|---|---|---|---|---|
| Score | 0.418601 | 0.514660 | 0.603464 | 0.941983 | 0.944530 | 0.184863 | 0.486022, 0.518638 |

| Variable | N for Mode | Skewness | Kurtosis |
|---|---|---|---|
| Score | 4 | -0.20 | -0.13 |

Table 4.9. Statistical Summary of the Similarity Scores per Case Number (All Scores)

| Case Number | count | mean | std | min | max |
|---|---|---|---|---|---|
| 1 | 1440 | 0.564804 | 0.114941 | 0.094066 | 0.874322 |
| 2 | 1080 | 0.541473 | 0.134865 | 0.091872 | 0.831022 |
| 3 | 1800 | 0.489556 | 0.142424 | -0.002546 | 0.857963 |
| 4 | 900 | 0.519060 | 0.105636 | 0.176521 | 0.823524 |
| 5 | 1170 | 0.499052 | 0.131592 | 0.091119 | 0.819052 |
| 6 | 1350 | 0.441817 | 0.144253 | 0.080196 | 0.810317 |
| 7 | 2610 | 0.515564 | 0.137191 | 0.000817 | 0.941983 |
| 8 | 2880 | 0.545585 | 0.117164 | 0.017705 | 0.868055 |
| 9 | 1170 | 0.481896 | 0.126989 | 0.079620 | 0.890803 |
| 10 | 1170 | 0.544601 | 0.137159 | 0.128692 | 0.892365 |
| 11 | 990 | 0.542251 | 0.135849 | 0.081111 | 0.896690 |
| 12 | 1350 | 0.481541 | 0.117399 | 0.135225 | 0.793191 |
| 13 | 990 | 0.527351 | 0.105120 | 0.182281 | 0.769201 |
| 14 | 1170 | 0.484332 | 0.132767 | 0.047236 | 0.886652 |
| 15 | 1800 | 0.448251 | 0.137292 | 0.005635 | 0.789976 |

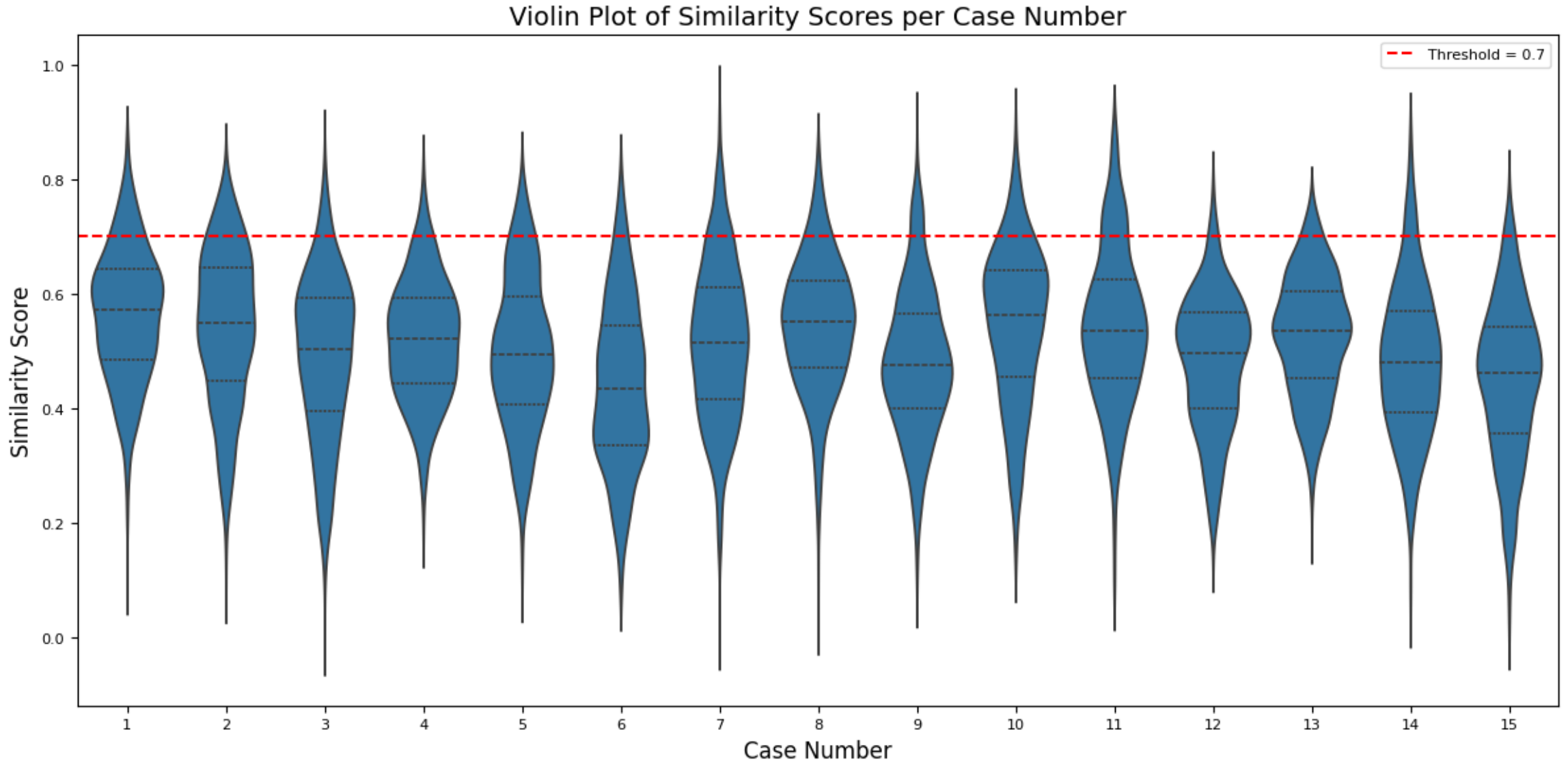

Figure 4.6. Violin Plot of the Similarity Scores per Case Number

**4.4.1.1 Descriptive Statistics.** The overall cosine similarity scores ranged from -0.0025 to 0.9420, with a mean score of 0.508789. The skewness value was -0.20, suggesting a slightly left-leaning but near-symmetrical skew. In contrast, the kurtosis value of -0.13 indicates a slightly flatter distribution, showing fewer extreme values at the high and low ends of the distribution. Figure 4.6 above presents a violin plot of the cosine similarity scores across each case, illustrating their density and distribution. The wider sections indicate higher concentrations of scores within each case, reinforcing the consistent, moderate similarity levels across cases. Table 4.9 summarizes the statistical properties of the similarity scores for each case study, highlighting variations in mean scores, standard deviations, and ranges across cases. These results suggest that while the tool generated semantically similar threats across a broad range of case studies, the average score across all the threats remained below the established threshold of 0.7, indicating moderate similarity.

### *4.4.2 Similarity Scores Above the Threshold*

Focusing on the cosine similarity scores ≥ 0.7, 1,594 threat outputs across all 15 case studies met or exceeded the 0.7 threshold. These results highlight the significant semantic overlap between the two datasets.

Table A3 presents the top two cosine similarity scores for each case study. These examples show how well AegisShield performed across several threat types. In many cases, the descriptions from the tool matched very closely with those from the experts. As an example, Cases 7 (spoofing) and 11 (elevation of privilege) had scores of 0.9419 and 0.8967, respectively. These numbers show that the tool generated threats that closely matched expert descriptions. AegisShield can capture both the context and specific details of the threats. In addition, the repudiation threats in Case 9 also achieved high similarity (0.8908 and 0.8351), indicating that AegisShield was able to identify situations involving log manipulation or deletion intended to hide unauthorized activity.

While these examples highlight the tool's potential for high accuracy, they represent only the strongest matches from each case study. Looking at the full dataset, the output showed more variation. Some threats aligned closely with expert descriptions, while others fell short. Still, the presence of high-scoring matches across multiple cases suggests that the model can often generate semantically accurate threats. The quality, however, did not remain consistent. It shifted depending on the complexity of the scenario and the specificity of the domain.

As shown in Table 4.10, the proportion of successful batches was evaluated for each case study using the adjusted exact method (Adjusted Blaker) within Minitab, applying a 95% confidence level. In 14 out of 15 cases, the outcome was statistically significant ($p < 0.05$), with the lower bound of the confidence interval remaining above the 50% mark. Case 12 was the lone exception—despite a success rate of 63.3%, the associated p-value of 0.100 and a lower bound of 46.7% suggest that the result may not be reliably distinguishable from chance. Overall, these findings offer strong empirical support for Hypothesis 2, indicating that AegisShield was able to produce at least one high-similarity threat in the majority of batch runs across most tested scenarios.

Table 4.10. Proportion of Successful Batches with 95% CI per Case Study

| **Case Number** | **Successful Batches** | **Total Batches** | **Sample p (%)** | **95% Lower Bound** | **p < 0.05** |
|---|---|---|---|---|---|
| 1 | 30 | 30 | 100.0% | 90.4% | Yes |
| 2 | 28 | 30 | 93.3% | 80.4% | Yes |
| 3 | 29 | 30 | 96.7% | 85.1% | Yes |
| 4 | 20 | 30 | 66.7% | 50.0% | Yes |
| 5 | 27 | 30 | 90.0% | 76.1% | Yes |
| 6 | 24 | 30 | 80.0% | 64.3% | Yes |
| 7 | 30 | 30 | 100.0% | 90.4% | Yes |
| 8 | 28 | 30 | 93.3% | 80.4% | Yes |
| 9 | 30 | 30 | 100.0% | 90.4% | Yes |
| 10 | 29 | 30 | 96.7% | 85.1% | Yes |
| 11 | 29 | 30 | 96.7% | 85.1% | Yes |
| 12 | 19 | 30 | 63.3% | 46.7% | No |
| 13 | 27 | 30 | 90.0% | 76.1% | Yes |
| 14 | 21 | 30 | 70.0% | 53.5% | Yes |
| 15 | 26 | 30 | 86.7% | 72.0% | Yes |

Table 4.11 presents a statistical summary of the scores with a value ≥ 0.7 per case number.

Table 4.11. Statistical Summary of the Similarity Scores per Case (Score ≥ 0.7)

| **Case Number** | **count** | **mean** | **std** | **min** | **max** |
|---|---|---|---|---|---|
| 1 | 169 | 0.750147 | 0.040116 | 0.700393 | 0.874322 |
| 2 | 129 | 0.739586 | 0.029373 | 0.700025 | 0.831022 |
| 3 | 92 | 0.744656 | 0.034637 | 0.700026 | 0.857963 |
| 4 | 37 | 0.736260 | 0.031379 | 0.701329 | 0.823524 |
| 5 | 85 | 0.737683 | 0.029000 | 0.700132 | 0.819052 |
| 6 | 56 | 0.741064 | 0.029948 | 0.700338 | 0.810317 |
| 7 | 249 | 0.756191 | 0.048743 | 0.700013 | 0.941983 |
| 8 | 254 | 0.748516 | 0.040191 | 0.700224 | 0.868055 |
| 9 | 71 | 0.748808 | 0.034607 | 0.701639 | 0.890803 |
| 10 | 120 | 0.753545 | 0.043142 | 0.700131 | 0.892365 |
| 11 | 140 | 0.766271 | 0.050853 | 0.700984 | 0.896690 |
| 12 | 32 | 0.728748 | 0.025985 | 0.701233 | 0.793191 |
| 13 | 41 | 0.724338 | 0.017079 | 0.700287 | 0.769201 |
| 14 | 72 | 0.763116 | 0.046930 | 0.700847 | 0.886652 |
| 15 | 47 | 0.726297 | 0.021776 | 0.700355 | 0.789976 |

Cases 1, 7, and 9 all achieved 100% success. Each batch in those cases included at least one threat scoring over 0.7. Case 6 also passed the 50% mark, but at 80%, and did not perform as strongly. This could mean the tool struggled more with threats related to social networking. Case 12 had the lowest success rate at 63.3%. It focused on IoT threats in the energy sector and used a mix of CAPEC IDs and MACM modeling—two areas not directly handled by AegisShield. Even so, the tool still produced useful outputs. Since AegisShield follows STRIDE while the case used more rigid taxonomies, the mismatch likely affected the result. For threats scoring 0.7 or above, the average similarity ranged from 0.724 to 0.766.

### *4.4.3 Visualization of the Results*

Figure 4.7 highlights four graphs. It includes a series of visualizations illustrating the distribution and consistency of the cosine similarity scores across the case studies. These visuals provide an overview of how AegisShield performed semantically against the case studies.

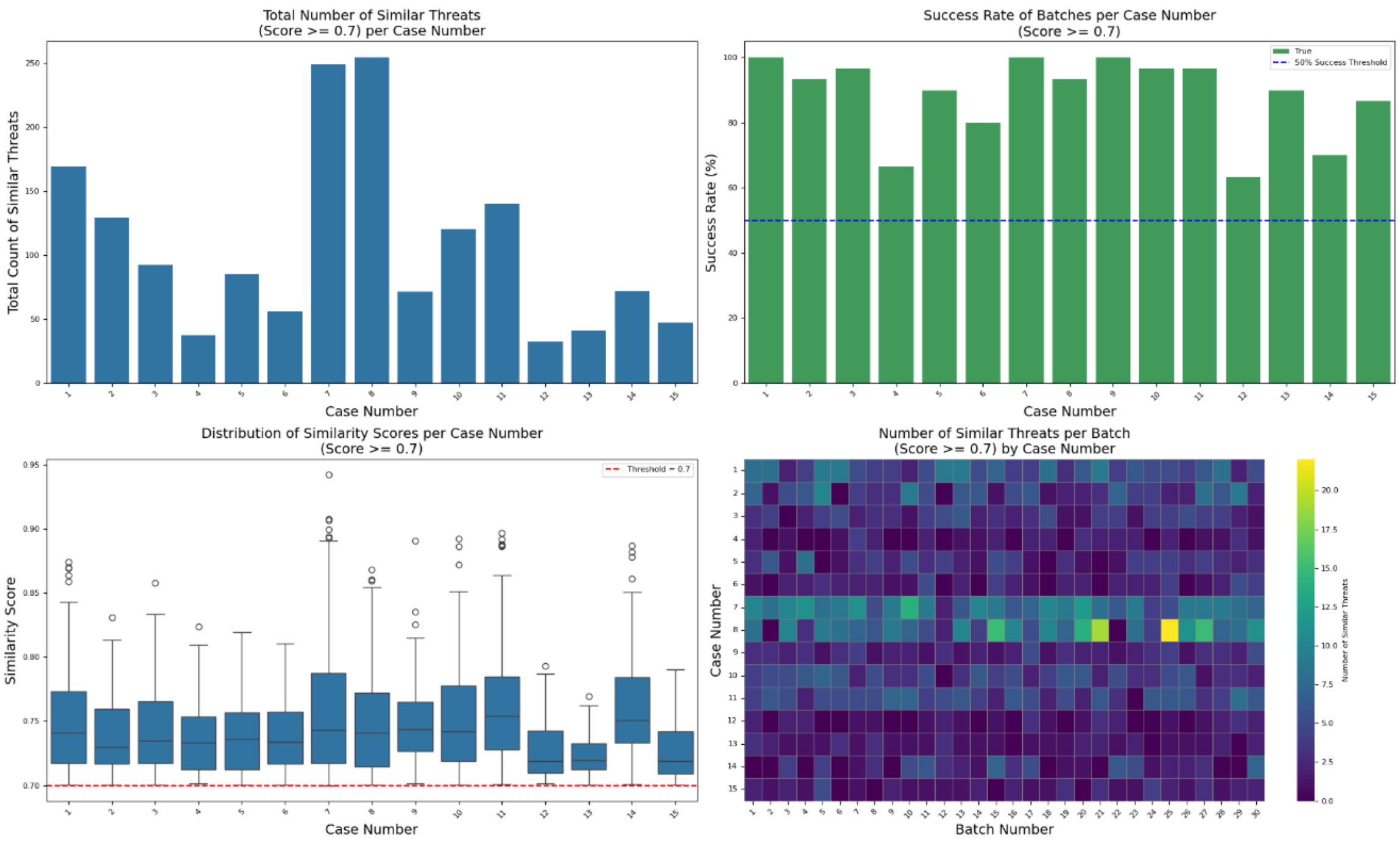

Figure 4.7. Visualization of the Cosine Similarity Scores Across the Case Studies

The bar chart in the top-left corner shows the total number of threats per case study that met the 0.7 cosine similarity threshold. There was considerable variability across the cases. Case 8 showed the highest number of similar threats at 254 and Case 12 produced the fewest, at 32. The content of the case studies and how the researcher presented the threats is the methodological reason we applied a 50% majority vote as the success threshold.

The top-right chart illustrates the success rate of the batches per case study, defined as the percentage of batches that contained at least one threat exceeding the 0.7 similarity threshold. All the cases showed success rates above the 50% threshold. Cases 1, 7, and 9 achieved a 100% success rate, while Cases 2, 3, 8, 10, 11, and 13 achieved a 90% success rate. In contrast, cases 4, 6, 12, 14, and 15 achieved success rates of 63–86%, indicating more challenges in producing highly similar threats in these cases.

The bottom-left box plot provides a detailed distribution of the similarity scores across the 15 case studies for threats meeting the 0.7 similarity threshold. Each case exhibits a range of scores, with several visual outliers reaching above 0.85. While many case studies have a tightly clustered distribution of scores near the threshold, there were exceptionally high similarity scores in some instances, validating the tool's effectiveness.

Finally, the heatmap in the bottom-right corner shows the number of similar threats per batch across the case studies, providing a more granular view of the tool's performance. The color intensity reflects the number of threats per batch, with the lighter yellow areas indicating higher counts. Cases 7 and 8 had consistently high counts of similar threats across batches, unlike Case 12, which had substantially fewer threats meeting the threshold, resulting in sparser areas on the heatmap.

Some cases performed better than the others, depending on how their author modeled the system's architectural complexity or represented the domain characteristics. In threat modeling, the quality and specificity of the system's architectural representation and domain-specific factors (e.g., IoT, healthcare, automotive) directly affect the AI's ability to generate relevant and semantically accurate threats. The tool may struggle to produce high-quality matches if the system's architecture or domain-specific details are underrepresented or oversimplified.

#### *4.4.4 Rubric Score Correlation Analysis*

A correlation analysis was undertaken to determine whether any of the system attributes (e.g., application type, data sensitivity) in the rubric influenced the tool's ability to generate semantically similar threats. This was done at two levels. First, at the threat level, each AI-generated threat scenario was matched with its corresponding expert-developed threat, and the resulting similarity scores were correlated with the rubric criteria. Second, at the case level, the similarity scores for each case study were aggregated to produce an overall average, which was then correlated with the same set of rubric attributes. Tables 4.12 (Threat Level) and 4.13 (Case Level) summarize the correlation coefficients. Across both approaches, the relationships proved weak, with correlation values near zero or only moderately negative. Even the highest positive coefficient (0.0361) remained well below the thresholds typically considered "strong" (≥ 0.5) (Cohen, 1988). Factors such as Application Type (–0.42) and Internet Facing (–0.42) also showed weak negative correlations, indicating that they did not meaningfully impact semantic alignment.

Table 4.12. Correlation between the Similarity and Rubric Scores (Threat Level)

| **Metric** | **Value** |
|---|---|
| Score | 1.000000 |
| Application/System Description or DFD | -0.055503 |
| Application Type | -0.110803 |
| Industry Sector | -0.008403 |
| Data Sensitivity | -0.051699 |
| Internet-facing Status | -0.110803 |
| Compliance Requirements | 0.036070 |

| Metric | Value |
|---|---|
| Authentication Methods | -0.062650 |
| Technical Details | -0.115046 |
| Threat Details | 0.024244 |
| Threat Count | 0.050639 |

Table 4.13. Correlation between the Similarity and Rubric Scores (Case Level)

| Metric | Value |
|---|---|
| Score | 1.000000 |
| Total Rubric Score | -0.131753 |
| Application/System Description or DFD | -0.246912 |
| Application Type | -0.421492 |
| Industry Sector | -0.012571 |
| Data Sensitivity | -0.163344 |
| Internet-facing Status | -0.421492 |
| Compliance Requirements | 0.133741 |
| Authentication Methods | -0.184383 |
| Technical Details | -0.346325 |
| Threat Details | -0.085967 |
| Threat Count | 0.022568 |

Heatmaps (Figures A15.1 and A15.2 in the Appendix) visually confirmed the minimal correlations observed. A plausible explanation for this outcome is the generalized and high-level scope of the case studies (e.g., limited specificity regarding programming languages or operating systems), which constrained the extent to which the attributes could meaningfully influence the AI-generated threats. As a result, the tool primarily generated generalized threats that demonstrated weak alignment with specific rubric categories. These findings indicate that, under similar data conditions, the tool exhibits limited sensitivity to variations in system or compliance characteristics.

However, future research incorporating more granular technical data may determine whether richer domain-specific details could yield stronger correlations.

#### *4.4.5 Conclusion*

The findings in this section support H2, confirming that AegisShield can generate threat descriptions with moderate to high semantic similarity to expert-developed models. While the tool achieved cosine similarity scores exceeding 0.7 in many cases, a significant portion fell within the accepted range of 0.5 to 0.7, as outlined in Chapter 3. This indicates consistent performance, with 14 out of 15 case studies meeting the 50% success rate threshold based on statistical validation. The study focused on assessing the tool's performance at the higher end of this range; however, even across the broader distribution of scores, the tool delivered meaningful and semantically accurate threat models. The examples provided in Table A3 illustrate the tool's ability to generate highly similar threats in specific instances, particularly for common attack patterns such as spoofing and tampering.

While the tool can produce accurate matches, the overall variability in scores indicates that certain domains or system architectures present more challenges for it to achieve consistently high similarity. These results underscore AegisShield's potential for practical application in threat modeling across diverse systems while highlighting areas for refinement.

### 4.5 MITRE ATT&CK Mapping (H3)

H3: AegisShield can systematically map STRIDE-categorized threats to relevant MITRE ATT&CK Tactics, Techniques, and Procedures (TTPs), ensuring comprehensive alignment with established cybersecurity frameworks.

This section evaluates the tool’s effectiveness in aligning generated threats with established MITRE ATT&CK TTPs across multiple domains and case studies.

### *4.5.1 Overview of the Mapping Results*

AegisShield generated 8,100 threats across the 15 case studies, of which 6,921 were successfully mapped to relevant MITRE ATT&CK TTPs, resulting in an overall mapping success rate of 85.4%. A one-proportion test was performed to determine whether this mapping rate was statistically greater than the 80% benchmark. The result was statistically significant ($p < 0.001$), with a 95% lower confidence bound of 84.8%, supporting Hypothesis 3. Figure 4.8 presents two visualizations of the mapped threats’ distribution and success rates across various STRIDE categories.

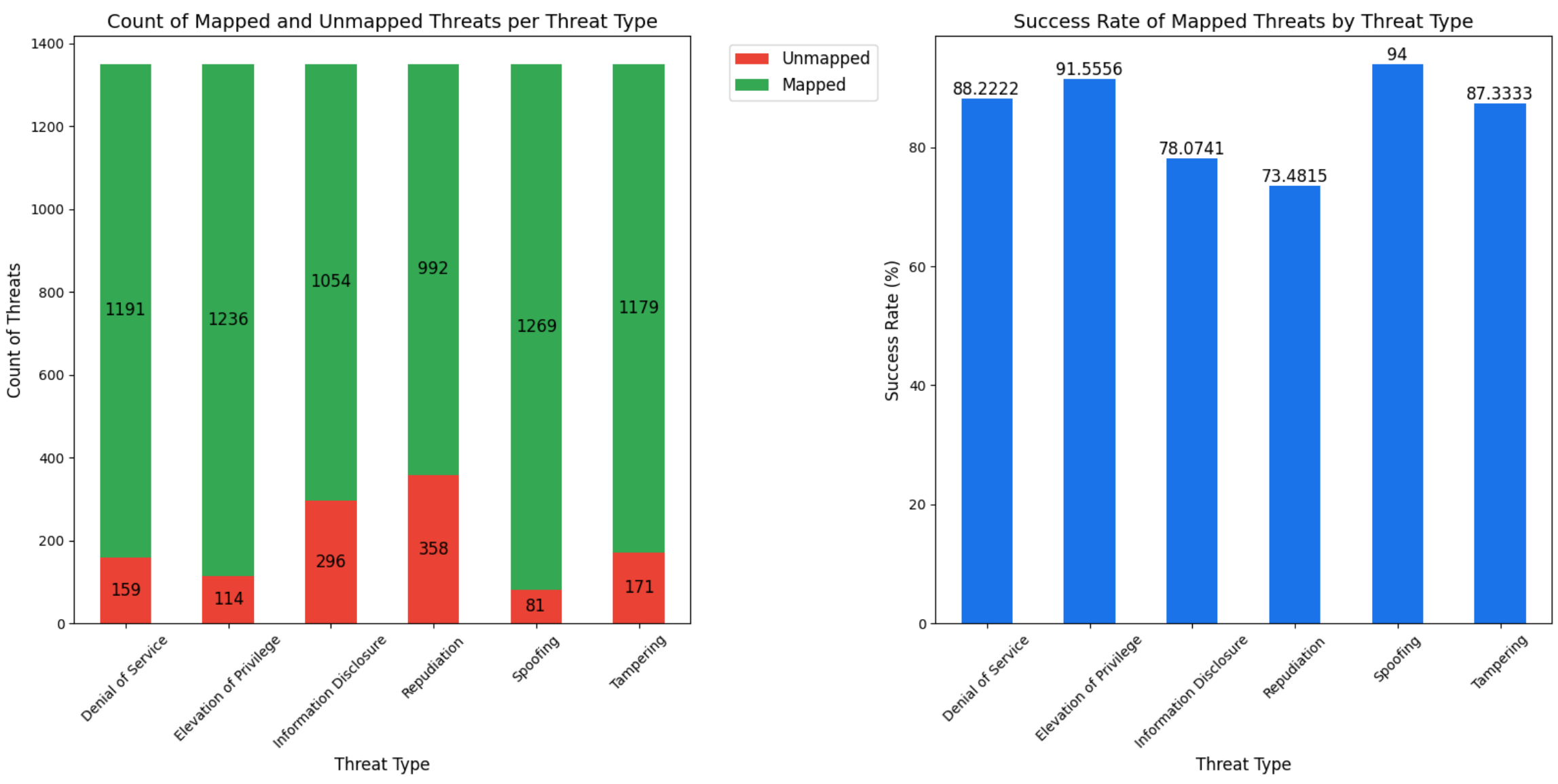

Figure 4.8. Successful MITRE ATT&CK Mappings by Type

The bar chart on the left shows the number of mapped and unmapped threats for each STRIDE category. Spoofing and elevation of privilege had the highest counts, at 1,269 and 1,236 mapped threats respectively. This identifies the categories where the tool performed best. Meanwhile, the bar chart on the right illustrates the percentage of successfully mapped threats by category. Here, as well, spoofing and elevation of privilege showed high success rates, with 94.0% and 91.6% of threats mapped to the corresponding MITRE ATT&CK TTPs, respectively. In contrast, repudiation had the lowest mapping success rate, at 73.5%, indicating potential challenges in aligning these threats with the MITRE framework.

To contextualize these findings, Figure A16 provides a radar chart that visually compares the success rates across different threat categories. It emphasizes AegisShield's consistent performance with respect to spoofing and elevation of privilege and confirms that repudiation consistently underperformed in mapping success. It also reinforces the bar chart insights but adds a comparative dimension across categories, highlighting the areas of improvement.

Figure 4.9 breaks down the mapping results across the 15 case studies, showing the proportion of mapped versus unmapped threats for each case. Consistently high mapping rates were observed across most cases with a slight variability. Cases 1 (voice application) and 6 (social media) had more unmapped threats than the other case studies, which may suggest domain-specific challenges or more specialized threats that do not yet have established mappings in the MITRE ATT&CK database based on the keywords generated by the AI.

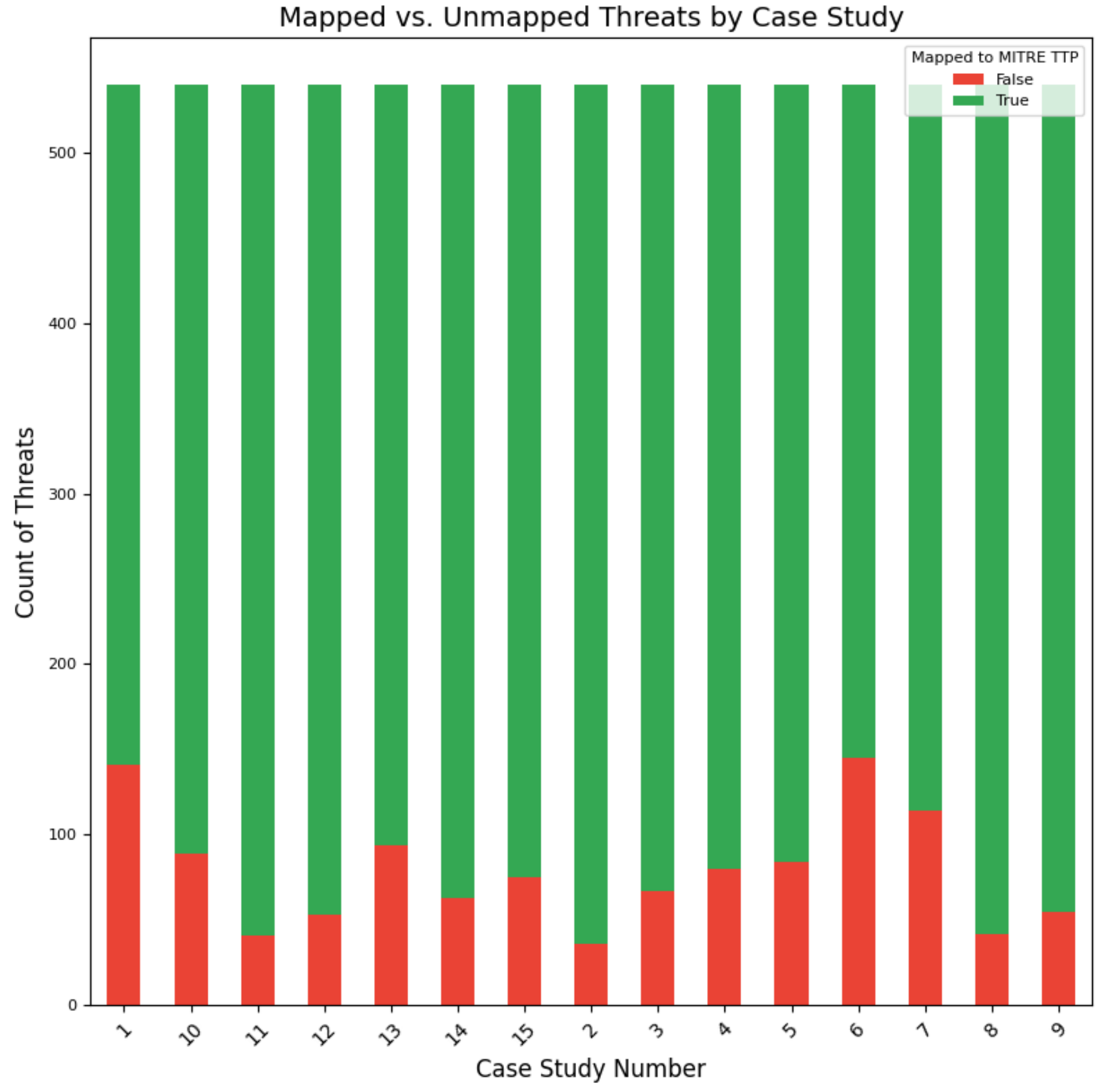

Figure 4.9. MITRE ATT&CK Mapped vs Unmapped Threats by Case Study

In summary, while the tool performed well in mapping STRIDE-categorized threats to MITRE ATT&CK TTPs, certain categories (e.g., repudiation) and specific case studies (Cases 1 and 6) showed lower mapping success rates. This indicates areas for potential improvement or compliance with evolving threat frameworks. Addressing these would ensure that the generative AI prompts return more standardized and searchable keywords that align closely with the MITRE ATT&CK STIX data structures, improving the tool's ability to map threats effectively to the MITRE framework. The next section explores why some threats were not mapped and provides insights into areas of improvement for future iterations of the tool.

### *4.5.3 Analysis of the Unmapped Threats*

The analysis of the unmapped threats revealed multiple reasons why specific threats generated by AegisShield were not successfully mapped to the MITRE ATT&CK framework. Due to the evolving nature of cyber threats, these unmapped threats often involved keywords or descriptions that were either too specific, too broadly conceptual, or not yet covered by the framework. Table 4.14 provides representative examples of the unmapped threats from Cases 9, 12, 14, and 15. It highlights the keywords generated by the tool for each threat and briefly explains why these keywords may have hindered mapping to the MITRE ATT&CK framework.

Table 4.14. Examples of the Unmapped Threats

| Case | Threat Description | Keywords Searched | Reason for Being Unmapped |
|---|---|---|---|
| Case 9 | Repudiation – "A user denies having accessed or viewed sensitive data from the TOREADOR platform" | "repudiation," "access denial," "log evasion," "data confidentiality compromise" | Repudiation is often a legal/transactional concern, not a technical tactic; the keywords focus on behavior rather than adversary actions, which limits the alignment with MITRE |
| Case 9 | Information Disclosure – "Metadata associated with sensor data exposed to unauthorized users" | "metadata exposure," "access control weakness," "sensitive information leak," "data confidentiality breach" | Keywords are too broad, lacking specificity in adversary methods; MITRE focuses on concrete actions rather than general weaknesses. |
| Case 12 | Tampering – "An attacker modifies MQTT messages in transit to alter device behavior" | "message tampering," "data integrity attack," "network manipulation" | Keywords are too broad and lack specific adversary actions typically mapped in MITRE ATT&CK, such as protocol manipulation or traffic interception. |
| Case 12 | Denial of Service – "Exploitation of known vulnerabilities in | "Database Crash," "Exploitable | Keywords are too general; MITRE focuses on attack vectors (e.g., |

| Case | Threat Description | Keywords Searched | Reason for Being Unmapped |
|---|---|---|---|
| | MySQL to crash the database server" | Vulnerability," "Service Disruption" | "Resource Exhaustion") rather than outcomes such as service crashes. |
| Case 14 | Spoofing – "An attacker uses malicious firmware updates disguised as legitimate updates to compromise UAVs" | "malware injection," "firmware compromise," "update mechanism," "embedded systems" | Keywords represent emerging attack vectors not fully integrated into MITRE ATT&CK, which typically focuses on more established methods and tactics. |
| Case 15 | Spoofing – "Impersonation of IoT devices to send incorrect data to the system" | "IoT attack," "impersonation," "protocol exploitation" | Emerging technologies such as IoT may lack comprehensive mappings in MITRE ATT&CK due to the rapid evolution of associated threats and protocols. |

The unmapped threats identified in the analysis can be attributed to several factors, each reflecting challenges inherent in aligning the generated threats with the MITRE ATT&CK framework. First, keyword matching presents significant obstacles. In some cases, the tool generated broad terms such as "metadata exposure" or "message tampering," which, while relevant, lacked some specificity necessary for alignment with the framework. These were very broad terms. MITRE relies on detailed descriptions of attack techniques, and these general terms fail to capture the technical details that the framework requires. Ultimately, threats that were described in a broader, more conceptual language were left unmapped.

Next, the tool relied on conceptual descriptions. In a similar way that broad terms were an issue, high-level terms such as "Data confidentiality breach" or "service disruption" reflect generalized cybersecurity concerns. They do not correspond directly to the concrete tactics or techniques documented in the MITRE ATT&CK matrix. While these descriptions capture a threat's intent, they lack the detail needed for precise mapping. This shows a gap between abstract threat descriptions and the framework's highly technical approach.

Finally, within the rapid evolution of the technology in domains such as IoT, MQTT, and 5G, we see challenges emerge when trying to map threats in these areas. While MITRE ATT&CK includes techniques for threats such as "firmware compromise" and "IoT attacks," which include T1495 (Firmware Corruption) and T0839 (Module Firmware), it may not fully capture evolving attack vectors specific to new IoT applications and device protocols. AegisShield identified these threats, but they often fell outside the MITRE framework's scope due to their novelty. Because of this, some novel techniques in these rapidly advancing fields may lack precise mapping within the current MITRE ATT&CK framework (attack.mitre.org). The framework itself will need to evolve to accommodate emerging threats in areas such as IoT, firmware vulnerabilities, and 5G security to address these gaps  (Pell et al., 2021). As a result, domain-specific or cutting-edge attack vectors will remain unmapped. This highlights the need for continuous updates to both the tool and the MITRE framework to stay aligned with cybersecurity trends.

Notably, 93 instances—only 1.15% of the 8,100 threats generated—received unique attack-pattern IDs that deviated from the MITRE standard unmapped identifier and were placed into the unmapped category. These mappings were the AI model hallucinating. This low occurrence highlights the tool's generally strong alignment but also suggests areas for refinement in keyword precision and contextual filtering. Future improvements may reduce these outliers, enhancing AegisShield's consistency with the MITRE ATT&CK framework.

### 4.6 Time Efficiency of AegisShield vs. Conventional STRIDE Methodology

AegisShield processed 8,100 threats across 30 batches, each containing 540 threats, by integrating OSINT data sources such as AlienVault's OTX and NVD and automating MITRE ATT&CK mapping. The entire process required approximately six hours to complete, but running a single threat model after receiving an application description took less than 10 minutes. This encompassed tasks such as threat creation, DREAD risk assessment, mitigation, test cases, and compiling these details in a downloadable PDF report. Figure B15 provides a sample PDF artifact for the DaaS case study and illustrates the comprehensive report format of the generated threat models consolidating all the relevant outputs from the threat modeling process.

In contrast, traditional implementations of the STRIDE methodology are known to be more time intensive. The manual effort needed to construct and deconstruct detailed DFDs and categorize threats into STRIDE's six categories can require several hours or even days, particularly in complex systems (Das et al., 2024). The time required reflects the need for cross-referencing threat categories with risk assessment frameworks such as DREAD, which increases the resources and expertise needed to execute these tasks efficiently (Shevchenko et al., 2018). The ability of AegisShield to automate these processes in minutes rather than hours or days underscores its potential to accelerate threat identification and supports the hypothesis that integrating AI with traditional threat modeling methods considerably reduces both time and complexity.

### 4.7 Conclusion and Implications

The results in this chapter helped to confirm AegisShield's effectiveness in automating and standardizing threat identification across various scenarios. The tool performed surprisingly well. It successfully met the core research hypotheses by reducing the complexity of threat descriptions (H1), generating semantically similar outputs to expert-developed models (H2), and effectively mapping threats to the MITRE ATT&CK framework (H3).

In hypothesis 1, the tool demonstrated its ability to simplify threat descriptions without sacrificing coverage. The Flesch-Kincaid readability scores for the AI-generated threats averaged at 12.81, compared to 13.47 for the expert-generated threats, showing a reduction in complexity. Meanwhile, the Mann-Whitney U test confirmed the statistical significance ($p < 0.001$) and demonstrated that AegisShield consistently produced more accessible descriptions. This is useful for a broader audience.

The results for hypothesis 2 highlight the tool's capability at replicating expert-level threats. The average cosine similarity scores across cases remained moderate but effective. The average cosine similarity score between AI- and expert-generated threat descriptions ranged from -0.0025 to 0.9420. In Cases 7 and 11, the cosine similarity scores reached as high as 0.9419 and 0.8967. This highlighted AegisShield's proficiency in generating highly similar descriptions in many instances. This is significant because the threats were generated by the tool without prior exposure to the case study threats they were being compared to, showing the model's ability to blindly replicate patterns and insights. The results demonstrate that while the tool may not consistently reach the 0.7 similarity threshold across all batches, its descriptions can closely align with the experts.

In terms of Hypothesis 3, the tool achieved an 85.4% success rate in mapping threats to the MITRE ATT&CK framework. A one-proportion test confirmed this result was statistically significant ($p < 0.001$), with a 95% lower confidence bound of 84.8%, supporting H3. Out of the 8,100 threats, 6,921 were successfully mapped. The tool was highly effective in handling spoofing and elevation of privilege threats, with mapping success rates of 94.0% and 91.6%, respectively. This high accuracy shows its alignment with common industry standards. This ensures the generated threat models are actionable and framework compliant.

All these findings confirm that AegisShield is a highly effective solution for democratizing threat modeling. It reduces the expertise and resource barriers in traditional threat modeling processes by generating less complex, semantically accurate, framework-aligned outputs. Its standardized and scalable outputs help speed up the process. This helps enable secure-by-design principles without the high costs typically incurred through traditional methods.

# Chapter 5—Discussion and Conclusions

## 5.1 Discussion

This praxis explored how generative AI can help democratize threat modeling, especially for small organizations that do not have dedicated cybersecurity resources. The threat modeling tool developed in this study addresses these challenges by cutting down the complexity of threat descriptions, as demonstrated in Chapter 4. It achieved statistically significant improvements in readability ($p < 0.001$), making it more accessible for organizations with limited technical expertise.

AegisShield also showed substantial advantages in both efficiency and accessibility when compared to traditional methods. For example, it lets organizations generate comprehensive models within minutes instead of days or weeks. This cost-effective, fast solution empowers teams to proactively address security concerns, all without the heavy investment usually needed for traditional threat modeling.

The tool demonstrated a high mapping success rate of 85.4%, which was statistically validated through a one-proportion test ($p < 0.001$) with a 95% lower confidence bound of 84.8%. This level of consistency contrasts with the variability that can occur in human-driven processes. By automating the mapping process, the tool establishes that the outputs are more reliable, completed more often, and consistent with established industry frameworks.

Unlike traditional methods that require frequent manual updates, AegisShield employs machine learning to adapt to evolving threats. This adaptability makes it resilient in rapidly changing environments, as it provides a scalable solution for fast changing cybersecurity challenges.

The tool also demonstrated the ability to generate threat descriptions with high semantic similarity to expert-developed models, as confirmed by batch-level proportion testing. In 14 out of 15 cases, the similarity results were statistically significant ($p < 0.05$). And because AegisShield makes threat modeling easier, it may lead to broader usage. Jevons Paradox suggests that increased efficiency leads to higher resource consumption; as a result, a greater efficiency in threat modeling could drive higher demand, potentially straining real-time intelligence systems (Giampietro & Mayumi, 2018). While this poses challenges for resource management and scalability, it also highlights the potential for the tool to expand threat modeling practices and contributes positively to the broader cybersecurity landscape.

From a cost perspective, running a threat model using a generative AI tool (e.g., 40,000 tokens on a GPT-4o model for about $3.60) is much less expensive than employing a senior cybersecurity architect (median salary $157,000) who might spend one day to a week (8–40 hours) on threat modeling at approximately $75/hour (Salary.com, 2024). Although these figures are high-level estimates and future research could refine the exact cost calculations, this basic comparison highlights how the AI-enhanced approach can reduce financial barriers. By lowering costs, the tool encourages a broader and more frequent application of threat modeling.

**5.2 Conclusion**

This research validated the hypothesis that AegisShield can effectively automate and standardize the identification and assessment of cybersecurity threats. By generating three threats per STRIDE category, the tool ensured a balance between comprehensiveness and simplicity

AegisShield's ability to generate outputs that exhibit semantic similarity to expert-developed models while reducing complexity supports its role in democratizing the process. Even though it was not tested in this praxis, AegisShield's alignment with secure-by-design principles, as evidenced by its ability to generate DREAD risk assessments and mitigations, emphasizes its contribution to proactive security measures. The overall test results support the three hypotheses as follows:

1. AegisShield produced a median Flesch-Kincaid readability score of 12.7 compared to 13.5 for expert-developed threats ($p = 0.001$). The results clearly show a reduced grade level.
2. AegisShield also generated threats with similar meaning to those of expert-developed models. It achieved a similarity score of 0.7 or higher in at least 50% of batch runs for 14 out of 15 case studies, with statistically significant results ($p < 0.05$) in each of those cases.
3. We demonstrated systematic alignment with established cybersecurity frameworks by mapping 85.4% of the threats to MITRE ATT&CK techniques. This mapping rate was statistically significant ($p < 0.001$), with a 95% lower bound of 84.8%.

While all these findings are encouraging, several limitations highlight areas for improvement. First, variability in mapping performance was observed across different application domains, which may reflect uneven alignment in frameworks like MITRE ATT&CK. Second, we saw that output variability inherent to generative AI may affect the depth and consistency of threat descriptions. This suggests the need for alternative models or more modification to the prompt-engineering. Finally, we acknowledged that potential biases could exist in the AI's training data, and that it could hinder the accurate detection of emerging threats. While these limitations do not negate the study's overall contributions, they show the importance of domain-specific validation, AI model improvements, and targeted data collection strategies. We explore these topics in Section 5.4's recommendations for future research. Notably, this study primarily focused on complete systems rather than smaller, more focused parts of applications.

### *5.2.1 Reflection on Research Questions*

This praxis investigated four research questions to evaluate AegisShield's potential in democratizing and enhancing threat modeling. The findings are summarized below:

- RQ1: AegisShield reduces the time required for threat identification. It completed tasks in minutes compared to the days or weeks needed with traditional methods. This efficiency highlights its role in streamlining security processes.
- RQ2: AegisShield simplifies threat modeling. It does this through automation and enhances readability by producing structured, easy-to-read PDF reports that compile results like STRIDE analysis and mitigation strategies. This makes the results accessible to a broad range of users.
- RQ3: AegisShield enables faster and more informed response to cyber threats. It helps automate early-stage threat identification and generates structured outputs. Its cost-effectiveness lowers barriers, helping smaller organizations take timely action instead of delaying or skipping the threat modeling process.
- RQ4: AegisShield demonstrated strong performance in identifying and incorporating emerging threats. We show that high semantic similarity scores and mapping rates to MITRE ATT&CK can validate its adaptability across diverse domains.

Together, these findings show how AegisShield fills gaps in traditional threat modeling by providing an accessible, efficient, and cost-effective approach.

**5.3 Contributions to the Body of Knowledge**

This praxis contributes to the advancement of threat modeling by demonstrating the feasibility of a scalable, accessible, and automated threat modeling solution. AegisShield addresses major gaps in traditional methods, particularly for organizations with limited resources, by streamlining and standardizing the threat identification process. The tool's focus on standardization and accessibility enables broader adoption of cybersecurity practices. In that way, it supports the goal of democratizing threat modeling.

From an academic standpoint, this research provides a framework for integrating, assessing, and automating AI into cybersecurity workflows. The method of combining the three hypotheses to evaluate the quality of generative AI results could serve as a future framework for future studies. It highlights how generative AI can simplify complex tasks while maintaining technical rigor, offering a foundation for future exploration into AI-enhanced tools. At the same time, this praxis also stresses the need to examine ethical considerations, such as AI bias and data security, to ensure responsible implementation.

When it comes to cybersecurity, this research shows how AI can enhance efficiency and accessibility while supporting alignment with industry standards. AegisShield's integration with frameworks like MITRE ATT&CK supports its practical value in addressing threats and enabling secure-by-design practices. The findings show the potential of AI-driven tools to scale and standardize threat modeling.

Finally, this research contributes to the broader understanding of AI's role in cybersecurity. We highlight its capacity to address resource constraints while noting challenges, namely, data variability and output consistency. These contributions provide a foundation for ongoing advancements in this area.

### 5.4 Recommendations for Future Research

There are many areas where future research could prove helpful. It should address scalability challenges, explore the effect of adjusting the complexity levels on readability and comprehensibility, and evaluate AegisShield's supplemental outputs (e.g., DREAD risk assessments, attack trees, mitigations, and test cases). Future research should also address the external data limitations by improving the tool's ability to process substantially more threat intelligence. At the time of writing, OpenAI offers the largest context capacity in the LLM space. This will help expand the scope of contextual analyses to include a wider variety of technologies and historical data. Retrieval-Augmented Generation (RAG) may offer a practical way to retrieve and use these threat intelligence datasets efficiently. These types of improvements would help expand the tool's reach across different contexts. Meanwhile, a long-term study should assess the tool's benefits across different industries in real world tests. Additionally, as new generative AI models emerge, we need to test their performance with the tool, particularly in terms of scalability and accuracy. This is important to ensure it remains relevant as technology changes.

Finally, integrating AI-specific risk intelligence into automated threat modeling will be needed as AI-driven cyber threats evolve. MIT's AI Risk Repository introduced a database of over 1000 AI-related risks, categorized by causal factors, domain relevance, and timing (Slattery et al., 2024). We should explore how this repository might enhance AegisShield's adaptability in detecting AI-specific attack vectors. Specifically, areas around adversarial ML exploits, AI-generated misinformation, and automation vulnerabilities. Mapping AI risks from MIT's taxonomy to frameworks like MITRE ATT&CK and STRIDE could also help bridge gaps between traditional threat modeling and newly exposed AI risks. This will help deliver stronger cybersecurity solutions down the road.

## Appendix A

Figure A1. MITRE ATT&CK OpenAI Prompt Text

You are to respond in a very specific format. Do not include any additional text, explanations, or context. Only output the JSON array as specified below.
Act as a cybersecurity expert in the {app_details['industry_sector']} sector with more than 20 years of experience using the STRIDE threat modeling methodology.
Your task is to analyze the following threat scenario and select the single most relevant MITRE ATT&CK attack pattern from the provided list of 25.

APPLICATION TYPE: {app_details['app_type']}
INDUSTRY SECTOR: {app_details['industry_sector']}
AUTHENTICATION METHODS: {app_details['authentication']}
INTERNET FACING: {app_details['internet_facing']}
SENSITIVE DATA: {app_details['sensitive_data']}
APPLICATION DESCRIPTION: {app_details['app_input']}
Threat Scenario:
{Json. Dumps(threat, indent=2)}
MITRE ATT&CK Techniques:
{Json. Dumps(technique_descriptions, indent=2)}

Your response should **ONLY** include the single most relevant MITRE ATT&CK Attack Pattern ID from the above MITRE ATT&CK Techniques, in a JSON array format like this:

["attack-pattern--xxxxxxxx-xxxx-xxxx-xxxx-xxxxxxxxxxxx"]

Select the closest one if none of the provided techniques are a
perfect match. If there truly is no relevant match, respond with ["attack-pattern--00000000-0000-0000-0000-000000000000"].

Figure A2. Threat Identification OpenAI Prompt Text

Act as a cybersecurity expert in the {industry_sector} sector with more than 20 years of experience using the STRIDE threat modeling methodology to produce comprehensive threat models for a wide range of applications. Your task is to use the application description and additional provided data to produce a list of specific threats for the application.

1. On a scale of Low, Medium, or High, the user's technical ability is: {technical_ability}. Simplify explanations for lower abilities without omitting details. For higher abilities, include all technical aspects; for lower abilities, provide clear, more readable explanations despite their lack of technical experience.

2. For each of the STRIDE categories (Spoofing, Tampering, Repudiation, Information Disclosure, Denial of Service, and Elevation of Privilege), list a mandatory multiple (3) credible threats if applicable. Each threat scenario should provide a credible scenario in which the threat could occur in the context of the application. It is very important that your responses are tailored to reflect the details you are given.

3. For each threat scenario, assess the potential impact on data confidentiality, integrity, and availability. Describe how the threat could lead to unauthorized disclosure of sensitive information, corruption or tampering of data, and disruption to system or data access. Not every threat scenario will impact all three aspects, but you should consider each in your analysis.

4. Threat models always have assumptions. For each threat scenario, provide a list of assumptions that must be true for the threat to be realized. Each assumption should include a description of the assumption, the role of the actor making the assumption, and the condition under which the assumption is valid.

5. When providing the threat model, use a JSON-formatted response with the keys "threat_model" and "improvement_suggestions". Under "threat_model", include an array of objects with the keys "Threat Type", "Scenario", "Potential Impact", and "MITRE ATT&CK Keywords".

6. Under "MITRE ATT&CK Keywords", include an array of relevant keywords that accurately represent the threat scenario. These should be a mix of specific and broad terms that capture relevant MITRE ATT&CK techniques. Avoid overly narrow terms and consider including related actions (e.g., "injection," "spoofing") and targets (e.g., "network," "device"). Do NOT include STIX IDs, ATT&CK Reference IDs, or Technique IDs.

7. Ensure that the "Potential Impact" is a concise summary string, not a nested object.

8. Under "improvement_suggestions", include an array of strings with suggestions on how the threat modeler can improve their application description to allow the tool to produce a more comprehensive threat model.

APPLICATION TYPE: {app_type}
INDUSTRY SECTOR: {industry_sector}
AUTHENTICATION METHODS: {authentication}
INTERNET FACING: {internet_facing}
SENSITIVE DATA: {sensitive_data}
APPLICATION DESCRIPTION: {app_input}
HIGH RISK NVD CVE VULNERABILITIES BELOW BASED ON TECHNOLOGIES USED IN THE APPLICATION:
{nvd_vulnerabilities}
ALIENVAULT OTX PULSE DATA FOR THE INDUSTRY SECTOR:
{otx_data}

Example of expected JSON response format:

{{"threat_model": [{{"Threat Type": "Spoofing","Scenario": "Example Scenario 1","Assumptions": [{{"Assumption": "Example Assumption 1", "Role": "Example Role 1", "Condition": "Example Condition 1"}},{{"Assumption": "Example Assumption 2", "Role": "Example Role 2", "Condition": "Example Condition 2"}}],"Potential Impact": "Example Potential Impact 1","MITRE ATT&CK Keywords": ["Example Keyword 1", "Example Keyword 2", "Example Keyword 3"]}},{{"Threat Type": "Spoofing","Scenario": "Example Scenario 2","Assumptions": [{{"Assumption": "Example Assumption 3", "Role": "Example Role 3", "Condition": "Example Condition 3"}},{{"Assumption": "Example Assumption 4", "Role": "Example Role 4", "Condition": "Example Condition 4"}}],"Potential Impact": "Example Potential Impact 2","MITRE ATT&CK Keywords": ["Example Keyword 1", "Example Keyword 2", "Example Keyword 3", "Example Keyword 4"]}} // ...

more threats],"improvement_suggestions": ["Example improvement suggestion 1.","Example improvement suggestion 2." // ... more suggestions]}}

Figure A3. DREAD Risk Assessment OpenAI Prompt Text

Act as a cyber security expert with more than 20 years of experience in threat modeling using STRIDE and DREAD methodologies.
Your task is to produce a DREAD risk assessment for the threats identified in a threat model.
Below is the list of identified threats (This should be your primary focus):
{threats}
Below is how they map to the MITRE ATT&CK framework (This is supplemental information for context):
{mitre_mapping}
Below are potential vulnerabilities found in the National Vulnerability Database (NVD) that could be exploited by attackers (This is supplemental information for context:
{nvd_vulnerabilities}
When providing the risk assessment, use a JSON formatted response with a top-level key "Risk Assessment" and a list of threats, each with the following sub-keys:
- "Threat Type": A string representing the type of threat (e.g., "Spoofing").
- "Scenario": A string describing the threat scenario.
- "Damage Potential": An integer between 1 and 10.
- "Reproducibility": An integer between 1 and 10.
- "Exploitability": An integer between 1 and 10.
- "Affected Users": An integer between 1 and 10.
- "Discoverability": An integer between 1 and 10.

Assign a value between 1 and 10 for each sub-key based on the DREAD methodology. Use the following scale:
- 1-3: Low
- 4-6: Medium
- 7-10: High

Ensure the JSON response is correctly formatted and does not contain any additional text. Here is an example of the expected JSON response format:
{{ "Risk Assessment": [{{"Threat Type": "Spoofing","Scenario": "An attacker could create a fake OAuth2 provider and trick users into logging in through it.","Damage Potential": 8,"Reproducibility": 6,"Exploitability": 5,"Affected Users": 9,"Discoverability": 7}},{{"Threat Type": "Spoofing","Scenario": "An attacker could intercept the OAuth2 token exchange process through a Man-in-the-Middle (MitM) attack.","Damage Potential": 8,"Reproducibility": 7,"Exploitability": 6,"Affected Users": 8,"Discoverability": 6}}]}}

Figure A4. Mitigation Strategies OpenAI Prompt Text

Act as a cybersecurity expert with more than 20 years of experience using the STRIDE threat modeling methodology. Your task is to provide potential mitigations for the threats identified in the threat model. It is crucial that your responses are tailored to reflect the details of the threats.

Please output the results in a markdown table format using the following columns:

- Column A: Threat Type
- Column B: Scenario
- Column C: Suggested Mitigation(s)

Do not use '<br>' or any other HTML tags in your response as a line break and do not use bullet points in a table cell.

Below is the list of identified threats:

{threats}

Below is how they map to the MITRE ATT&CK framework:

{mitre_mapping}

Below are potential vulnerabilities found in the National Vulnerability Database (NVD) that could be exploited by attackers:

{nvd_vulnerabilities}

YOUR RESPONSE (do not wrap in a code block):

Figure A5. Test Cases OpenAI Prompt Text

Act as a cyber security expert with more than 20 years experience of using the STRIDE threat modelling methodology.

Your task is to provide Gherkin test cases for the threats identified in a threat model. It is very important that your responses are tailored to reflect the details of the threats.

Below is the list of identified threats:

{threats}

Use the threat descriptions in the 'Given' steps so that the test cases are specific to the threats identified.

Put the Gherkin syntax inside triple backticks (```) to format the test cases in Markdown. Add a title for each test case.

For example:

```gherkin
Given a user with a valid account
When the user logs in
Then the user should be able to access the system
```

YOUR RESPONSE (do not add introductory text, just provide the Gherkin test cases):

Figure A6.1 Attack Tree OpenAI Prompt Text

Act as a cyber security expert with more than 20 years of experience using the STRIDE threat modelling methodology to produce comprehensive threat models for a wide range of applications. Your task is to use the application description provided to you to produce an attack tree in Mermaid syntax.

The attack tree should reflect the potential threats for the application based on all the details given. You should create multiple levels in the tree to capture the hierarchy of threats and sub-threats, ensuring a very detailed and comprehensive representation of the attack scenarios. Use subgraphs to group related threats for better readability.

You MUST only respond with the Mermaid code block. See below for an example of the required format and syntax for your output.

Please utilize proper terminology and structure to ensure the attack tree is clear, organized, and informative. If a MITRE ATT&CK pattern is mentioned, include the relevant details in the attack tree.

````
  ```mermaid
graph LR
  A["Compromise of Application (CIA)"] --> B(Spoofing)
  A --> C(Tampering)
  A --> D(Repudiation)
  A --> E["Information Disclosure"]
  A --> F["Denial of Service (DoS)"]
  A --> G["Elevation of Privilege"]

  %% Subgraph for Spoofing Threats
  subgraph Spoofing Threats
    B[Sub-threat 1: Spoofing]
    B --> B1[Detailed Threat 1.1]
    B --> B2[Detailed Threat 1.2]
    B1 --> B1a[Specific Attack Vector 1.1]
    B2 --> B2a[Specific Attack Vector 1.2]
    ...
    ...
  end
%% Subgraph for Tampering Threats
  subgraph Tampering Threats
    C[Sub-threat 2: Tampering]
    C --> C1[Detailed Threat 2.1]
    C --> C2[Detailed Threat 2.2]
````
````

Figure A6.2 Continued – Attack Tree OpenAI Prompt Text

```
    ...
    ...
    C1 --> C1a[Specific Attack Vector 2.1]
    C2 --> C2a[Specific Attack Vector 2.2]
    ...
    ...
end

%% Subgraph for Repudiation Threats
subgraph Repudiation Threats
    D[Sub-threat 3: Repudiation]
    D --> D1[Detailed Threat 3.1]
    D --> D2[Detailed Threat 3.2]
    ...
    ...
    D1 --> D1a[Specific Attack Vector 3.1]
    D2 --> D2a[Specific Attack Vector 3.2]
    ...
    ...
end
%% Subgraph for Information Disclosure Threats
subgraph Information Disclosure Threats
    E[Sub-threat 4: Information Disclosure]
    E --> E1[Detailed Threat 4.1]
    E --> E2[Detailed Threat 4.2]
    ...
    ...
    E1 --> E1a[Specific Attack Vector 4.1]
    E2 --> E2a[Specific Attack Vector 4.2]
    ...
    ...
end
```

Figure A6.3 Continued – Attack Tree OpenAI Prompt Text

````
    %% Subgraph for Denial of Service Threats
    subgraph Denial of Service Threats
      F[Sub-threat 5: Denial of Service]
      F --> F1[Detailed Threat 5.1]
      F --> F2[Detailed Threat 5.2]
      F1 --> F1a[Specific Attack Vector 5.1]
      F2 --> F2a[Specific Attack Vector 5.2]
    end
    %% Subgraph for Elevation of Privilege Threats
    subgraph Elevation of Privilege Threats
      G[Sub-threat 6: Elevation of Privilege]
      G --> G1[Detailed Threat 6.1]
      G --> G2[Detailed Threat 6.2]
      ...
      ...
      G1 --> G1a[Specific Attack Vector 6.1]
      G2 --> G2a[Specific Attack Vector 6.2]
      ...
      ...
    end
```
````

IMPORTANT: Round brackets are special characters in Mermaid syntax. If you want to use round brackets inside a node label you MUST wrap the label in double quotes. For example, ["Example Node Label (ENL)"].

Application description: {application_description}
APPLICATION TYPE: {app_type}
AUTHENTICATION METHODS: {authentication}
INTERNET FACING: {internet_facing}
SENSITIVE DATA: {sensitive_data}
APPLICATION DESCRIPTION: {app_input}
#STRIDE AND MITRE ATT&CK TTPs:
#{mitre_data}
#VD VULNERABILITIES:
#{nvd_vulnerabilities}
#ALIENTVAULT OTX CYBER THREAT INTELLIGENCE:
#{otx_vulnerabilities}
````

Table A1. Tool Questions Overview

| Question# | Question | Purpose | Expect Input |
|---|---|---|---|
| 1 | Upload a Dataflow Diagram | This can help a user get started on pre-filling out and writing their description. | An image file |
| 2 | Describe the application to be modeled | To collect a detailed description of the application, including its purpose, technologies, and other relevant information. | Text input (Descriptive text) |
| 3 | Application Type | To classify the application and understand its specific threat landscape. | Drop-down selection (e.g., Web application, Mobile application, etc.) |
| 4 | Industry Sector | To tailor the threat model to industry specific threats and compliance requirements. | Drop-down selection (e.g., Healthcare, Finance, etc.) |
| 5 | Data Sensitivity | To assess the potential impact and prioritization of threats based on data sensitivity. | Drop-down selection (None, Low, Medium, High) |
| 6 | Internet Facing | To determine if the application is exposed to the internet, influencing the threat landscape. | Drop-down selection (Yes or No) |
| 7 | Number of Employees | To gauge the potential scale of the organization and its exposure to internal threats. | Drop-down selection (Unknown, 0-10, 11-100, 101-1000, Over 1000) |
| 8 | Compliance Requirements | To ensure the model considers applicable regulatory standards. | Multiple selection (e.g., HIPAA, PCI DSS, GDPR, etc.) |
| 9 | Authentication Methods | To identify potential weaknesses in the system's authentication mechanisms. | Multiple selection (e.g., MFA, Passwords, API Key, etc.) |
| 10 | User’s Technical Level | To understand the user's technical background, which can impact how the threat model is presented. | Drop-down selection (Static) (Medium) |

| Question# | Question | Purpose | Expect Input |
|---|---|---|---|
| 11 | Database Technology Used | To identify specific threats related to the database technologies employed. | Drop-down selection, plus text input for version numbers. |
| 12 | Operating Systems Used | To assess the risks related to specific operating systems in use. | Drop-down selection, plus text input for version numbers. |
| 13 | Programming Languages Used | To determine potential vulnerabilities associated with the programming languages used. | Drop-down selection, plus text input for version numbers. |
| 14 | Web Frameworks Used | To identify potential risks associated with the web frameworks employed. | Drop-down selection, plus text input for version numbers. |

Table A2. Public APIs, External Resources, and Code Structure of AegisShield

| Resource/File | Description | Link/Reference |
| --- | --- | --- |
| National Vulnerability Database (NVD) API | Free resource that provides real-time data on known vulnerabilities. | https://nvd.nist.gov/vuln/data-feeds#APIS |
| MITRE ATT&CK Framework | Free curated knowledge base of adversary tactics and techniques. | https://github.com/mitre-attack/attack-stix-data |
| AlientVault Open Threat Exchange (OTX) API | Free community-powered threat intelligence platform. | https://otx.alienvault.com/api |
| main.py | The entry point and main UI for AegisShield. | N/A |
| alienvault_search.py | The module that handles the integration with AlienVault's OTX. | N/A |
| dread.py | Generates the DREAD risk assessment generative AI prompt model for prioritizing identified threats. | N/A |
| Mitigations.py | Generates mitigation strategies generative AI prompt for identified threats based on risk assessment and threat modeling. | N/A |
| mitre_attack.py | Interfaces with the local MITRE ATT&CK framework files to map identified threats to known tactics, techniques, and procedures. | N/A |
| nvd_search.py | Searches and retrieves vulnerability data from the National Vulnerability Database (NVD). | N/A |
| test_cases.py | Generates test cases generative AI prompt based on identified threats to validate mitigation strategies. | N/A |

| Resource/File | Description | Link/Reference |
|---|---|---|
| threat_model.py | Core logic for generating and analyzing the generative AI prompt threat models for various cybersecurity frameworks. | N/A |
| AegisShield Code Repository | Full extracted case study descriptions, rubric scoring details, and implementation code, JSON samples and source files are available in the AegisShield repository. | https://github.com/mgrofsky/AegisShield |

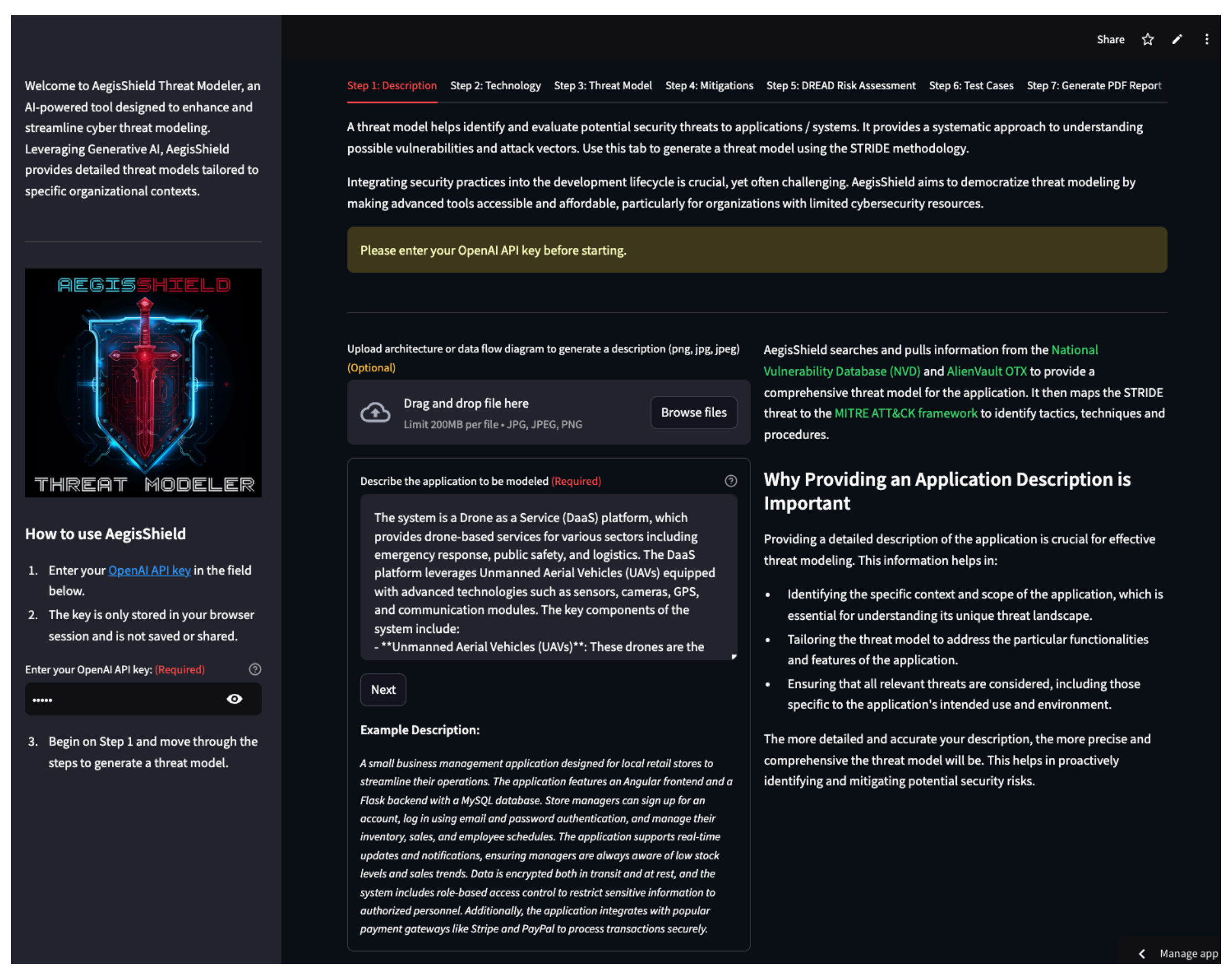

Figure A7. Main UI Tool Web page

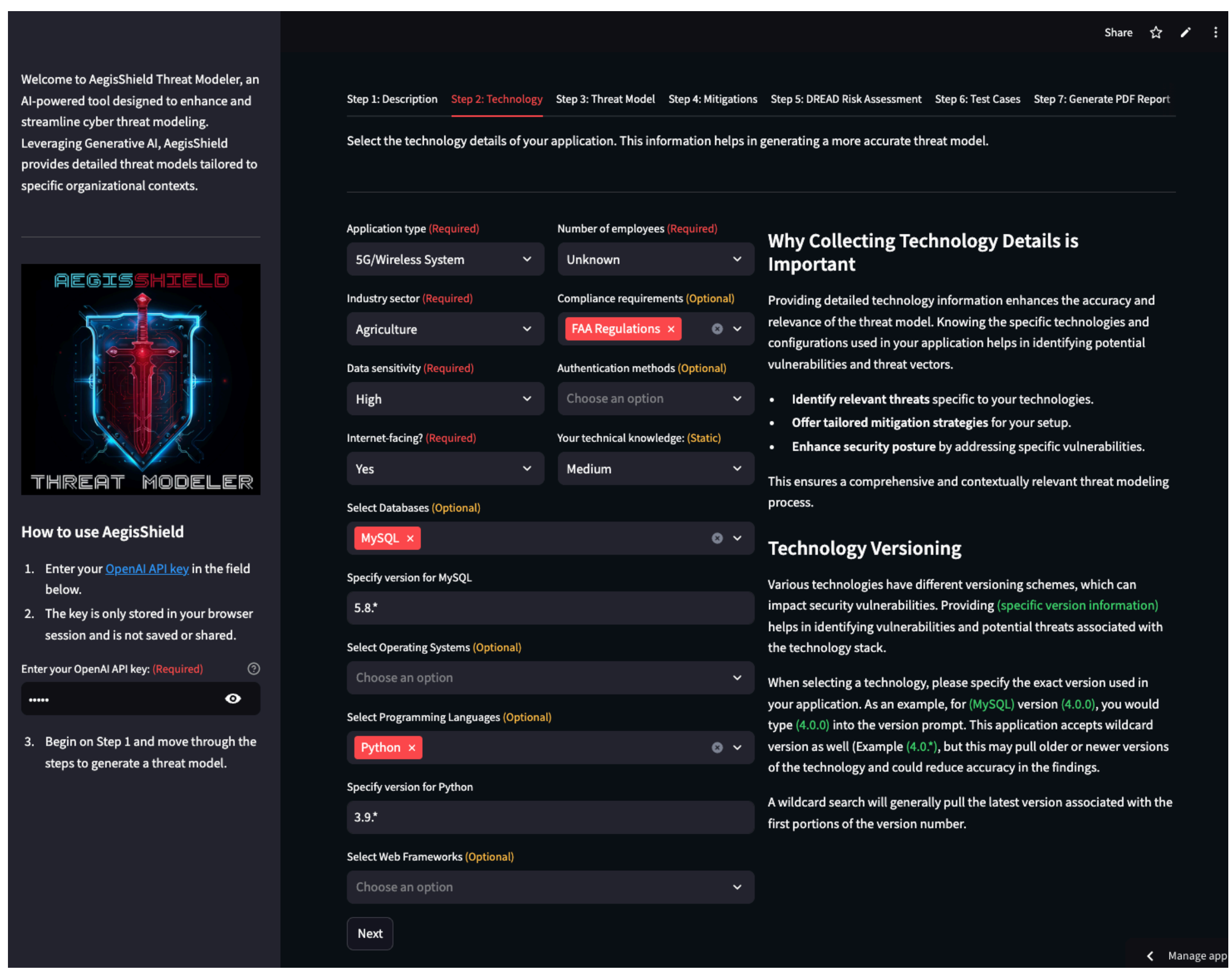

Figure A8. Technology and Compliance UI Tool Web page

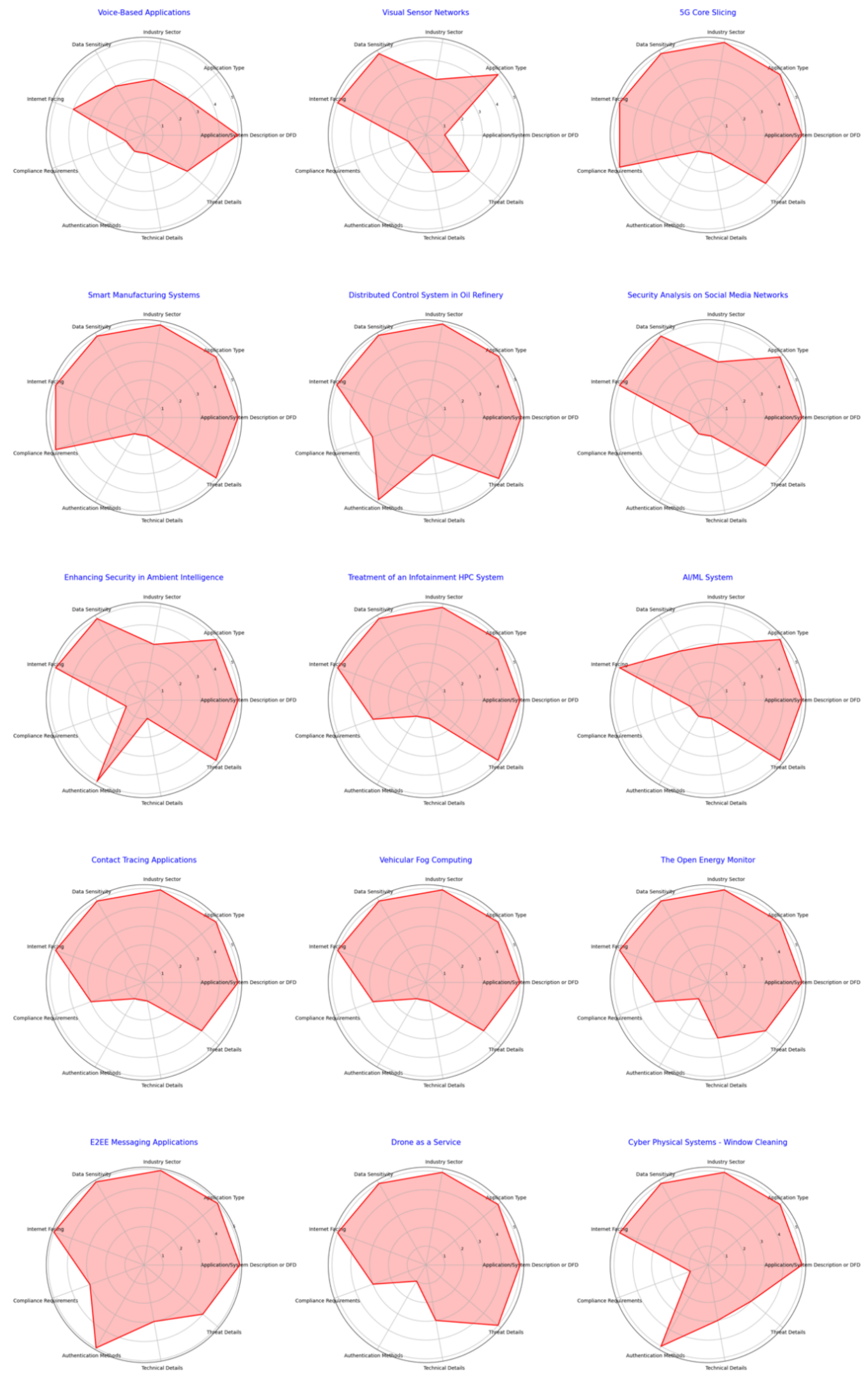

Figure A9. Case Study Rubric Radar Chart

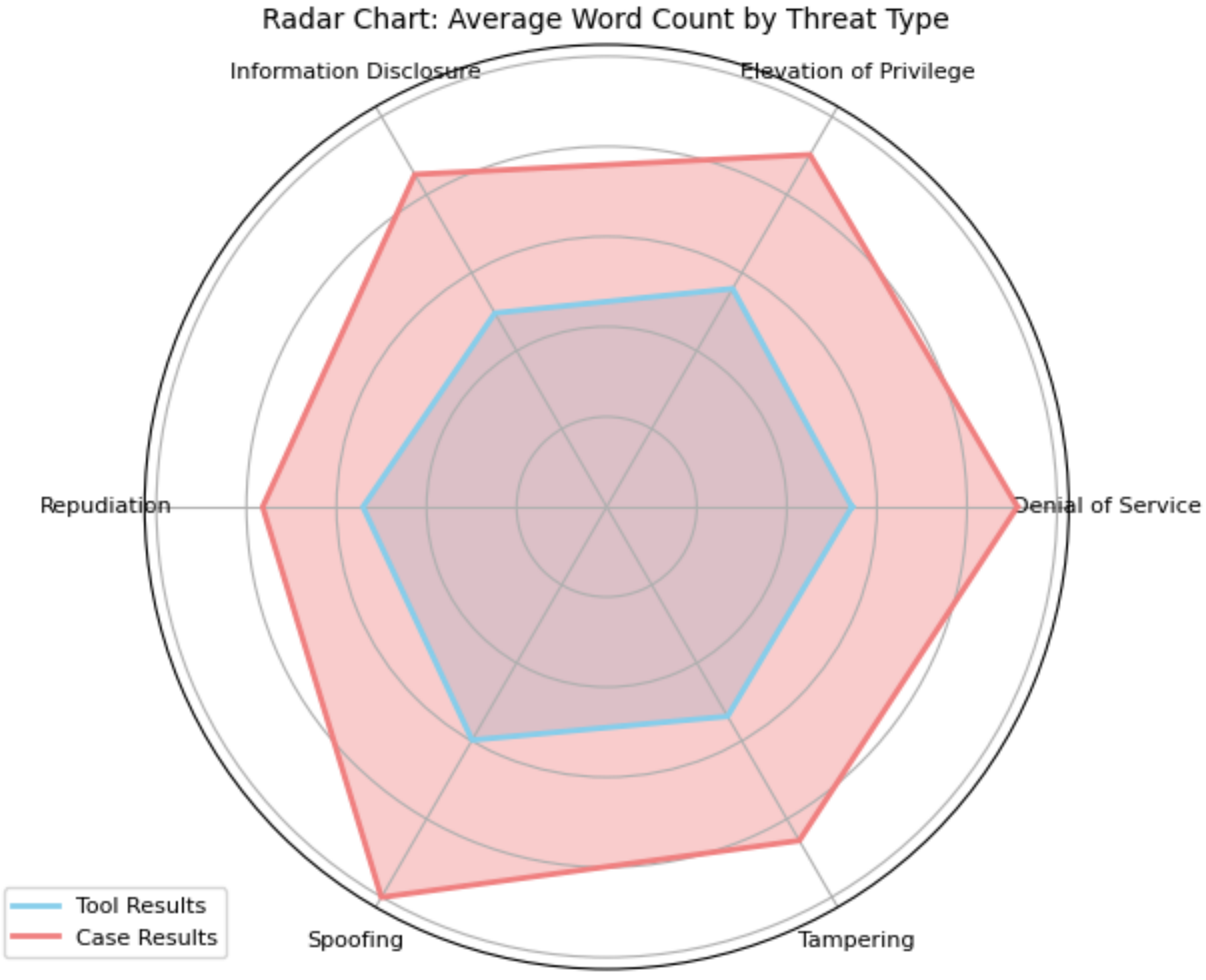

Figure A10. Radar Chart of Word Count by Threat Type and Source

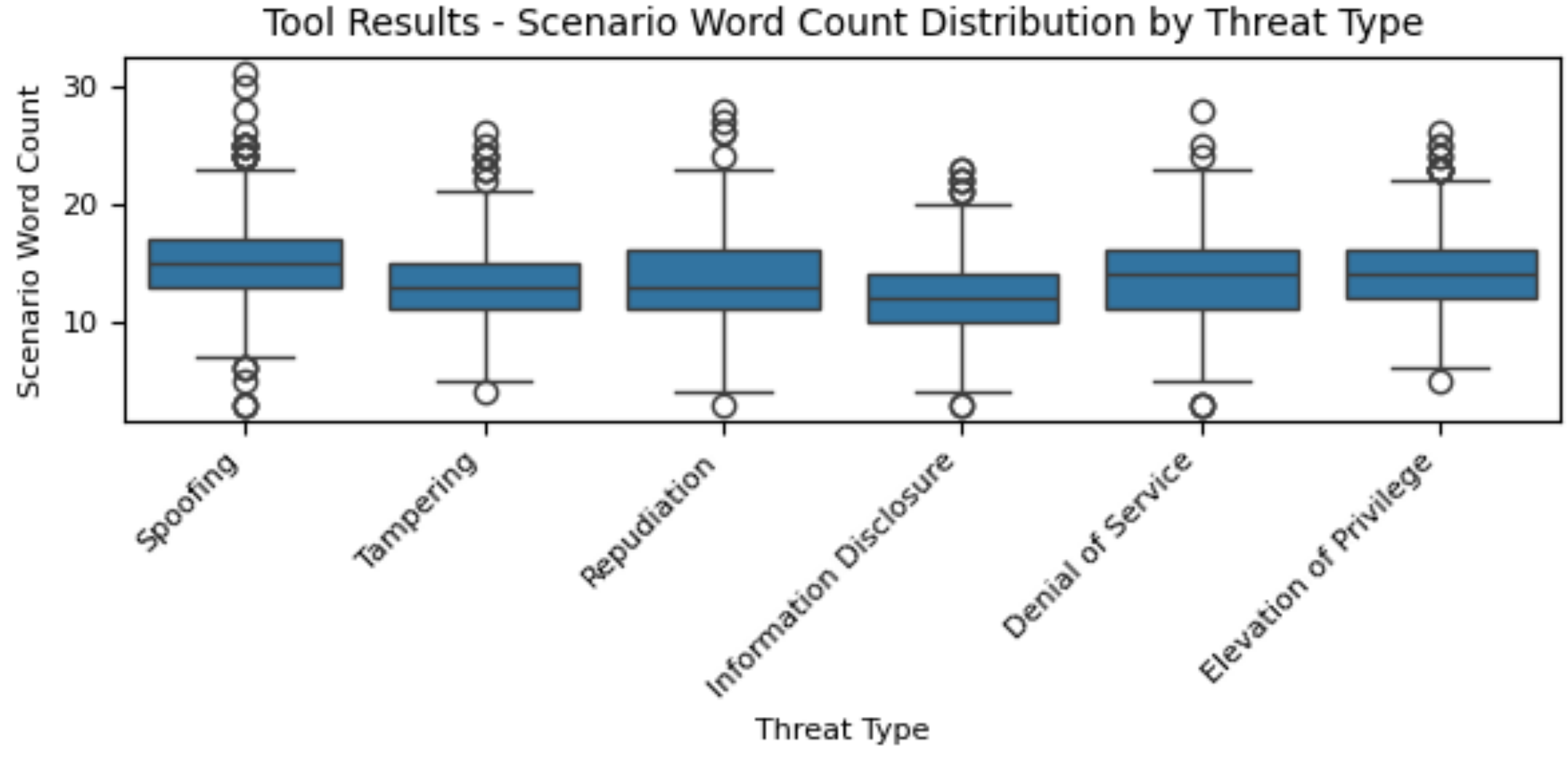

Figure A11. Tool Results - Scenario Word Count Distribution by Threat Type

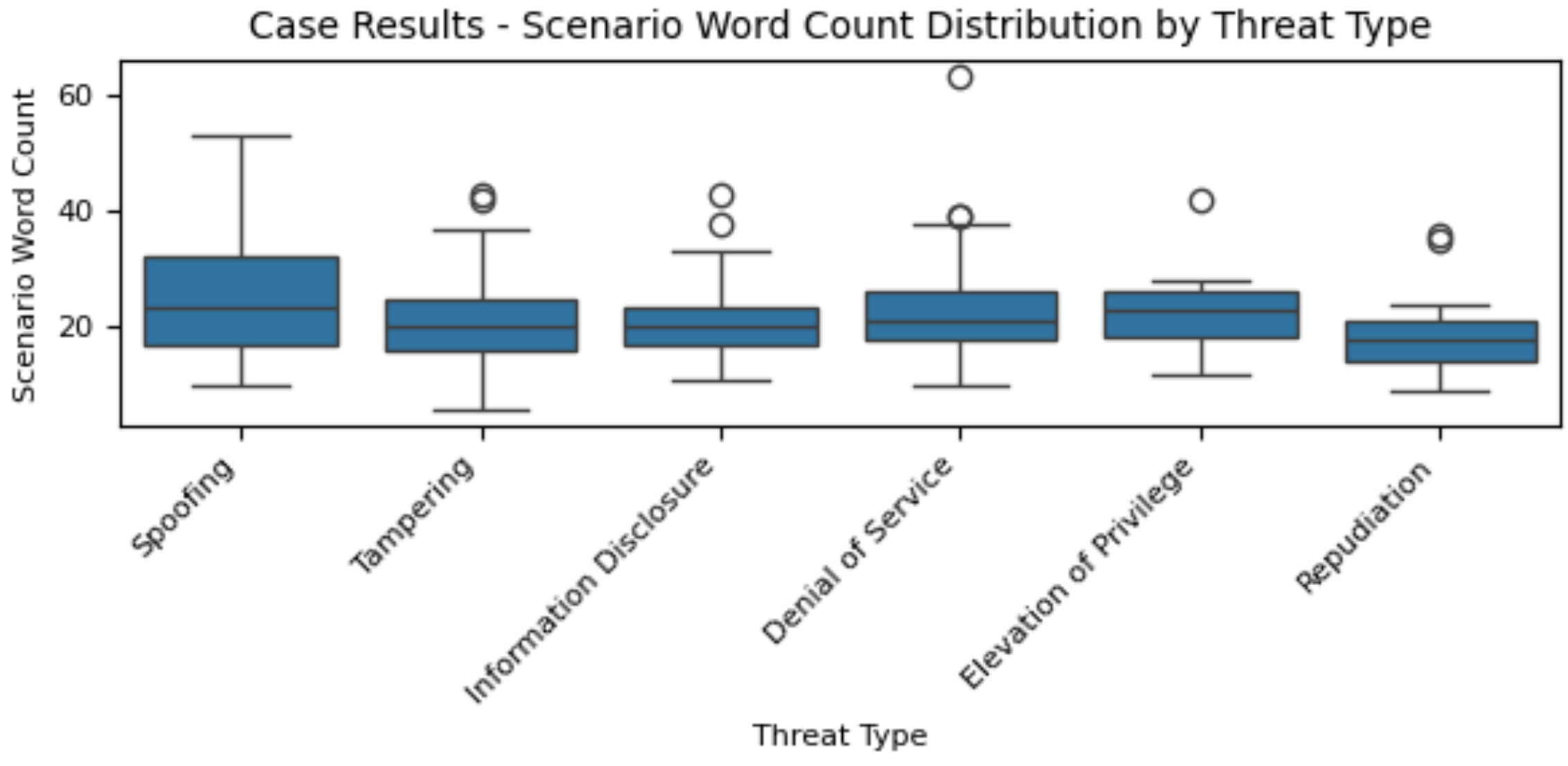

Figure A12. Case Results: Scenario Word Count Distribution by Threat Type

Word Cloud for Case Results

Figure A13. Word Cloud of the Case Study Threats

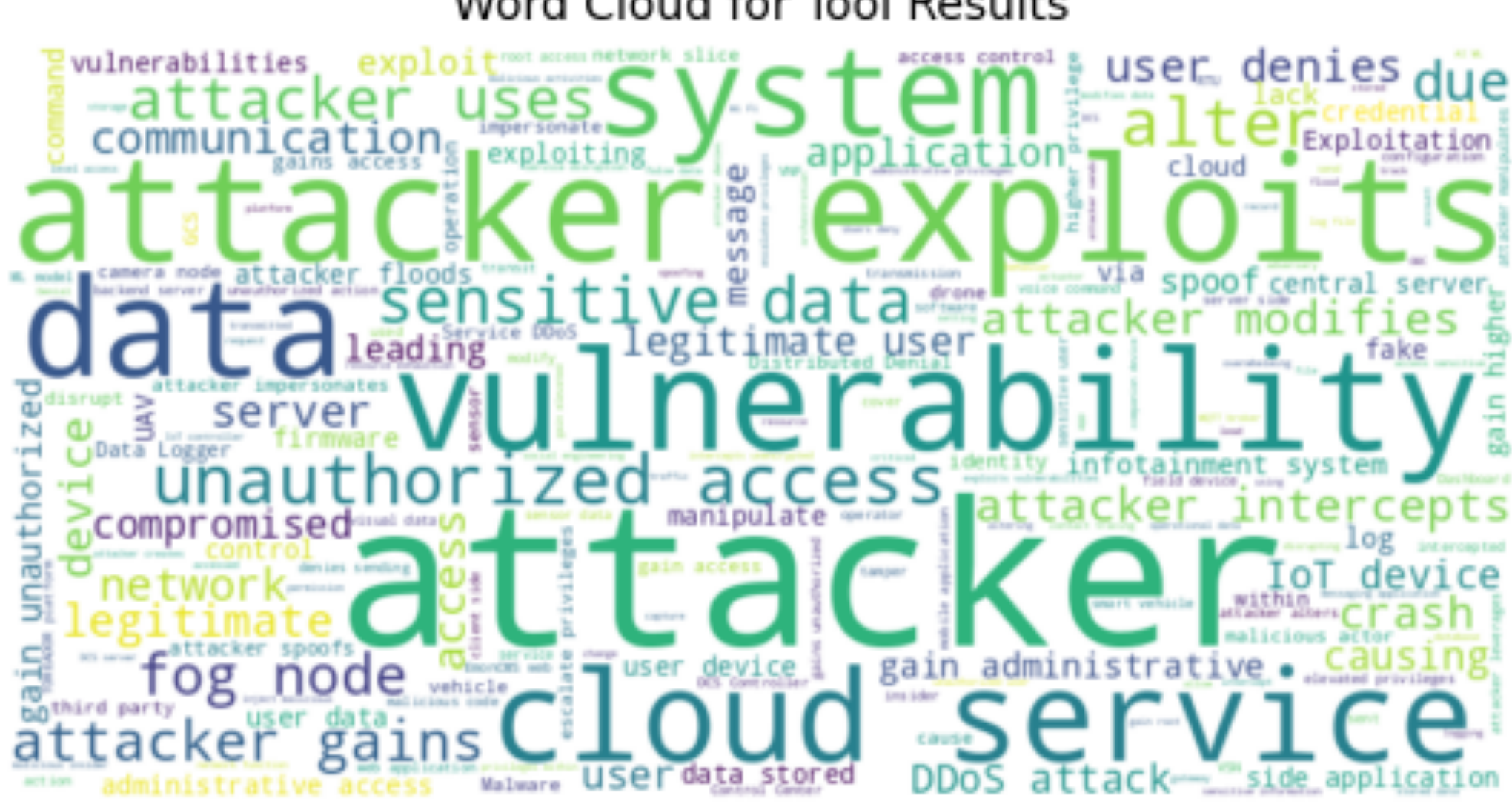

Figure A14. Word Cloud of the Tool-generated Threats

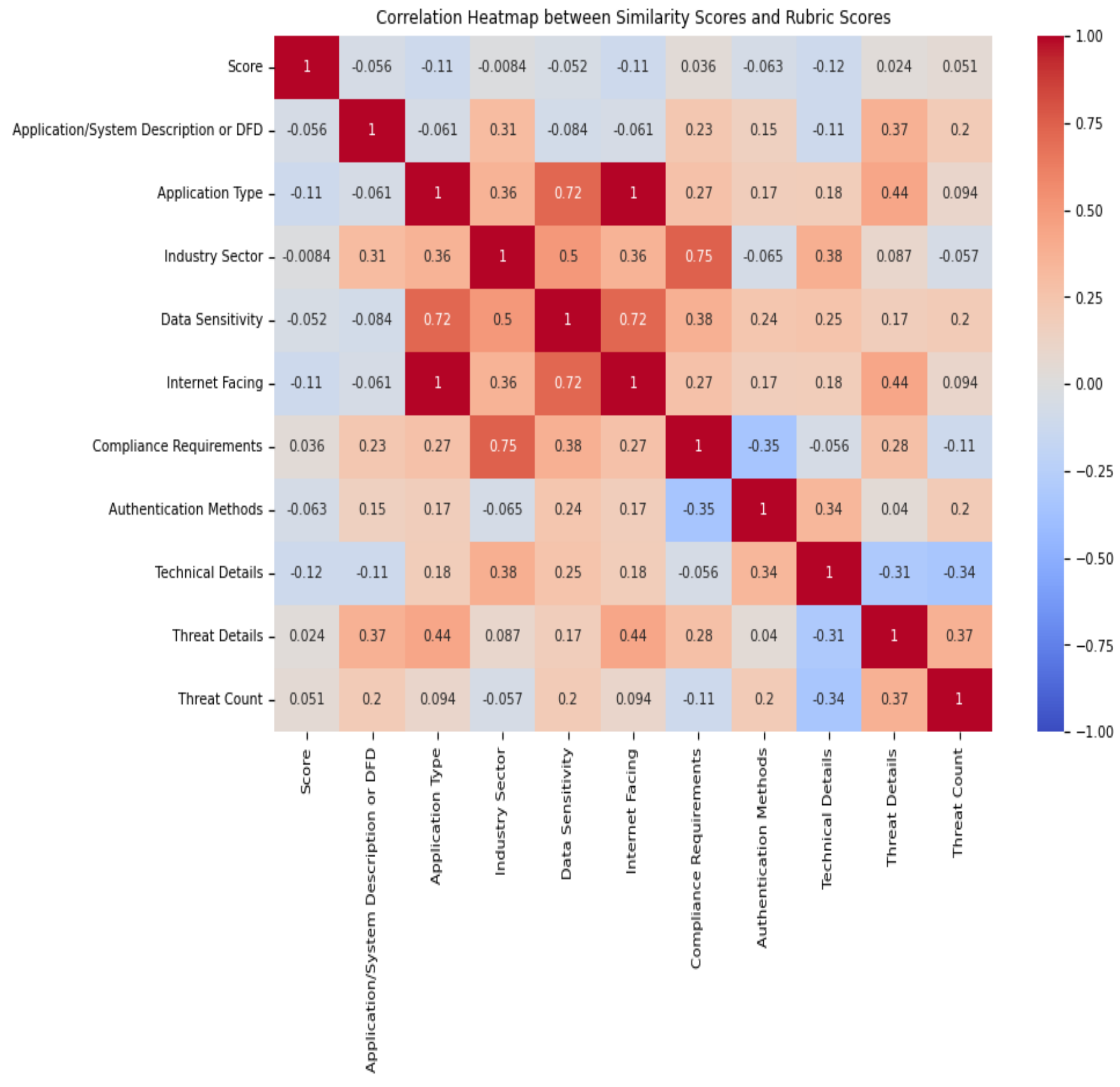

Figure A15.1. Correlation Heatmap Between the Similarity and Rubric Scores (Threat Level)

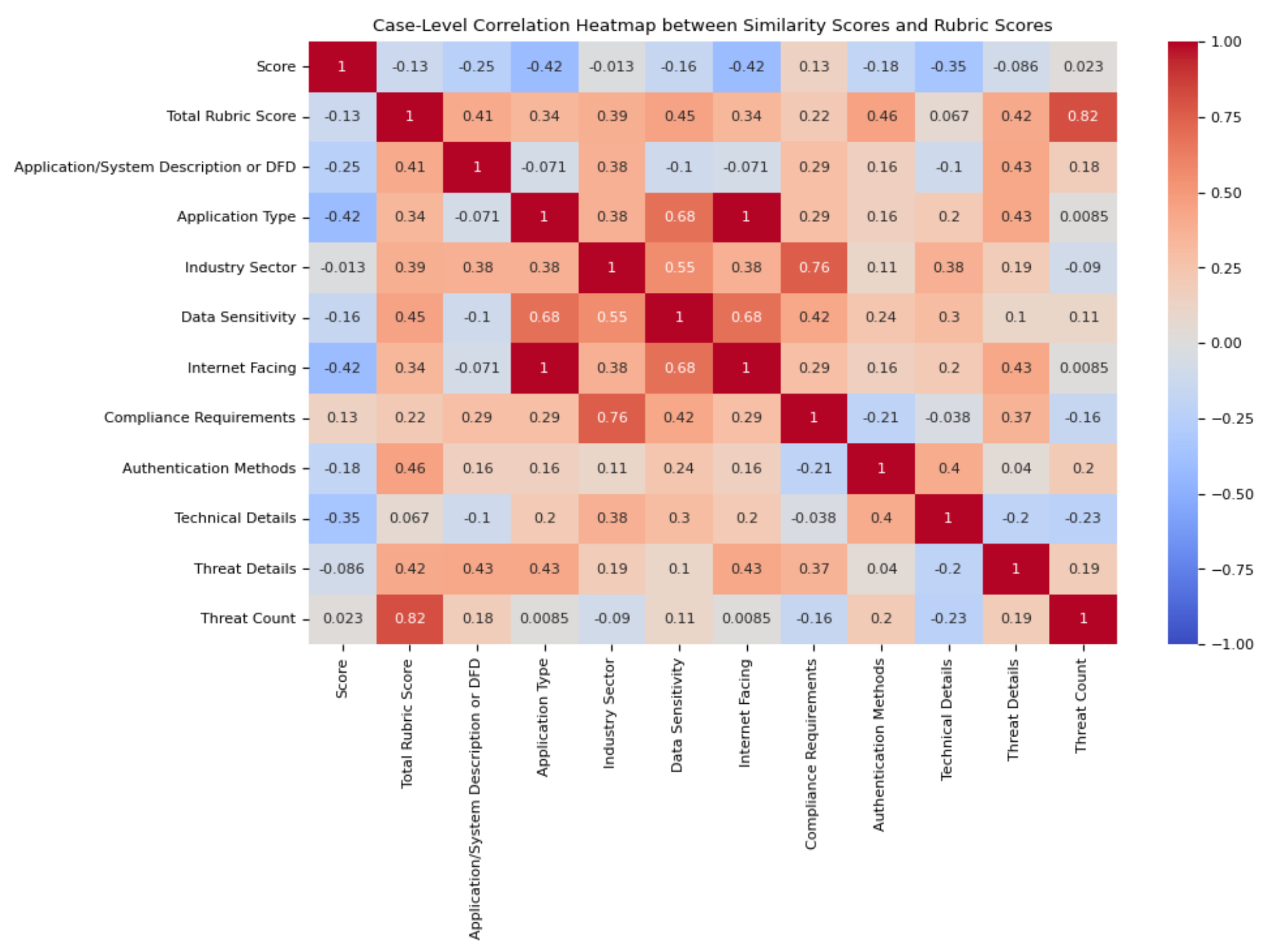

Figure A15.2. Correlation Heatmap Between the Similarity and Rubric Scores (Case Level)

Table A3. Top Two Similarity Scores per Case Study

Expert-generated threats are quoted or paraphrased from the 15 case studies cited in Section 2.4.

| Case | Threat Type | Tool-Generated Threat | Expert-Generated Threat | Score |
|---|---|---|---|---|
| 1 | Spoofing | An attacker could impersonate a legitimate user by recording and replaying voice commands to the microphone. | Voice command: An attacker can record the voice of a real user and submit it to the system, that is, it can be faked. | 0.8743225 |
| 1 | Spoofing | An attacker could spoof the user's voice commands by recording and replaying them to the microphone. | Voice command: An attacker can record the voice of a real user and submit it to the system, that is, it can be faked. | 0.8700433 |
| 10 | Spoofing | An attacker could spoof the device ID to masquerade as a legitimate user, gaining access to the contact tracing system. | Source Device Spoofing: The attacker can steal the user’s device data and pretend to be a legitimate device with performing the authentication to the server. | 0.8923649 |
| 10 | Spoofing | An attacker pretends to be a legitimate user by exploiting weak or missing authentication methods during device pairing. | Source Device Spoofing: The attacker can steal the user’s device data and pretend to be a legitimate device with performing the authentication to the server. | 0.8864945 |
| 11 | Spoofing | An attacker spoofs a smart vehicle's identity to send false data to a fog node. | Impersonation Attack: In an impersonation attack, attackers masquerade themselves as another identity. In VFC, attackers could compromise another vehicle’s user account, authenticate themselves to a fog node and impersonate this vehicle. | 0.8966899 |

| Case | Threat Type | Tool-Generated Threat | Expert-Generated Threat | Score |
|---|---|---|---|---|
| 11 | Spoofing | An attacker spoofs a smart vehicle's identity to send false data to fog nodes. | Impersonation Attack: In an impersonation attack, attackers masquerade themselves as another identity. In VFC, attackers could compromise another vehicle’s user account, authenticate themselves to a fog node and impersonate this vehicle. | 0.891749 |
| 12 | Information Disclosure | An attacker eavesdrops on unencrypted MQTT communication to capture sensitive data. | Eavesdropping (Global): An adversary retrieve data accessing communication among multiple assets communicating through MQTT. | 0.79319096 |
| 12 | Denial of Service | An attacker launches a Denial of Service attack against the MQTT Broker, disrupting communication between IoT devices. | Device Isolation: An attacker can make the asset (an IoT Device acting as MQTT client) unable to send or receive messages. | 0.7871238 |
| 13 | Spoofing | An attacker creates a fake desktop client that mimics a legitimate messaging application to steal user credentials. | In Signal, attackers can capture credentials using a tls interceptor from the rooted device, restart the desktop client in the attacker’s device and operate as the victim. | 0.77903646 |
| 13 | Spoofing | An attacker creates a fake messaging application that mimics the legitimate ones (e.g., WhatsApp, Signal) to harvest user credentials. | For WhatsApp, an attacker with short-lived access can steal the credentials and communicate as the victim. | 0.7692005 |
| 14 | Tampering | An attacker could tamper with the drone firmware by exploiting vulnerabilities and uploading malicious firmware updates. | Firmware updates can be tampered with by attackers, enabling them to inject malicious code into the drone system. | 0.8866515 |
| 14 | Tampering | Firmware on the drones is altered by a threat actor to inject malicious code. | Firmware updates can be tampered with by attackers, enabling them to inject malicious code into the drone system. | 0.8818167 |
| 15 | Tampering | An attacker could intercept HTTP POST requests and manipulate data sent to the API gateways. | API Gateway: Spoofing since API keys are sent in the HTTP POST request URL and can be obtained. | 0.83956796 |

| Case | Threat Type | Tool-Generated Threat | Expert-Generated Threat | Score |
|---|---|---|---|---|
| 15 | Tampering | Attackers modify HTTP POST requests to manipulate data sent to API gateways. | API Gateway: Spoofing since API keys are sent in the HTTP POST request URL and can be obtained. | 0.7961138 |
| 2 | Spoofing | An attacker could spoof a camera node within the Visual Sensor Network (VSN) to gain unauthorized access. | Manipulate drivers: If an attacker manages to manipulate the sensor drivers of a VSN node, this may become a channel to inject forged image data without being recognized by the application. | 0.84395635 |
| 2 | Spoofing | An attacker spoofs the identity of legitimate camera nodes to inject false visual data into the network. | Fake/impersonate nodes: An attacker might create fake nodes or impersonate benign nodes through publishing forged data. | 0.8310221 |
| 3 | Tampering | An attacker could tamper with configuration files of Virtual Network Functions (VNFs). | An adversary with access to the network between the VNF image repository and host can alter image data (e.g., incorporate malware) as it is downloaded. | 0.8579632 |
| 3 | Tampering | Attacker modifies VNFs to introduce malicious code. | An adversary with access to the network between the VNF image repository and host can alter image data (e.g., incorporate malware) as it is downloaded. | 0.83344126 |
| 4 | Elevation of Privilege | An attacker exploits a software vulnerability in the Control Center's HMI to gain administrative privileges. | An attacker could remotely execute arbitrary code on the Human Machine Interface (HMI) and gain unauthorized privilege by tricking the operator into opening a crafted PowerPoint document. | 0.82352394 |
| 4 | Elevation of Privilege | An attacker exploits a software vulnerability in the HMI to gain admin privileges. | An attacker could remotely execute arbitrary code on the Human Machine Interface (HMI) and gain unauthorized privilege by tricking the operator into opening a crafted PowerPoint document. | 0.80959594 |
| 5 | Tampering | An attacker intercepts and alters the command data between the DCS controller and actuators to cause unsafe operating conditions. | Attackers can tamper with the process by modifying data or commands exchanged between controllers and actuators, creating unsafe conditions (DF-3, DF-9, DF-11). | 0.8190523 |

| Case | Threat Type | Tool-Generated Threat | Expert-Generated Threat | Score |
|---|---|---|---|---|
| 5 | Denial of Service | Attackers flood the DCS network with traffic, causing availability issues. | DoS attacks interrupt or interfere with regular operations by generating excessive traffic in a DCS network or exhausting system resources by calling a specific DCS process. | 0.80077785 |
| 6 | Spoofing | Attacker uses phishing techniques to obtain user login credentials. | Unauthorized party gain access to information. Example: Malicious links, e.g., phishing URL. | 0.8219571 |
| 6 | Spoofing | Attacker uses phishing techniques to obtain user login credentials. | Unauthorized party gain access to information. Example: Malicious links, e.g., phishing. | 0.8163984 |
| 7 | Elevation of Privilege | An attacker uses social engineering to trick an authorized user into revealing admin credentials, gaining elevated access. | Social Engineering: Through social engineering techniques, attackers may trick authorized users or administrators into disclosing their credentials or granting higher-level access. | 0.94198334 |
| 7 | Spoofing | An attacker uses a malicious device to spoof legitimate user devices and gain unauthorized access. | Identity Spoofing: In identity spoofing attacks, attackers mimic the identity of authorized users or devices to gain access to the system. | 0.90784574 |
| 8 | Tampering | Injection of malicious code via Bluetooth to alter system operations. | An adversary may tamper with the data flow from Bluetooth to on-board computer and make unauthorized manipulation to the system. | 0.86805516 |
| 8 | Spoofing | An attacker might spoof Bluetooth signals to connect unauthorized devices to the infotainment system. | An adversary may tamper with the data flow from Bluetooth to on-board computer and make unauthorized manipulation to the system. | 0.8617716 |
| 9 | Repudiation | An attacker deletes logs to obfuscate changes made to the data within the system. | An attacker deletes information about a transaction, such as login records, to deny that a certain action occurred. | 0.8908032 |
| 9 | Repudiation | An attacker alters log files to erase traces of unauthorized activities. | An attacker deletes information about a transaction, such as login records, to deny that a certain action occurred. | 0.8351402 |

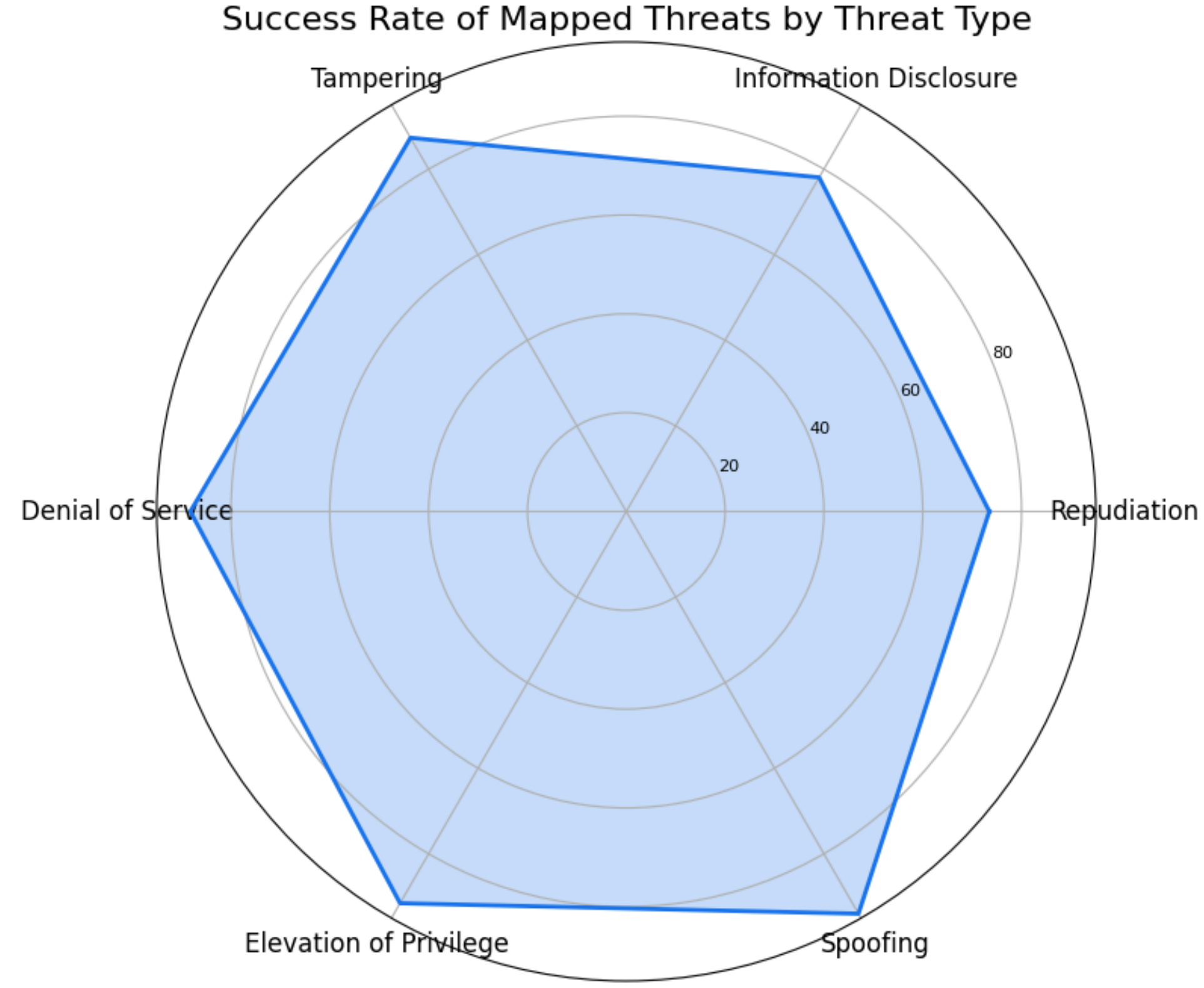

Figure A16. Radar Chart of the MITRE ATT&CK Mappings by Threat Type

## Appendix B

Figure B15. Example PDF Artifact: DaaS Case Study

**AegisShield Security Report**

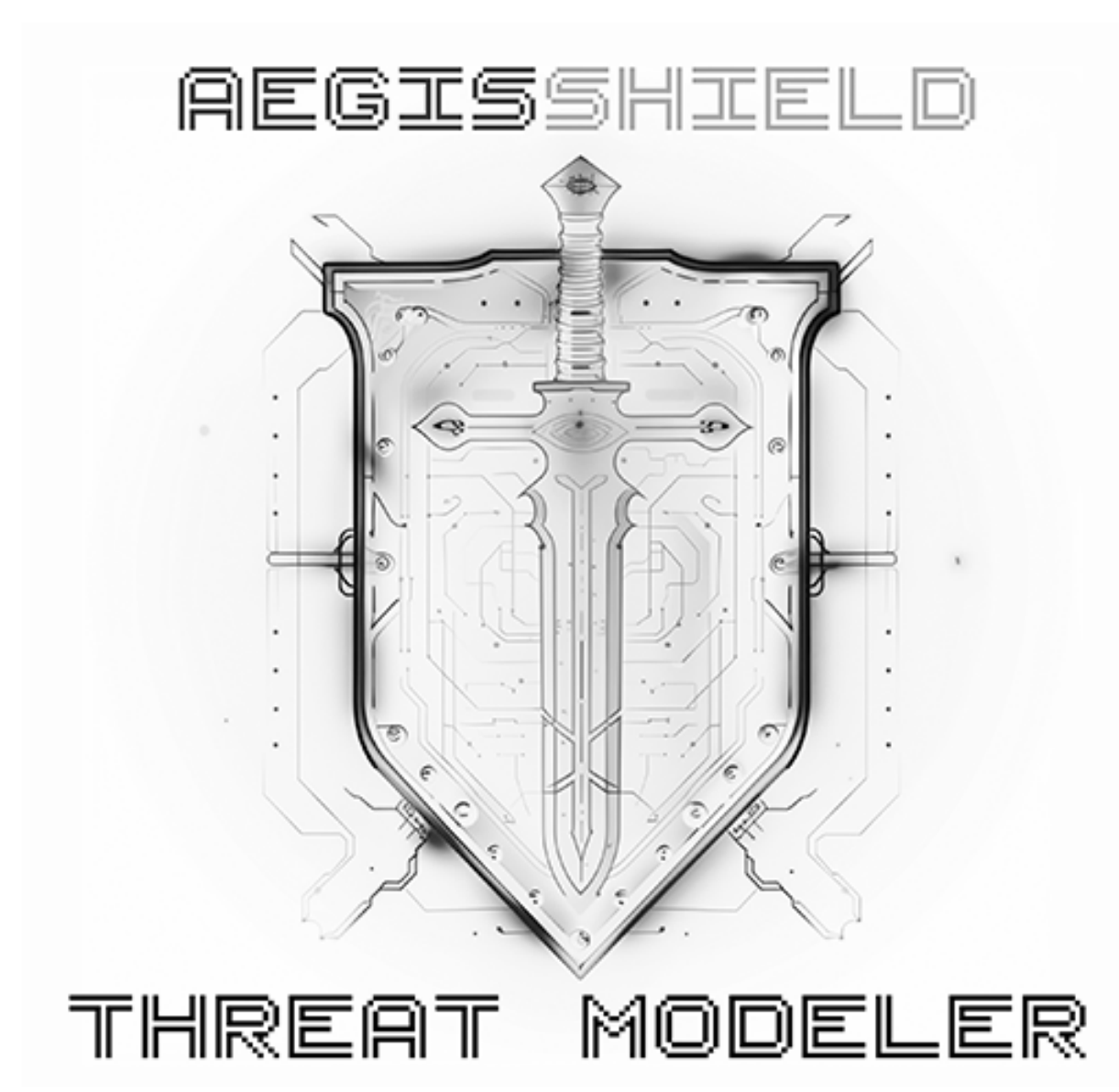

Created: 2024-10-31 22:05:49

## Application Description

- **Application Type:** Drone as a Service (DaaS) Application
- **Industry Sector:** Aerospace
- **Sensitive Data:** High
- **Internet Facing:** Yes
- **Number of Employees:** Unknown
- **Compliance Requirements:** ['FAA Regulations']
- **Technical Ability:** Medium
- **Authentication Method:** N/A
- **Selected Technologies:** Linux Kernel
- **Selected Versions:** Linux Kernel

The system is a Drone as a Service (DaaS) platform, which provides drone-based services for various sectors including emergency response, public safety, and logistics. The DaaS platform leverages Unmanned Aerial Vehicles (UAVs) equipped with advanced technologies such as sensors, cameras, GPS, and communication modules. The key components of the system include:

- **Unmanned Aerial Vehicles (UAVs)**: These drones are the primary devices used for various missions, equipped with embedded systems and firmware to control their operations. They may perform tasks such as surveillance, delivery, and data collection.
- **Firmware and Embedded Systems**: The drones operate using firmware that manages their hardware components and ensures proper functionality. Firmware security is critical as vulnerabilities in the software could expose the drones to remote hijacking or data tampering.
- **Ground Control Station (GCS)**: This is the central hub for managing and controlling the drones. Operators can monitor drone activities, issue commands, and manage data collected by the drones through the GCS. It is connected to the drones via secure communication channels.
- **Cloud Services**: The DaaS platform utilizes cloud-based services for data storage, processing, and analysis. The cloud infrastructure supports functions such as long-term data storage, large-scale analytics, and coordination between multiple drones and ground stations.
- **Data Communication**: The system relies on secure communication protocols to transmit data between drones, the GCS, and cloud services. The integrity and security of these communication channels are crucial to prevent interception and tampering of critical data. The DaaS platform is designed to offer flexible and scalable drone services across various sectors. It incorporates multiple layers of technology, including UAV hardware, embedded systems, cloud infrastructure, and secure communication protocols to deliver reliable and efficient services.

### Improvement Suggestions

- Provide detailed list of authentication methods including credentials, multi-factor authentication, and security tokens.
- Specify version details and configurations for all software including drone firmware and cloud services.
- Include descriptions of any security policies or measures in place for data access, encryption standards, and communication protocols.
- Detail the security measures implemented for each component, such as encryption methods, authentication protocols, and physical security controls.
- State whether penetration tests or security audits have been conducted and summarized their findings.

## STRIDE Threat Model

| Threat Type | Scenario | Assumptions | Potential Impact |
|---|---|---|---|
| Spoofing | An attacker impersonates a drone operator to gain control of UAVs. | • No strict authentication for operators. (Role: Attacker, Condition: Operators have weak or no authentication mechanisms.)<br>• Access to communication channels. (Role: Attacker, Condition: Attacker intercepts communication or uses social engineering.) | Unauthorized control of UAVs affects service integrity and confidentiality. |
| Spoofing | An attacker spoofs GPS signals to redirect a drone. | • Drones rely solely on GPS for navigation. (Role: Attacker, Condition: No additional verification of location data is implemented.)<br>• Proximity to drone's operational space. (Role: Attacker, Condition: Physical proximity to the area of drone operation.) | Disruption of navigation leads to lost drones and service interruption. |
| Spoofing | Phishing attack leading to credential theft from GCS operators. | • Operators use unvetted third-party email clients. (Role: Attacker, Condition: Lack of email security training for the staff.)<br>• Sensitive data accessible via email credentials. (Role: Attacker, Condition: Critical system access through compromised credentials.) | Credential theft leads to unauthorized data access and mission disruption. |
| Tampering | Firmware on drones is maliciously altered to change drone behavior. | • Access to drone firmware update process. (Role: Attacker, Condition: Weak protection of the update process.)<br>• Firmware is not signed or verified post-update. (Role: Attacker, Condition: Lacking digital signature verification.) | Compromised firmware causes performance failures and data breaches. |
| Tampering | Data communication channels are altered leading to misinformation. | • Unsecured communication protocols. (Role: Attacker, Condition: Insufficient encryption or integrity checks during transmission.)<br>• Access to network traffic. (Role: Attacker, Condition: Proximity and ability to intercept the communication.) | Corrputed data misguides decision-making disrupting operations. |
| Tampering | An attacker modifies stored data in the cloud infrasture. | • Weak access controls on cloud services. (Role: Attacker, Condition: No stringent identity and access management policies.)<br>• Presence of exploitable cloud vulnerabilities. (Role: Attacker, Condition: Vulnerabilities in cloud configurations or software.) | Altered cloud data results in misinformation and potential data breaches. |
| Repudiation | An operator's command logs are missing due to inadequate logging. | • Lack of comprehensive logging mechanisms. (Role: Application, Condition: Insufficient audits trail for critical actions.)<br>• Log tampering goes undetected. (Role: Attacker, Condition: Absence of tamper-proof log systems.) | Difficulties in forensic investigation and operational accountability. |
| Repudiation | Attackers perform unauthorized actions and deny responsibility. | • No digital signature on critical commands. (Role: Application, Condition: Command execution without verification.)<br>• Inadequate user action monitoring. (Role: System Administrator, Condition: Absence of robust monitoring tools.) | Compromised accountability leads to untraceable unauthorized actions. |
| Information Disclosure | Sensitive data is exposed through unsecured cloud storage. | • Default cloud storage settings are insecure. (Role: Cloud Provider, Condition: Inadequate securing of storage permissions.)<br>• Sensitive data not encrypted at rest. (Role: Data Handler, Condition: Encryption policies are not enforced.) | Exposure of sensitive operation data breaches confidentiality. |
| Information Disclosure | Intercepted communication leads to leakage of operational data. | • Weak encryption on communication channels. (Role: Application Architect, Condition: Inadequate encryption protocols for transmission.)<br>• Advancement in interception tools. (Role: Attacker, Condition: High sophistication tools present in threat landscape.) | Operational plans leak to adversaries affecting mission integrity. |
| Denial of Service | Overloading the GCS with traffic prevents normal operational access. | • GCS lacks robust anti-DDoS protections. (Role: Defender, Condition: Absence of traffic analysis or limiting measures.)<br>• Sufficient resources to launch large-scale attack. (Role: Attacker, Condition: Attacker has access to botnet resources.) | Disruption of drone operations due to unavailability of GCS. |
| Denial of Service | Attackers exploit firmware vulnerabilities to crash UAVs. | • Known firmware vulnerabilities present. (Role: Attacker, Condition: Exploitable vulnerabilities without patches.)<br>• Remote access to UAV systems. (Role: Attacker, Condition: Access through unauthorized channels or compromised nodes.) | Operational drones crash leading to mission failure and potential hazards. |
| Elevation of Privilege | Exploitation of Linux Kernel vulnerabilities for unauthorized control access. | • Vulnerabilities unpatched in kernel system. (Role: Attacker, Condition: System updates neglected or irregularly applied.)<br>• System privileges inadequately segregated. (Role: System Administrator, Condition: Lack of least privilege enforcements.) | Unauthorized privilege gain resulting in full system control. |

### MITRE ATT&CK

### Threat: Spoofing **Scenario**: An attacker impersonates a drone operator to gain control of UAVs.
**Potential Impact**: Unauthorized control of UAVs affects service integrity and confidentiality.

**MITRE ATT&CK Techniques**

**Name**: Adversary-in-the-Middle

- **URL**: https://attack.mitre.org/techniques/T1557/
- **Technique ID**: T1557
- **Attack Pattern ID**: attack-pattern--035bb001-ab69-4a0b-9f6c-2de8b09e1b9d

---

### Threat: Spoofing

**Scenario**: An attacker spoofs GPS signals to redirect a drone.
**Potential Impact**: Disruption of navigation leads to lost drones and service interruption.

**MITRE ATT&CK Techniques**

**Name**: Adversary-in-the-Middle

- **URL**: https://attack.mitre.org/techniques/T1557/
- **Technique ID**: T1557
- **Attack Pattern ID**: attack-pattern--035bb001-ab69-4a0b-9f6c-2de8b09e1b9d

---

### Threat: Spoofing

**Scenario**: Phishing attack leading to credential theft from GCS operators.
**Potential Impact**: Credential theft leads to unauthorized data access and mission disruption.

**MITRE ATT&CK Techniques**

**Name**: Spearphishing Attachment

- **URL**: https://attack.mitre.org/techniques/T1566/001/
- **Technique ID**: T1566.001
- **Attack Pattern ID**: attack-pattern--2e34237d-8574-43f6-aace-ae2915de8597

---

### Threat: Tampering

**Scenario**: Firmware on drones is maliciously altered to change drone behavior.
**Potential Impact**: Compromised firmware causes performance failures and data breaches.

**MITRE ATT&CK Techniques**

**Name**: System Firmware

- **URL**: https://attack.mitre.org/techniques/T1542/001/
- **Technique ID**: T1542.001
- **Attack Pattern ID**: attack-pattern--16ab6452-c3c1-497c-a47d-206018ca1ada

---

### Threat: Tampering

**Scenario**: Data communication channels are altered leading to misinformation.
**Potential Impact**: Corrputed data misguides decision-making disrupting operations.

**MITRE ATT&CK Techniques**

**Name**: Transmitted Data Manipulation

- **URL**: https://attack.mitre.org/techniques/T1493/
- **Technique ID**: T1493
- **Attack Pattern ID**: attack-pattern--cc1e737c-236c-4e3b-83ba-32039a626ef8

---

### Threat: Tampering

**Scenario**: An attacker modifies stored data in the cloud infrasture.
**Potential Impact**: Altered cloud data results in misinformation and potential data breaches.

**MITRE ATT&CK Techniques**

**Name**: Data Destruction

- **URL**: https://attack.mitre.org/techniques/T1485/
- **Technique ID**: T1485
- **Attack Pattern ID**: attack-pattern--d45a3d09-b3cf-48f4-9f0f-f521ee5cb05c

---

### Threat: Repudiation

**Scenario**: An operator's command logs are missing due to inadequate logging.
**Potential Impact**: Difficulties in forensic investigation and operational accountability.

**MITRE ATT&CK Techniques**

**Name**: Clear Linux or Mac System Logs

- **URL**: https://attack.mitre.org/techniques/T1070/002/
- **Technique ID**: T1070.002
- **Attack Pattern ID**: attack-pattern--2bce5b30-7014-4a5d-ade7-12913fe6ac36

---

### Threat: Repudiation

**Scenario**: Attackers perform unauthorized actions and deny responsibility.
**Potential Impact**: Compromised accountability leads to untraceable unauthorized actions.

**MITRE ATT&CK Techniques**

**Name**: Valid Accounts

- **URL**: https://attack.mitre.org/techniques/T1078/
- **Technique ID**: T1078
- **Attack Pattern ID**: attack-pattern--b17a1a56-e99c-403c-8948-561df0cffe81

---

### Threat: Information Disclosure

**Scenario**: Sensitive data is exposed through unsecured cloud storage.
**Potential Impact**: Exposure of sensitive operation data breaches confidentiality.

**MITRE ATT&CK Techniques**

**Name**: Unknown

- **URL**: https://attack.mitre.org/techniques/N/A/
- **Technique ID**: N/A
- **Attack Pattern ID**: attack-pattern--00000000-0000-0000-0000-000000000000

---

### Threat: Information Disclosure

**Scenario**: Intercepted communication leads to leakage of operational data.
**Potential Impact**: Operational plans leak to adversaries affecting mission integrity.

**MITRE ATT&CK Techniques**

**Name**: Traffic Duplication

- **URL**: https://attack.mitre.org/techniques/T1020/001/
- **Technique ID**: T1020.001
- **Attack Pattern ID**: attack-pattern--7c46b364-8496-4234-8a56-f7e6727e21e1

---

**Name**: Unknown

- **URL**: https://attack.mitre.org/techniques/N/A/
- **Technique ID**: N/A
- **Attack Pattern ID**: attack-pattern--c0eb592e-31f3-4fa2-8d3f-523471df9c64

---

### Threat: Denial of Service

**Scenario**: Overloading the GCS with traffic prevents normal operational access.
**Potential Impact**: Disruption of drone operations due to unavailability of GCS.

**MITRE ATT&CK Techniques**

**Name**: Network Denial of Service

- **URL**: https://attack.mitre.org/techniques/T1498/
- **Technique ID**: T1498
- **Attack Pattern ID**: attack-pattern--d74c4a7e-ffbf-432f-9365-7ebf1f787cab

---

### Threat: Denial of Service

**Scenario**: Attackers exploit firmware vulnerabilities to crash UAVs.
**Potential Impact**: Operational drones crash leading to mission failure and potential hazards.

**MITRE ATT&CK Techniques**

**Name**: Unknown

- **URL**: https://attack.mitre.org/techniques/N/A/
- **Technique ID**: N/A
- **Attack Pattern ID**: attack-pattern--00000000-0000-0000-0000-000000000000

---

### Threat: Elevation of Privilege

**Scenario**: Exploitation of Linux Kernel vulnerabilities for unauthorized control access.
**Potential Impact**: Unauthorized privilege gain resulting in full system control.

**MITRE ATT&CK Techniques**

## Mitigations

| Threat Type | Scenario | Suggested Mitigation(s) |
|---|---|---|
| Spoofing | An attacker impersonates a drone operator to gain control of UAVs. | Implement multi-factor authentication (MFA) for all operators. Encrypt communication channels between operators and drones to prevent interception. Regularly update and patch systems to protect against known vulnerabilities. |
| Spoofing | An attacker spoofs GPS signals to redirect a drone. | Integrate redundant navigation systems (e.g., inertial measurement units or ground-based signals) to verify GPS data. Employ GPS signal authentication mechanisms and continually monitor GPS signal integrity. |
| Spoofing | Phishing attack leading to credential theft from GCS operators. | Conduct regular phishing awareness training for operators. Implement email filtering solutions to detect and block phishing attempts. Use MFA for access to critical systems. |
| Tampering | Firmware on drones is maliciously altered to change drone behavior. | Require digital signature verification for all firmware updates. Harden the firmware update process with encryption and access controls. Regularly audit firmware integrity through automated and manual checks. |
| Tampering | Data communication channels are altered leading to misinformation. | Utilize end-to-end encryption and integrity checks, such as hashing, for all communications. Set up intrusion detection systems to alert on suspicious activities in communication channels. |

| Tampering | An attacker modifies stored data in the cloud infrastructure. | Enforce strict identity and access management policies, including role-based access controls. Regularly audit access logs and cloud configurations for anomalies. Apply data integrity checks and employ encryption at rest and in transit. |
|---|---|---|
| Repudiation | An operator's command logs are missing due to inadequate logging. | Implement comprehensive logging policies to capture all critical actions. Use tamper-evident log solutions, such as secure, centralized logging servers with checksum validation. Conduct regular log audits to ensure completeness and accuracy. |
| Repudiation | Attackers perform unauthorized actions and deny responsibility. | Digitally sign all commands and use strong authentication methods to ensure command authenticity. Enhance user activity monitoring and provide regular accountability reviews. |
| Information Disclosure | Sensitive data is exposed through unsecured cloud storage. | Apply strict access controls and encryption for cloud storage. Use cloud security posture management tools to ensure storage configurations comply with security policies. Conduct periodic penetration tests and vulnerability assessments. |
| Information Disclosure | Intercepted communication leads to leakage of operational data. | Use strong encryption protocols, such as TLS, for all data communications. Conduct regular threat assessments to identify emerging interception technologies and adapt defensive measures. |
| Denial of Service | Overloading the GCS with traffic prevents normal operational access. | Employ proper network segmentation and DDoS protection services to absorb and mitigate attack traffic. Implement rate limiting and anomaly detection to monitor and respond to abnormal traffic patterns. |
| Denial of Service | Attackers exploit firmware vulnerabilities to crash UAVs. | Regularly update and patch UAV firmware to address vulnerabilities. Establish an incident response plan specific to UAV operations to quickly isolate and recover compromised systems. |
| Elevation of Privilege | Exploitation of Linux Kernel vulnerabilities for unauthorized control access. | Keep the Linux kernel and associated packages up-to-date with the latest patches. Enforce the principle of least privilege for all user accounts. Perform regular security audits and assessments to identify and rectify privilege escalation vectors. |

### DREAD Risk Assessment

| Threat Type | Scenario | Damage Potential | Reproducibility | Exploitability | Affected Users | Discoverability | Risk Score |
|---|---|---|---|---|---|---|---|
| Spoofing | An attacker impersonates a drone operator to gain control of UAVs. | 9 | 6 | 8 | 8 | 7 | 7.60 |
| Spoofing | An attacker spoofs GPS signals to redirect a drone. | 8 | 5 | 7 | 9 | 6 | 7.00 |
| Spoofing | Phishing attack leading to credential theft from GCS operators. | 8 | 7 | 8 | 9 | 5 | 7.40 |
| Tampering | Firmware on drones is maliciously altered to change drone behavior. | 9 | 6 | 7 | 8 | 6 | 7.20 |
| Tampering | Data communication channels are altered leading to misinformation. | 8 | 5 | 6 | 7 | 5 | 6.20 |
| Tampering | An attacker modifies stored data in the cloud infrastructure. | 8 | 6 | 7 | 8 | 5 | 6.80 |
| Repudiation | An operator's command logs are missing due to inadequate logging. | 6 | 4 | 5 | 6 | 4 | 5.00 |
| Repudiation | Attackers perform unauthorized actions and deny responsibility. | 7 | 6 | 6 | 7 | 6 | 6.40 |
| Information Disclosure | Sensitive data is exposed through unsecured cloud storage. | 8 | 5 | 6 | 9 | 7 | 7.00 |
| Information Disclosure | Intercepted communication leads to leakage of operational data. | 7 | 6 | 7 | 8 | 5 | 6.60 |
| Denial of Service | Overloading the GCS with traffic prevents normal operational access. | 9 | 8 | 8 | 8 | 5 | 7.60 |
| Denial of Service | Attackers exploit firmware vulnerabilities to crash UAVs. | 9 | 6 | 7 | 8 | 6 | 7.20 |
| Elevation of Privilege | Exploitation of Linux Kernel vulnerabilities for unauthorized control access. | 9 | 5 | 7 | 7 | 6 | 6.80 |

## Attack Tree

**Attack Tree diagram instructions**: Copy the below code and paste it into https://mermaid.live/

```
graph LR
   A["Compromise of DaaS Application (CIA)"] --> B(Spoofing)
   A --> C(Tampering)
   A --> D(Repudiation)
   A --> E["Information Disclosure"]
   A --> F["Denial of Service (DoS)"]
   A --> G["Elevation of Privilege"]

   %% Subgraph for Spoofing Threats
   subgraph Spoofing Threats
       B(Operator Spoofing)
       B --> B1["Spoofing UAV Operators"]
       B --> B2["GPS Spoofing"]
       B --> B3["Credential Theft via Phishing"]

       %% Detailed threats for Spoofing UAV Operators
       B1 --> B1a["Adversary-in-the-Middle (T1557): Intercept communication"]
       B1a --> B1b["Network Sniffing"]
       B1a --> B1c["Credential Access"]

       %% Detailed threats for GPS Spoofing
       B2 --> B2a["Adversary-in-the-Middle (T1557): Spoof GPS signals"]
       B2a --> B2b["Network Traffic Manipulation"]

       %% Detailed threats for Credential Theft via Phishing
       B3 --> B3a["Spearphishing Attachment (T1566.001)"]
       B3a --> B3b["Credential Theft"]
   end

   %% Subgraph for Tampering Threats
   subgraph Tampering Threats
       C(UAV Firmware and Data Tampering)
       C --> C1["Firmware Tampering"]
       C --> C2["Communication Alteration"]
       C --> C3["Cloud Data Manipulation"]

       %% Detailed threats for Firmware Tampering
       C1 --> C1a["System Firmware (T1542.001): Alter firmware to change behavior"]

       %% Detailed threats for Communication Alteration
       C2 --> C2a["Transmitted Data Manipulation (T1493)"]
       C2a --> C2b["Insufficient encryption or integrity checks"]
```

```
    %% Detailed threats for Cloud Data Manipulation
    C3 --> C3a["Data Destruction (T1485): Modify stored data']
end

%% Subgraph for Repudiation Threats
subgraph Repudiation Threats
    D(Command and Action Repudiation)
    D --> D1["Missing Command Logs"]
    D --> D2["Denial of Unauthorized Actions"]

    %% Detailed threats for Missing Command Logs
    D1   > D1a["Clear Linux or Mac System Logs (T1070.002)"]

    %% Detailed threats for Denial of Unauthorized Actions
    D2 --> D2a["Valid Accounts (T1078): Perform actions with stolen credentials"]
end

%% Subgraph for Information Disclosure Threats
subgraph Information Disclosure Threats
    E['Leakage of Sensitive Data']
    E --> E1["Unsecured Cloud Storage"]
    E --> E2["Intercepted Communication"]

    %% Detailed threats for Unsecured Cloud Storage
    E1 --> E1a["Storage Exposure']

    %% Detailed threats for Intercepted Communication
    E2 --> E2a["Traffic Duplication (T1020.001)']
end

%% Subgraph for Denial of Service Threats
subgraph Denial of Service Threats
    F('Service Disruption")
    F --> F1["Network DDoS on GCS']
    F --> F2["Firmware Vulnerability Exploitation"]

    %% Detailed threats for Network DDoS on GCS
    F1 --> F1a["Network Denial of Service (T1498): Overwhelm GCS"]

    %% Detailed threats for Firmware Vulnerability Exploitation
    F2 --> F2a["Exploit firmware vulnerabilities"]
end

%% Subgraph for Elevation of Privilege Threats
subgraph Elevation of Privilege Threats
    G(Unlawful Access)
```

```
    G --> G1["Linux Kernel Exploitation"]

    %% Detailed threats for Linux Kernel Exploitation
    G1 --> G1a["Exploit kernel vulnerabilities for privilege escalation"]
end
```

## Test Cases

For the history of Behavior Driven Development (BDD) and Gherkin syntax, see: https://cucumber.io/docs/bdd/history/

---

```
# Test Case: UAV Operator Authentication

Feature: Spoofing Authentication
Scenario: An attacker impersonates a drone operator to gain control of UAVs
Given operators have weak or no authentication mechanisms
And an attacker intercepts communication or uses social engineering
When the attacker attempts to impersonate a drone operator
Then the attacker gains unauthorized control of the UAVs
And the service integrity and confidentiality are compromised

# Test Case: GPS Signal Spoofing
Feature: GPS Navigation Spoofing
Scenario: An attacker spoofs GPS signals to redirect a drone
Given drones rely solely on GPS for navigation
And the attacker is in proximity to the drone's operational space
When the attacker spoofs GPS signals
Then the drone's navigation is disrupted
And this leads to lost drones and service interruption

# Test Case: Phishing for Credentials
Feature: Operator Credential Phishing
Scenario: Phishing attack leading to credential theft from GCS operators
Given operators use unvetted third-party email clients
And critical system access is possible through compromised credentials
When an attacker successfully conducts a phishing attack
Then credentials are stolen
And unauthorized data access and mission disruption occur

# Test Case: Firmware Tampering
Feature: Drone Firmware Alteration
Scenario: Firmware on drones is maliciously altered
Given weak protection of the firmware update process
And lack of digital signature verification
When the attacker alters the firmware
Then the drone behavior changes
And performance failures and data breaches occur

# Test Case: Data Communication Tampering
Feature: Communication Channel Tampering
```

```gherkin
Scenario: Data communication channels are altered
Given insufficient encryption or integrity checks during transmission
And the attacker can intercept communication
When the attacker alters the data in communication channels
Then the data becomes corrupted
And operational decision-making is disrupted

# Test Case: Cloud Data Tampering
Feature: Cloud Data Modification
Scenario: An attacker modifies stored data in the cloud infrastructure
Given no stringent identity and access management policies
And exploitable vulnerabilities in cloud configurations
When the attacker modifies the stored data
Then misinformation occurs
And potential data breaches are possible

# Test Case: Missing Command Logs
Feature: Inadequate Command Logging
Scenario: Operator's command logs are missing
Given insufficient audits trail for critical actions
And absence of tamper-proof log systems
When command logs are needed for investigation
Then forensic difficulties occur
And operational accountability is compromised

# Test Case: Unauthorized Actions and Repudiation
Feature: Action Accountability
Scenario: Attackers perform unauthorized actions and deny responsibility
Given command execution without verification
And inadequate user action monitoring
When unauthorized actions are performed
Then accountability is compromised
And actions remain untraceable

# Test Case: Cloud Storage Security
Feature: Unsecured Cloud Storage
Scenario: Sensitive data is exposed through unsecured cloud storage
Given inadequate securing of storage permissions
And sensitive data is not encrypted at rest
When an attacker gains access to cloud storage
Then sensitive operation data is exposed
And confidentiality is breached

# Test Case: Communication Interception
Feature: Communication Data Leakage
Scenario: Intercepted communication leads to data leakage
Given inadequate encryption protocols
```

```
And sophisticated interception tools present
When communication is intercepted
Then operational plans leak
And mission integrity is affected

# Test Case: GCS Denial of Service
Feature: GCS Overloading
Scenario: Overloading the GCS with traffic prevents operational access
Given absence of traffic analysis or limiting measures
And attacker has access to botnet resources
When an attacker launches a DDoS attack on GCS
Then drone operations are disrupted
And GCS becomes unavailable

# Test Case: UAV Firmware Exploitation
Feature: UAV Crash Exploitation
Scenario: Attackers exploit firmware vulnerabilities to crash UAVs
Given exploitable firmware vulnerabilities without patches
And access through unauthorized channels
When the attacker exploits these vulnerabilities
Then operational drones crash
And mission failure and hazards result

# Test Case: Privilege Elevation via Kernel Exploit
Feature: Linux Kernel Vulnerability
Scenario: Exploitation of Linux Kernel vulnerabilities for unauthorized control access
Given system updates are irregularly applied
And lack of least privilege enforcements
When the attacker exploits kernel vulnerabilities
Then unauthorized privilege gain occurs
And full system control is achieved
```